\documentclass{aa}  
\newcommand{\ratio}{0.9}
\newcommand{\wi}{0.4}
\usepackage{graphicx}
\usepackage{txfonts}
\usepackage{siunitx}
\usepackage{natbib}
\usepackage{lipsum}
\usepackage{rotating}
\usepackage[breaklinks=true,colorlinks=true,linkcolor=blue,citecolor=blue]{hyperref}
\usepackage{placeins}
\usepackage{soul} 
\usepackage{graphicx}
\bibpunct{(}{)}{;}{a}{}{,}

\newcommand{\li}[3]{\ion{#1}{#2}~$\lambda$#3\,$\angstrom$}
\newcommand{\liu}[3]{\underline{\ion{#1}{#2}~$\lambda$#3\,$\angstrom$}}
\newcommand{\Teff}{\ensuremath{T_\mathrm{eff}}}
\newcommand{\msun}{\ensuremath{\mathrm{M}_\sun}}
\newcommand{\lsun}{\ensuremath{\mathrm{L}_\sun}}

\newcommand{\rsun}{\ensuremath{\mathrm{R}_\sun}}

\newcommand{\rstar}{\ensuremath{R_*}}

\newcommand{\mstar}{\ensuremath{M_*}}
\newcommand{\lstar}{\ensuremath{L_*}}

\newcommand{\msunyr}{{\ensuremath{\msun}/\mathrm{yr}}}

\newcommand{\mdot}{\ensuremath{\dot M}}
\newcommand{\second}{\mbox{s}}
\newcommand{\kms}{\ensuremath{\mbox{km}/\second}}
\newcommand{\angstrom}{\text{\normalfont\AA}}
\newcommand{\cmss}{\ensuremath{\mbox{cm}/\second^{2}}}
\newcommand{\Kelvin}{\ensuremath{\mbox{K}}}

\newcommand{\PoWR}{{\tt PoWR}}

\newcommand{\YC}{\ensuremath{Y_\mathrm{C}}}
\newcommand{\YS}{\ensuremath{Y_\mathrm{S}}}
\newcommand{\mini}{\ensuremath{M_\mathrm{ini}}}

\newcommand{\hzav}[1]{\left[#1\right]}
\defcitealias{Szecsi:2015}{Paper~I}
\defcitealias{Kubatova:2019}{Paper~II}
\usepackage[usenames,dvipsnames,svgnames,table]{xcolor}
\definecolor{gold}{HTML}{FFD700}
\definecolor{crimson}{HTML}{DC143C}
\definecolor{violet}{HTML}{FF1493}	
\definecolor{bestmatch}{HTML}{fff9d6}

\begin{document}

   \title{Low-metallicity massive single stars with rotation}

   \subtitle{III. Source of ionization and \ion{C}{IV} emission in I~Zw~18}

	\titlerunning{Low-metallicity massive single stars with rotation. III. Source of ionization and \ion{C}{IV} emission in I~Zw~18}

\author{Dorottya Sz\'ecsi\inst{\ref{inst1}}
	\and
	Frank Tramper\inst{\ref{cab}, \ref{inst2}}
	\and
	Brankica Kub\'atov\'a\inst{\ref{inst3}}	
	\and
	Carolina Kehrig\inst{\ref{inst4},\ref{newinst}}
	\and
	Ji\v{r}\'{\i} Kub\'at\inst{\ref{inst3}}
	\and 
	Ji\v{r}\'{\i} Krti\v{c}ka\inst{\ref{inst5}}
	\and
	Andreas~A.C.~Sander\inst{\ref{inst6},\ref{inst:iwr}}
	\and		 
	Miriam Garcia\inst{\ref{cab}}
}

\institute{
	Institute of Astronomy – Faculty of Physics, Astronomy and Informatics – Nicolaus Copernicus University, Grudzi\k{a}dzka 5, 87-100 Toru\'n, Poland \email{dorottya.szecsi@gmail.com}\label{inst1}
	\and
        Centro de Astrobiolog{\'i}a (CAB), CSIC-INTA, Carretera de Ajalvir km 4, E-28850 Torrej{\'o}n de Ardoz, Madrid, Spain\label{cab}
        \and
	Institute of Astronomy, KU Leuven, Celestijnenlaan 200D, B-3001 Leuven, Belgium\label{inst2}
	\and
	Astronomick\'y \'ustav, Akademie v\v{e}d \v{C}esk\'e republiky,
Fri\v{c}ova 298, 251~65 Ond\v{r}ejov, Czech Republic
\email{brankica.kubatova@asu.cas.cz}\label{inst3}
	\and
	Instituto de Astrof\'isica de Andaluc\'ia (IAA/CSIC), Glorieta de la Astronom\'ia s/n Aptdo. 3004, E-18080 Granada, Spain\label{inst4}	
    \and 
    {Observat\'orio Nacional/MCTIC, R. Gen. Jos\'e Cristino, 77, 20921-400, Rio de Janeiro, Brazil\label{newinst}}
	\and
	{\'{U}stav teoretick\'e fyziky a astrofyziky, Masarykova univerzita, Kotl\'a\v rsk\'a 267/2, 611 37, Brno, Czech Republic\label{inst5}}
	\and
	{Zentrum f{\"u}r Astronomie der Universit{\"a}t Heidelberg, Astronomisches Rechen-Institut, M{\"o}nchhofstr. 12-14, 69120 Heidelberg, Germany\label{inst6}}
    \and
    {Interdisziplin{\"a}res Zentrum f{\"u}r Wissenschaftliches Rechnen, Universit{\"a}t Heidelberg, Im Neuenheimer Feld 225, 69120 Heidelberg, Germany\label{inst:iwr}}
}

   \date{Received 4 Oct 2024; accepted 1 Jul 2025}

 
\abstract{Chemically homogeneously evolving stars have been proposed to account for several exotic phenomena, including gravitational-wave emissions, gamma-ray bursts and certain types of supernovae.}{Here we study whether these stars can explain the observations of the metal-poor star-forming dwarf galaxy, I~Zwicky~18.}{We apply our synthetic spectral models from Paper~II to (i) establish a classification sequence for these hot stars, (ii) predict the photonionizing flux and the strength of observable emission lines from a I~Zw~18-like stellar population, and (iii) compare our predictions to all available observations of this galaxy.}{Adding two new models computed with \PoWR, we report that (i) these stars follow a unique sequence of classes: O $\rightarrow$ WN $\rightarrow$ WO (i.e. without ever being WC). From our population synthesis with standard assumptions, we predict that (ii) the source of the UV \li{C}{IV}{1550}
and other emission bumps is a couple of dozen WO-type Wolf--Rayet stars (not WC as previously assumed) which are the result of chemically homogeneous evolution, while these, combined with the rest of the O-star population, account for the high \ion{He}{II} ionizing flux and the spectral hardness.
{Contrasting our results against published optical and UV data from the literature and accounting for different aperture sizes and spatial regions probed by the observations, we find that (iii) our models are highly consistent with existing measurements.}}{Since our “massive Pop~II stars” might just as well exist in early star-forming regions, our findings have implications for upcoming 
James Webb Space Telescope (JWST)
surveys: the first galaxies in the high-redshift Universe may also experience the extra contribution of UV photons and the kinds of exotic explosions that chemically homogeneous stellar evolution predicts. {Given that our results apply for binary populations too as long as the same fraction (10\%) of the systems evolves chemically homogeneously, we conclude that the stellar progenitors of gravitational waves may very well exist today in I~Zw~18.}
}


   \maketitle
%

\section{Introduction}\label{sec:intro}

Dwarf starburst galaxies form stars at a high rate, and thus harbour large populations of metal-poor massive stars \citep[e.g.][]{Zhao:2013}. Despite the recent burgeoning of interest in these stars' lives \citep[][]{Weisz:2014,Izotov:2016,Schootemeijer:2018,Evans:2019,Garcia:2019,Groh:2019,Senchyna:2019,Franeck:2022,Szecsi:2022ok,Gull:2022,Lahen:2023} and deaths as exotic explosions \citep{Perley:2016,AguileraDena:2018,Stevenson:2019,Agrawal:2022}, and `afterlife' as -- potentially merging -- compact objects \citep{Eldridge:2016,Marchant:2016,deMink:2016,Mandel:2016,Marchant:2017,VignaGomez:2018,Eldridge:2019,Romagnolo:2022}, theoretical models are far from being properly tested observationally \citep{Mintz:2025}. 

\begin{figure*}[!t]\centering
	\includegraphics[width=1.7\columnwidth]{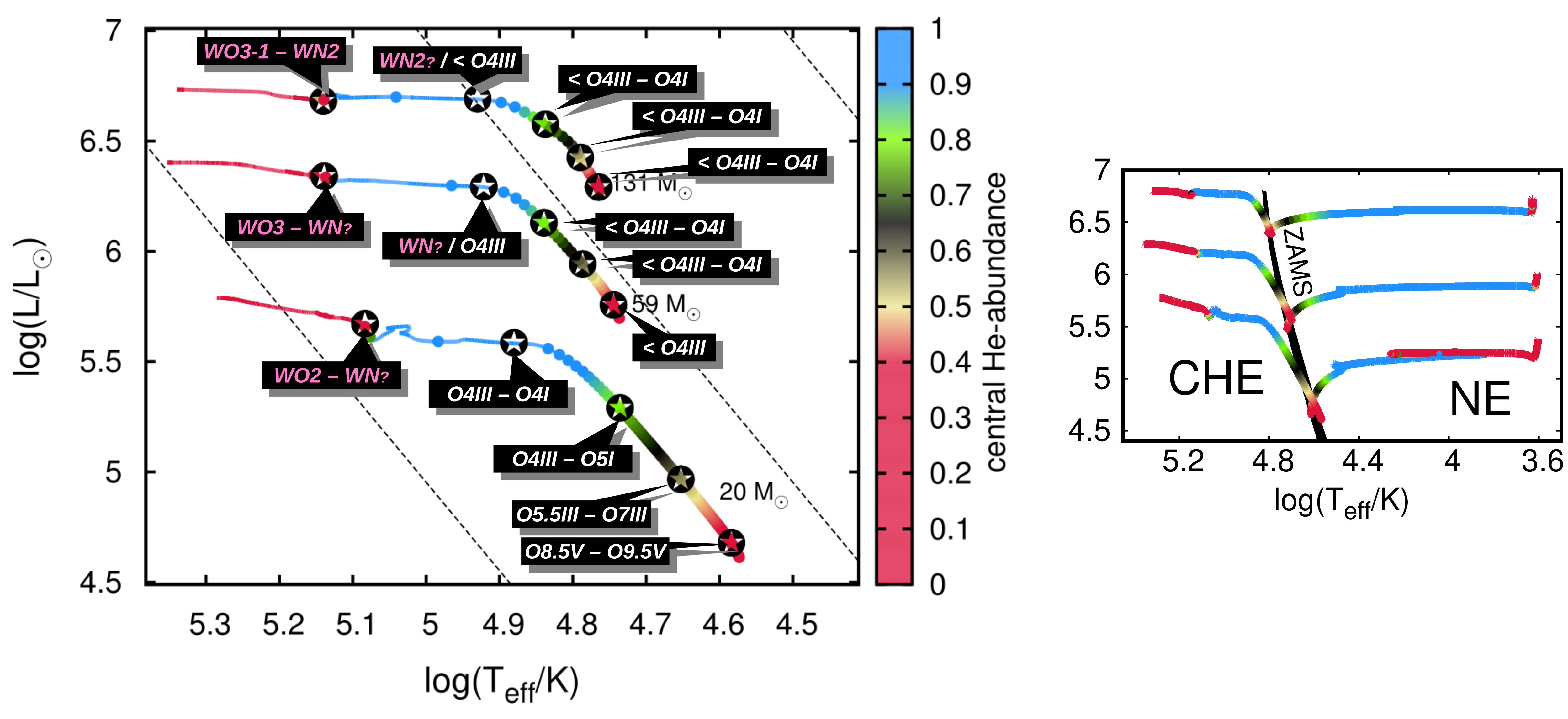}
	\caption{\textit{Left:} Hertzsprung--Russell diagram summarizing the findings of \citetalias{Kubatova:2019} (their Fig.~1 \& Table~4) and our Sect.~\ref{sec:newmodel}, showing that chemically homogeneous stellar evolution at I~Zw~18 composition proceeds from class~O via a shorter-lived class~WN to a longer-lived WO~phase -- i.e. without experiencing any long-term WC~phase. Evolutionary models are taken from \citetalias{Szecsi:2015} (main-sequence) and \citet[][post-main-sequence]{Szecsi:2022ok}, without correcting for the wind optical depth (see the last paragraph of Sect.~\ref{sec:newmodel}).
	Initial masses are labelled, showing where the tracks start
	their evolution, which proceeds towards the hot side of the diagram (i.e. CHE). Colours show the central helium mass fraction, and dots represent every	10$^5$~years of evolution. Dashes mark equiradial lines with 1, 10, and 100~R$_{\odot}$ from left to right. The black star symbols represent the models for which \PoWR\ synthetic spectra were computed in \citetalias{Kubatova:2019} and Sect.~\ref{sec:newmodel}. From right to left (note that the evolution progresses from red to blue, contrarily to classical massive-star evolution): main-sequence models with surface helium mass fractions of 0.28, 0.5, 0.75, and 0.98, while the fifth symbol on the very left corresponds to 
	the core-helium-burning phase (post-main-sequence, Sect.~\ref{sec:newmodel}).
    Spectral classification was performed in \citetalias{Kubatova:2019}; the results (for four different assumptions about mass loss and clumping) were presented in their Table~4 and Appendix~A. Here the labels summarize the ranges in class found there, except for the WN~and~WO~stars which are discussed in Sect.~\ref{sec:newmodel} (see also Fig.~\ref{fig:131evol}).  
    {\textit{Right:} HR diagram summarizing our findings from \citetalias{Szecsi:2015}. Stellar models computed with I~Zw~18's chemical composition are shown following normal evolution (NE) towards the red supergiant branch with slow rotation ($\lesssim$\,300\,km\,s$^{-1}$). With fast rotation, they are seen following CHE towards hot surface temperatures; that is, leftward from the ZAMS. This latter evolutionary path is what is elaborated upon in the left panel. For more details on how we construct a synthetic population out of both these kinds of models, see Sect~\ref{sec:population}.}
	}
	\label{fig:classification}
\end{figure*}

One convenient testbed for constraining our models is \textit{local} dwarf star-forming galaxies. These objects typically display low metallicities and high star formation rates \citep{Searle:1972,Hunter:1995,Vilchez:1998,Izotov:2002,Izotov:2004,Thuan:2005,Vaduvescu:2007,Shirazi:2012,Annibali:2013,Kehrig:2013,Kehrig:2018,Kehrig:2021,ArroyoPolonio:2024,Hirschauer:2024}, meaning that they are local equivalents for high-redshift star-forming regions. Studying them can shed light on the nature of metal-poor massive-star populations that presumably contribute to the structure of the galaxy with their radiative, chemical, and mechanical feedback. 

In the blue compact dwarf galaxy I~Zw~18 in particular, a high amount of photoionization has been derived from nebular line observations: \citet{Kehrig:2015} reported a \ion{He}{II} ionizing flux of Q$^{obs}_{\ion{He}{II}}$~$=$~1.33$\cdot$10$^{50}$~photons~s$^{-1}$ (see their Sect.\,3.2). I~Zw~18 has two large clusters: a north-western cluster and a south-western cluster. Of these, only the north-western cluster was spatially associated with this nebular emission; the total stellar mass of this region is about 300~000~M$_{\odot}$ \citep{Izotov:1998,Lebouteiller:2013}. The high 
\ion{He}{II} flux 
was puzzling because it could not be accounted for by the ionization from  conventional ionizing sources including standard hot massive stars (\citealt{Kehrig:2015}).

The hottest and most luminous massive stars are Wolf--Rayet (WR) stars: they are even stronger ionizing sources than O~stars \citep{Hamann:2015}. Due to their optically thick stellar winds, WR~stars also emit broad emission lines. One such line is the UV~line \li{C}{IV}{1550}, which has indeed been identified in the spectrum of I~Zw~18 \citep{Lebouteiller:2013}: in the north-western cluster, a line luminosity between L$^{obs}_{1550}$~$\sim$~2.2$-$5.5$\cdot$10$^{37}$~erg~s$^{-1}$ 
has been measured by \citet{Brown:2002}. 
The problem is that according to classical stellar theory, 
the number of WR stars derived from the carbon-line luminosity cannot account for the high \ion{He}{II} emission observed. As \citet{Kehrig:2015} report, to explain L$^{obs}_{1550}$ {(which they take to be to be 4.67$\cdot$10$^{37}$~erg~s$^{-1}$; see their Sect.~4.1)}, about nine classical Wolf--Rayet stars of type~WC (with standard WC-properties given by \citealt{Crowther:2006}, see also \citealt{LopezSanchez:2010a,LopezSanchez:2010b} and \citealt{Sander:2022IAU}) are required in the region: and the ionizing flux of only nine of them is $\sim$50~times lower than Q$^{obs}_{\ion{He}{II}}$.

Therefore, other candidates were suggested as the source of ionization. One option is that a population of peculiar (nearly) metal-free massive stars \citep[Pop~III-star siblings/Pop~III-like stars][]{Kehrig:2015,Kehrig:2015b,Heap:2015} formed recently from some leftover primordial gas in this galaxy. Population~III stars are expected to be hotter than metal-poor stars 
without developing carbon emission lines \citep{Yoon:2012}. This hypothesis relies on I~Zw~18 keeping, or somehow obtaining, pockets of primordial gas that have just recently started to form massive stars.

Another scenario was presented by \citet{Pequignot:2008}, \citet{Lebouteiller:2017}, and \citet{Schaerer:2019}, in which a population of X-ray binaries produce the ionizing radiation. 
However, \citet{Kehrig:2021} found that neither a high-mass X-ray binary nor the soft X-ray photons observed in the galaxy can account for the bulk of the nebular HeII emission, and \citet{Senchyna:2020} support this conclusion.
In turn, \citet{Oskinova:2022} suggest cluster winds and superbubbles as an alternative explanation.

\begin{figure*}[!t]
	\centering\includegraphics[height=\ratio\columnwidth,page=1,angle=270]{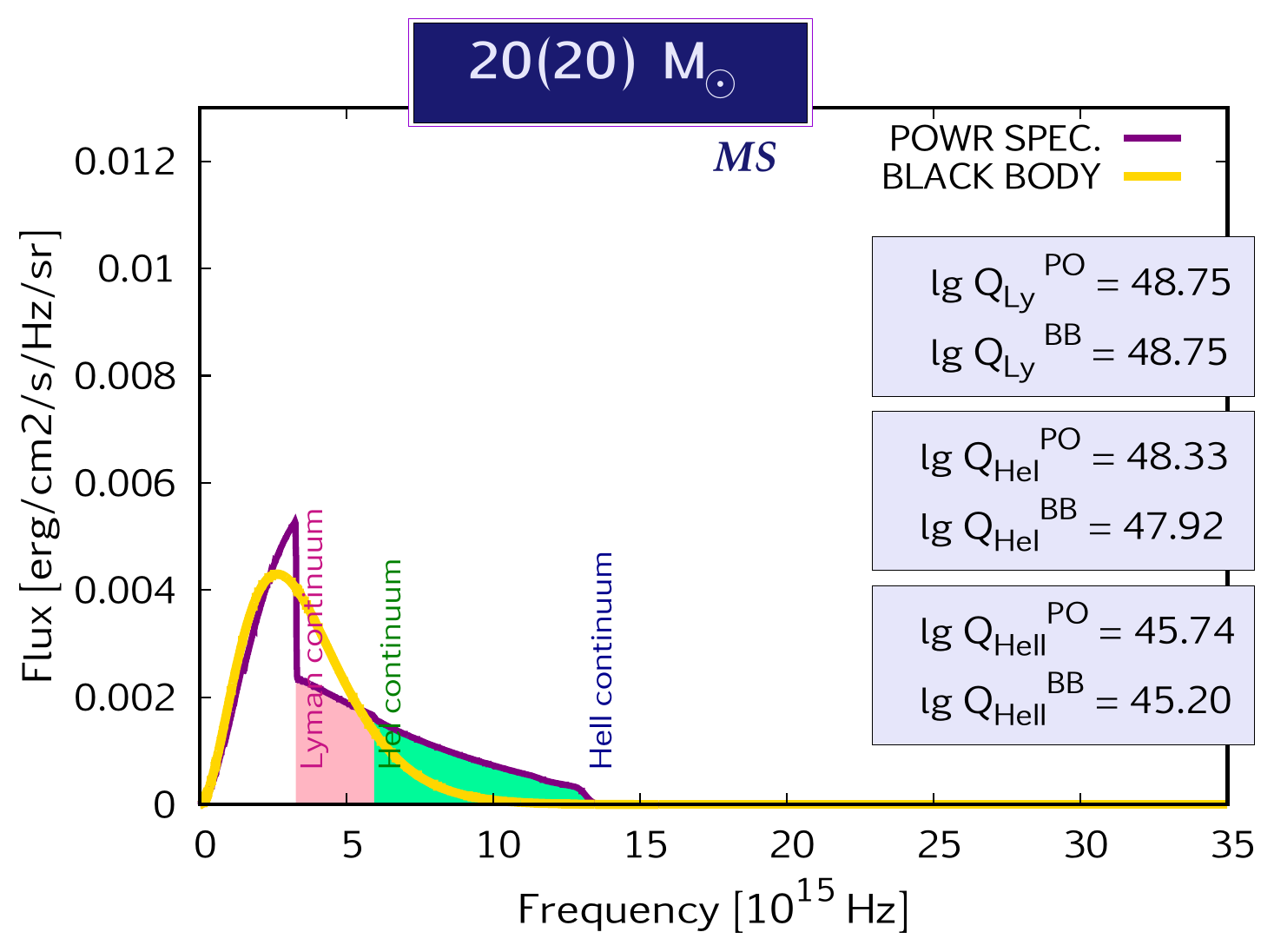}\hspace{25pt}
	\includegraphics[height=\ratio\columnwidth,page=2,angle=270]{pics/BBcompare}\\ \vspace{10pt}
	\includegraphics[height=\ratio\columnwidth,page=3,angle=270]{pics/BBcompare}\hspace{25pt}
	\includegraphics[height=\ratio\columnwidth,page=4,angle=270]{pics/BBcompare}\\\vspace{10pt}
	\includegraphics[height=\ratio\columnwidth,page=5,angle=270]{pics/BBcompare}\hspace{25pt}
	\includegraphics[height=\ratio\columnwidth,page=6,angle=270]{pics/BBcompare}
	\caption{Spectral energy distributions of our \PoWR~synthetic spectra of chemically homogeneously evolving massive stars published in \citetalias{Kubatova:2019} with the corresponding black body distributions overplotted for context. The models correspond to the second (MS) and last (pMS) phases on the HR~diagram in Fig.~\ref{fig:classification}; their main physical parameters and ionizing fluxes are listed in Table~\ref{tab:Q}.
	Titles indicate the initial mass of the evolutionary models with the {actual} mass (i.e. the mass when the synthetic spectra were computed) in parenthesis.
		The three main ionizing continua (Lyman:~$<$~912~$\angstrom$, \ion{He}{I}:~$<$~504~$\angstrom$, \ion{He}{II}:~$<$~228~$\angstrom$) are marked (note the X~scale showing the frequency instead of the wavelength, in order to spread out the relevant high-energy parts of the distributions).
		Framed boxes present the number of ionizing photons predicted by both the \PoWR\ models (unclumped, with nominal mass-loss rates) and the black body distributions, in units of log(s$^{-1}$); the values being close to each other is an interesting coincidence.
		\textit{Left column:} Models in the main-sequence (MS) phase with surface helium mass fraction of Y$_{\rm S}$~$=$~0.5. \textit{Right column:} Models in the post-main-sequence (pMS) phase (in the case of the 131~M$_{\odot}$ model, the late-pMS phase, see Sect.~\ref{sec:newmodel} and Fig.~\ref{fig:newmodel}). Note the order of magnitude difference in the Y~scale between the left column and the right column. 
	}\label{fig:BB}
\end{figure*}

Here we study yet another possibility. We simulated stellar evolutionary models of single massive stars with the exact composition of I~Zw~18 in \citet[][hereafter Paper~I]{Szecsi:2015}, using state-of-the-art physics. We found that these “massive Pop-II stars” mostly evolve the classical way (similar to how metal-rich massive stars in our Galaxy do) as long as their rotational rate is moderate. But those models that rotate fast follow another path: they evolve chemically homogeneously, and become hot and luminous stars -- similar to Wolf-Rayet stars but without the high wind-mass-loss and thus without the broad emission features during most of their lives. To analyse these objects further, we computed synthetic spectra for them in \citet[][hereafter Paper~II]{Kubatova:2019}. We dubbed them transparent wind UV-intense stars (in short TWUIN~stars) to distinguish them from traditional Wolf--Rayet stars with optically thick winds.

Below we investigate whether a metal-poor massive star population based on our evolutionary models from \citetalias{Szecsi:2015} and our spectral models from \citetalias{Kubatova:2019} can \textit{simultaneously} account for both the \ion{He}{II} ionizing flux and the strength of emission lines observed in I~Zw~18. 
Section~\ref{sec:theory} explains our modelling approach: from evolutionary models (Sect.~\ref{sec:models}) to synthetic spectra (Sect.~\ref{sec:newmodel}, including the addition of new \PoWR\ spectra and the newly established classification sequence for chemically homogeneously evolving stars) to population synthesis (Sect.~\ref{sec:population}) and emission line synthesis (Sect.~\ref{sec:luminosity}). Section~\ref{sec:results} compares our theoretical predictions to observations in terms of ionizing radiation and \li{C}{IV}{1550} emission. Then in Section~\ref{sec:literature} we dive into the existing literature to figure out if other emission lines observed so far in I~Zw~18 (both optical and UV) support or contradict our theory. Section~\ref{sec:disc} discusses our explanation's place amongst alternative ones 
as well as the necessary caveats. In Section~\ref{sec:conc} we conclude that, as far as the currently available observational data allows, I~Zw~18 may -- similarly to high-redshift galaxies \citep{Liu:2025} -- very well harbour
chemically homogeneously evolving stars.


\section{Stellar evolution, synthetic spectra and population synthesis}\label{sec:theory}

\subsection{The models: Predicting O and WO stars}\label{sec:models}

Metal-poor massive single stars have been modelled using the `Bonn' stellar evolution code
in \citetalias{Szecsi:2015} with the measured composition
of I~Zw~18 (totalling in a metallicity of Z=0.0002\footnote{
	Metallicities of dwarf galaxies are usually inferred from their nebular \ion{O}{III} lines, which can be problematic since the ratio of iron to oxygen (and all other elements) does not necessarily follow solar patterns -- on this, see Sect.~2.2.1 of \citet{Chruslinska:2019}. Different works report different metallicities for I~Zw~18, such as 1/40~Z$_{\odot}$ in \citet{Kehrig:2016} and 1/30~Z$_{\odot}$ in \citet{Izotov:1998}; this may depend on, for example, the calibration used to translate \ion{O}{III} equivalent width into abundances. \citetalias{Szecsi:2015} modelled I~Zw~18 relying on element abundances reported in \citet{Lebouteiller:2013} corresponding to 1/50~Z$_{\odot}$.
}, or [Fe/H]~$=$~$-$1.7) and standard input physics. Mass-loss rate prescription from \citet{Hamann:1995} scaled down by a factor of 10 and Z-scaling from \citet{Vink:2001} was used for the hot helium-star phase, which yields consistent predictions with \citet[][for details, see \citealt{Yoon:2005} and \citealt{Szecsi:2022ok}]{Nugis:2000}. 
These evolutionary sequences predicted that fast-rotating stars evolve \textit{directly} towards hot surface temperatures (so-called chemically homogeneous evolution, CHE) and thus serve as potential sources of photoionization in metal-poor dwarf galaxies. Applying simple estimations for the wind optical depth \citep[following][]{Langer:1989a} and black body radiation, \citetalias{Szecsi:2015} showed that during the long-lived main-sequence (core-hydrogen-burning) phase, these hot stars are expected to have transparent winds and therefore should not develop emission lines in their spectra (hence the name TWUIN~stars). \citetalias{Szecsi:2015} estimated that a population of them with a regular mass function can explain the observationally derived \ion{He}{II} ionizing flux, Q$^{obs}_{\ion{He}{II}}$, in I~Zw~18. However, these results were rather crude; a more reliable conclusion could be drawn from synthetic spectra created especially for these stars. 

In \citetalias{Kubatova:2019}, we computed such spectra, again with the composition of I~Zw~18, using the Potsdam~Wolf-Rayet (\PoWR) stellar atmosphere code \citep{Grafener:2002,Hamann:2003,Hamann:2004,Surlan:2013,Sander:2015,Sander:2020}. We found that chemically homogeneously evolving stars in the main-sequence phase indeed do not develop emission lines, just as suggested in \citetalias{Szecsi:2015}. According to standard stellar classification presented in Figure~\ref{fig:classification}, these TWUIN stars, if observed, would be classified as early O~type stars (see also Table~4 \& Appendix~A of \citetalias{Kubatova:2019}). 
As for the post-main-sequence, the evolutionary behaviour during this phase was not included into \citetalias{Szecsi:2015} but was later published by \citealt{Szecsi:2022ok} (their grid called `IZw18-CHE') and also studied in \citetalias{Kubatova:2019}.
We found that during core-helium burning, chemically homogeneously evolving stars do actually develop strong emission lines: they were classified as WR~stars of class~WO. Fig.~\ref{fig:classification} summarizes these findings in a HR~diagram, and Sect.~\ref{sec:newmodel} provides a deeper look into the post-main-sequence phase. 

\begin{figure}[!t]\centering
	\includegraphics[width=\columnwidth]{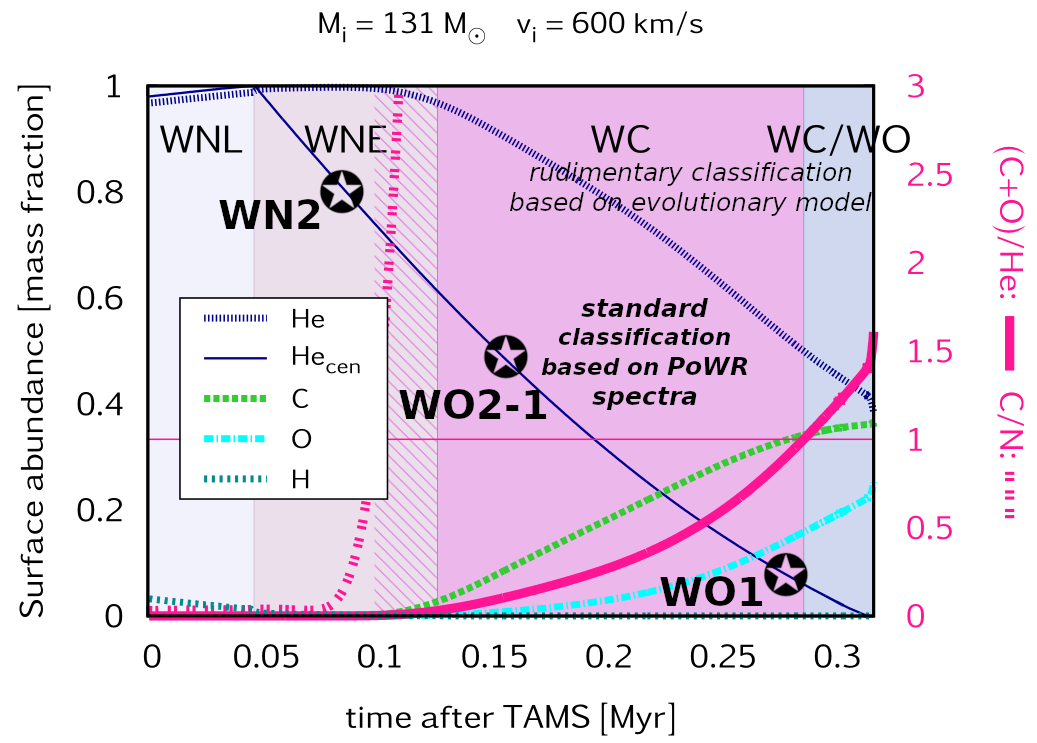}
	\caption{Time evolution of surface properties of the evolutionary model with M$_{\rm ini}$~$=$~131~M$_{\odot}$ after the terminal-age main sequence (TAMS). 	Reproduced from \citet[][their Fig~4.13]{Szecsi:2016}, adding information about our three post-main-sequence \PoWR\ models scrutinized in Sect.~\ref{sec:newmodel} and Fig~\ref{fig:newmodel}. Black star-symbols mark their positions on the central helium mass fraction: Y$_{\rm C}$~$=$~0.8 (early), Y$_{\rm C}$~$=$~0.5 (mid), Y$_{\rm C}$~$=$~0.1 (late). The rudimentary classification of \citet[][see~their Sect.~4.5.1, based on \citealt{Georgy:2012}]{Szecsi:2016} is surpassed by the spectral-line-based classification scheme (see also Fig.~\ref{fig:newmodel} showing all the optical emission lines), meaning that chemically homogeneously evolving stars at I~Zw~18 metallicity do not spend significant time in the WC phase, proceeding from their main-sequence O-star phase to WN and then directly to WO. 
	}\label{fig:131evol}
\end{figure}

The spectra published in \citetalias{Kubatova:2019} were created for three masses (M$_{\rm ini}$~$=$~20, 59 and 131~M$_{\odot}$, all with a high initial rotational rate of 300-600~km$^{-1}$), five evolutionary stages (four on the main-sequence and one post-main-sequence, see also Sect.~\ref{sec:newmodel}), two mass-loss rates (a nominal one taken from the evolutionary models, and a reduced one; actual values listed in Table~2 of \citetalias{Kubatova:2019}) and two types of assumptions about wind clumping (a smooth wind and a clumped wind with clumping factor D~$=$~10). 
The mass-loss rate values are supported by those that \citet{Hainich:2015} empirically predicted for the same stars. Additionally, more realistic mass-loss rates of these stars were calculated by \citet{Abdellaoui:2023} using the VH-1 code, which allows for a multi-dimensional treatment of the wind. The obtained results (listed in their Table~6.3) are somewhere between our nominal and reduced values. 

\begin{figure*}[!t]\centering
	\includegraphics[width=2\columnwidth]{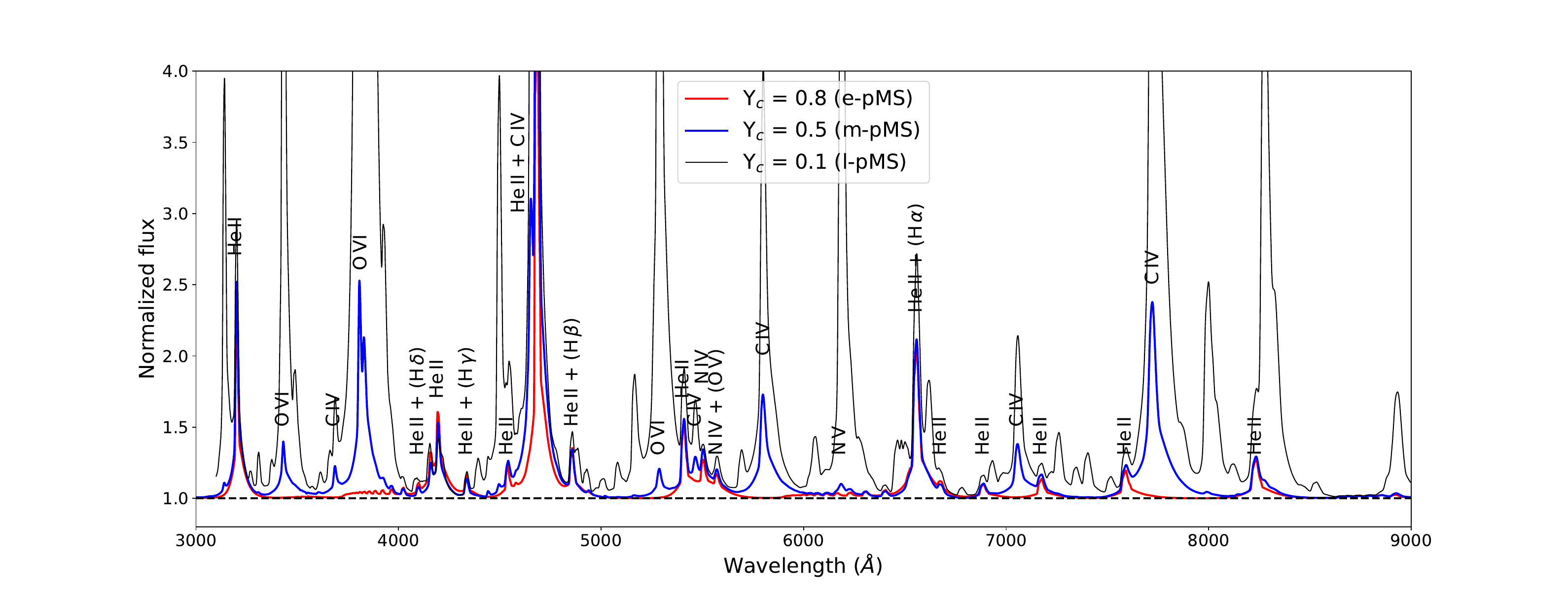}
    \includegraphics[width=1.7\columnwidth,angle=0]{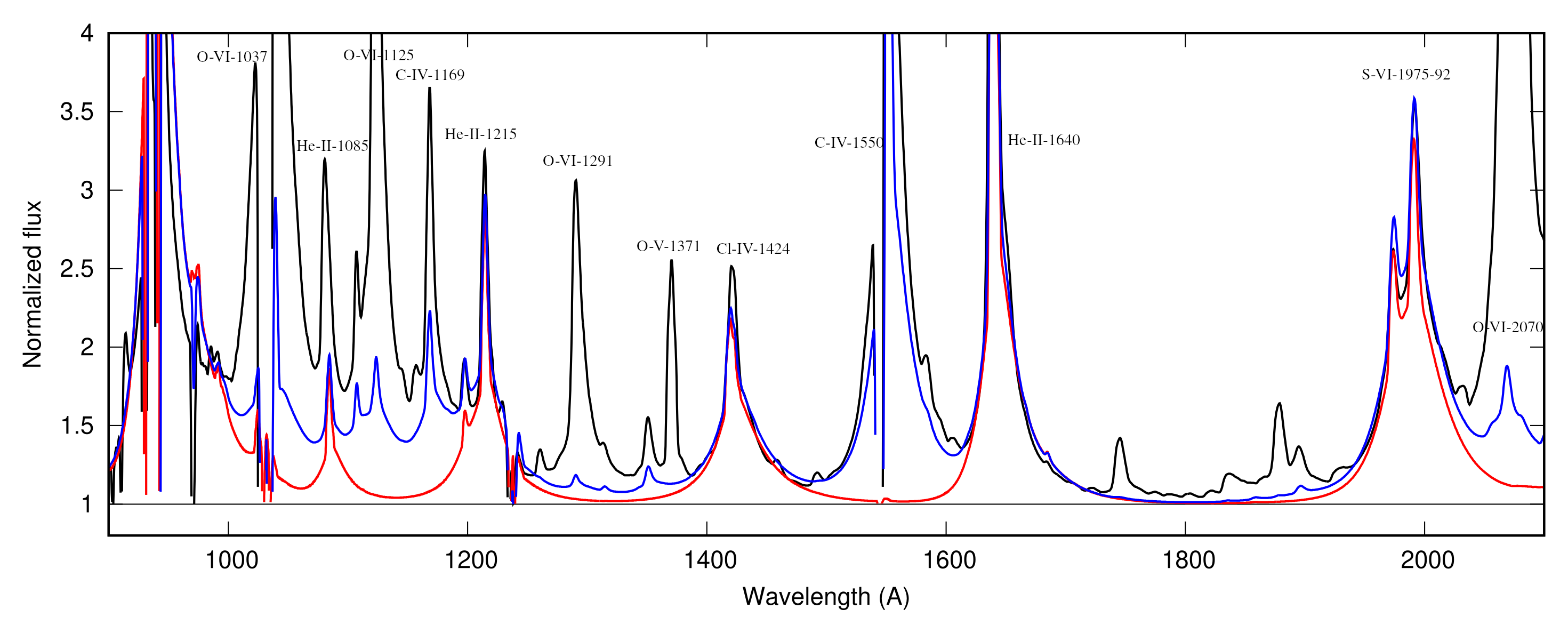}
	\caption{\PoWR\ model spectra of individual stars during the post-main-sequence (core-helium burning) phase of the same evolutionary model (M$_{\rm ini}$~$=$~131~M$_{\odot}$, {actual} mass between 93$-$106~M$_{\odot}$). The model's approximate position is shown on the HR~diagram in Fig.~\ref{fig:classification} by the last point (pMS phase) at around log(T$_{\mathrm{eff}}$/K)~$\sim$~5.14 and log(L/L$_{\odot}$)~$\sim$~6.7.
    \textit{Top}: Optical range. \textit{Bottom}: UV range. Nominal mass loss rate and no clumping was assumed, to be consistent with previous work (see \citetalias{Kubatova:2019}, and the discussion of the caveats in Sect.~\ref{sec:caveats}). Y$_{\rm C}$ indicates core-helium abundance and thus the evolutionary progress during the post-MS. The earliest model (red straight line, Y$_{\rm C}$~$=$~0.8, i.e. just burned about 20\% of its helium in the core) shows prominent emission only in helium, while the mid (blue straight line, Y$_{\rm C}$~$=$~0.5) and late (black straight line, Y$_{\rm C}$~$=$~0.1, corresponding to the bottom right panel of Fig.~\ref{fig:BB}) ones develop carbon and oxygen lines too. {(The Cl and S lines are modelling side-effects, we do not expect them to show up in observations.)} However, the distinguishing emission line \li{C}{iii}{5696} which serves as the basis of classifying a stars as WC \citep{Crowther:1998} is missing in all three models. So are the nitrogen emission lines that would categorize a star as late-type WN \citep{Crowther:1995,Smith:1996,Crowther:2011}. If observed,
	therefore, these stars would be identified as early-type WN (i.e. WN2) and then during most of the core-helium burning, as WO -- more precisely, WO~2 evolving to WO~1.
	}
	\label{fig:newmodel}
\end{figure*}

Fig.~\ref{fig:BB} shows the spectral energy distribution (SED) of some of our models (those with nominal mass-loss rates and smooth wind), as well as the corresponding black body spectrum for context (\textit{without} applying wind optical depth correction from \citealt{Langer:1989a}). 
As Fig.~\ref{fig:BB} attests, the SED of these stars
do not follow the distribution of black body radiation, even though the \textit{total} amount of ionizing photons in the three main ionizing continua (Lyman, \ion{He}{I} and \ion{He}{II} continua) are on the same order of magnitude as in the (uncorrected) black body.  

Having these models allows us to build a synthetic population that includes spectral synthesis of various emission lines. Before we do that (Sect.~\ref{sec:population}), we present two new model spectra to properly account for the post-main-sequence, a phase crucially relevant for our goals.

\subsection{New \PoWR \ spectra: WN and WO stars (no WC)}\label{sec:newmodel}

In our \citetalias{Kubatova:2019}, the post-main-sequence phase was accounted for by models with central helium mass fraction of Y$_{\rm C}$~$=$~0.1; that is, nearly the end of core-helium burning. The spectra were classified as type~WO according to the standard spectral classification scheme (for details on this, see Table~4 \& Appendix~A of \citetalias{Kubatova:2019}, as well as our Fig.~\ref{fig:classification}), and showed very prominent emission lines in both the optical and the UV regimes. {However,} according to Figure~\ref{fig:131evol},
the evolutionary model predicts significant change in the surface abundances while core-helium burning is in progress, with the Y$_{\rm C}$~$=$~0.1 stage being quite different from earlier times. {Therefore, before using these late-stage models in our population synthesis (Sect.~\ref{sec:population}), we should justify the extent to which they are} representative of the entirety of the post-main-sequence, and discuss their evolutionary behaviour and spectral classification.

\begin{figure}[!t]\centering
	\includegraphics[height=0.95\columnwidth,angle=270]{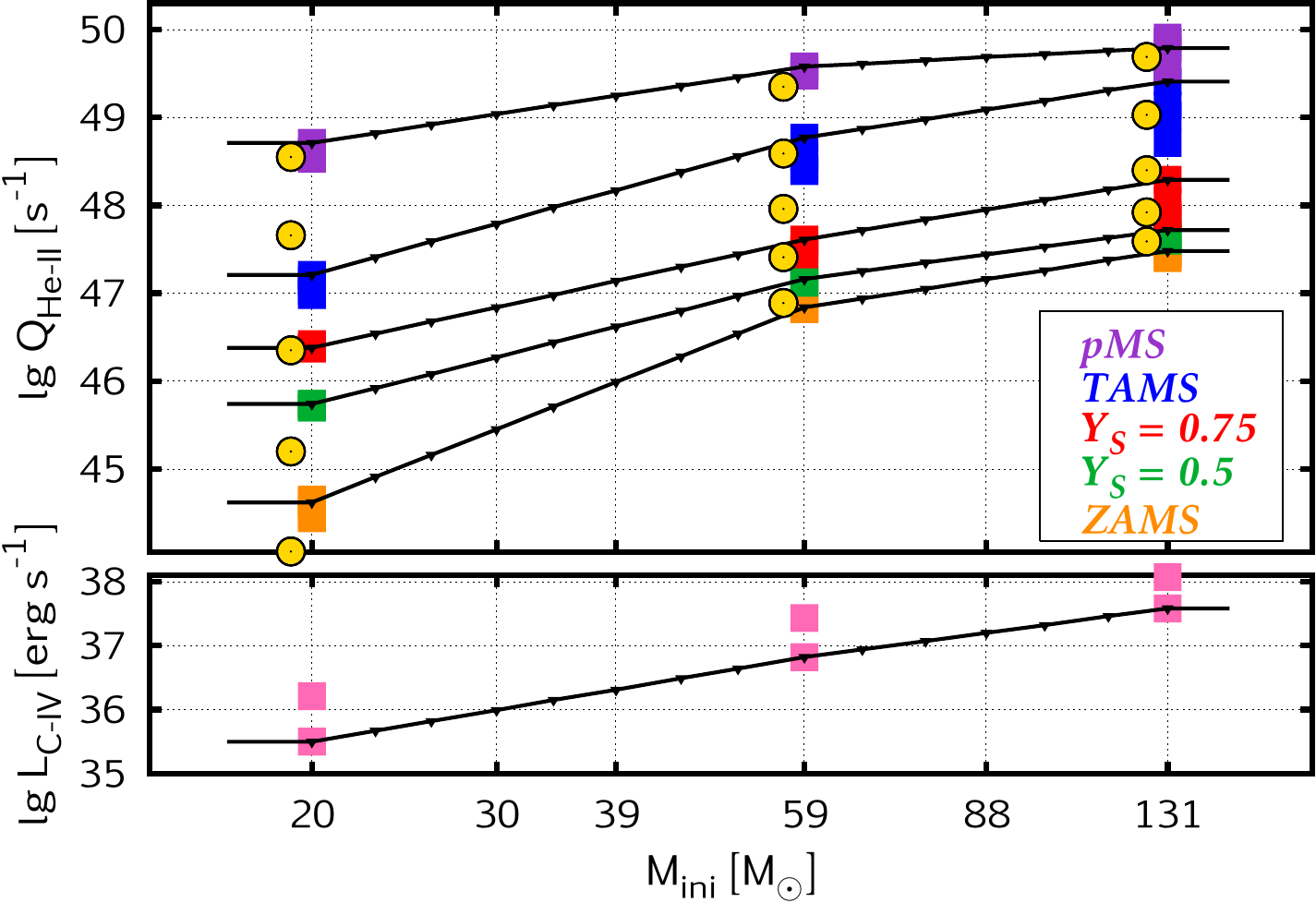}
	\caption{\textit{Top panel:} Number of ionizing photons in the \ion{He}{II} continuum in all our spectral models from \citetalias{Kubatova:2019}. Evolutionary stages are indicated by colours (ZAMS by yellow; two main sequence models with surface helium mass fractions Y$_{\rm S}$~$=$~0.5 and~0.75 by green and red, respectively; terminal-age-main-sequence model, TAMS, by blue; and post-main-sequence model, pMS, by purple). For every evolutionary stage, Q$_{\ion{He}{II}}$ from all four spectral models of \citetalias{Kubatova:2019} are shown (hence the spread in the coloured stripes).
	Yellow circles represent the \ion{He}{II} ionizing photon number computed from corresponding black body distributions (see Fig.~\ref{fig:BB}). 
		To account for a \textit{population} of massive stars, we interpolate linearly between the logarithm of the three simulated masses. Models with nominal mass-loss rate and unclumped wind are used; interpolated values are shown with the solid black lines, small triangles marking the bin sizes. Above and below our highest and lowest mass models, the values of those models are applied (down to 9~M$_{\odot}$). 
		\textit{Bottom panel:} Same as the top panel, but for the line luminosities in \li{C}{IV}{1550} of the pMS model. The second trio of pink rectangles with higher values correspond to clumped wind models (see also Fig.~\ref{fig:Civ}). Other line luminosities are treated the same way (see Sect.~\ref{sec:luminosity}). 
	}\label{fig:compare}
\end{figure} 

{To this end}, we computed two more spectral models using \PoWR \ for Y$_{\rm C}$~$=$~0.8 and Y$_{\rm C}$~$=$~0.5 of a very massive star with initial mass M$_{\rm ini}$~$=$~131~M$_{\odot}$ ({actual} mass around 93$-$106~M$_{\odot}$, see the caption). 
The computational set-up was the same as in \citetalias{Kubatova:2019}. We apply nominal mass-loss rates taken from the evolutionary model \citep{Szecsi:2015,Szecsi:2016,Szecsi:2022ok}, and smooth winds (clumping factor D~$=$~1); we discuss in Sect.~\ref{sec:caveats} that these assumptions are actually quite robust. Figure~\ref{fig:newmodel} compares the optical band for these spectra, while Figure~\ref{fig:131evol} summarizes what we learned in terms of stellar classification. 

Applying the rudimentary scheme of classification defined by \citet{Georgy:2012} for evolutionary models, \citet[][in their Sect.~4.5.1]{Szecsi:2016} suggested that these stars are of the classes WN, WC, and close-to-WO, respectively (Fig.~\ref{fig:131evol}). However, the \PoWR\ spectra tell a different story. The standard classification scheme (working with spectral line intensity\footnote{This classification scheme follows the footsteps of the Harvard-system. The Harvard system actually included WR-stars, calling them Oa (now WC), Ob (now WNE), Oc (now WNL) -- see \citet{Cannon:1916} -- but efforts like \citet{Payne:1930}, \citet{Edlen:1932}, and \citet{Beals:1933} helped shape the version adopted by the IAU in 1938, which was modernized in the 1960s by Hiltner, Schild and Lindsey Smith, with WO stars first introduced by \citet{Barlow:1982}. Today, we rely on emission line intensities to assign classes to WR-stars, as summarized in Appendix~A of \citetalias{Kubatova:2019}, although switching between subtypes is not as clear as for the absorption-line-based classes OBAFGKM and the usage of specific primary and secondary criteria is also less sharp.} as summarized by Appendix~A of \citetalias{Kubatova:2019}) yields the following classes: WN2 for Y$_{\rm C}$~$=$~0.8, WO~2 or WO~1 (consistent with both) for Y$_{\rm C}$~$=$~0.5, and WO~1 for Y$_{\rm C}$~$=$~0.1. 
Based on what we learned in \citetalias{Kubatova:2019} (especially their Figs.~4 and~B6), we expect that reducing the mass loss rate and/or increasing the wind clumping may shift these spectra towards different subtypes -- that is, towards different line intensities -- but not towards a different class. In other words, varying mass loss and clumping will not make any of these models a WC~star.

We conclude therefore that chemically homogeneously evolving stars at I~Zw~18 metallicity would be seen, after spending their main-sequence as O~stars, as early-type WN~stars at the beginning, and WO stars in the \textit{majority} of their post-main-sequence lifetimes. This is surprising since typical evolution calculations for single
stars predict the WO~stage to be a rather late, short-term phase, preceded by a lengthier WC~phase \citep[see e.g.][]{Groh:2014}, the typical sequence of classes reading: 
\begin{center}
	O $\rightarrow$ RSG/LBV $\rightarrow$ WN $\rightarrow$ WC $\rightarrow$ WO.
\end{center}
Chemically homogeneous evolution does not follow that picture, as these stars complete their main-sequence as very hot O stars (Sect.~\ref{sec:models} and Fig.~\ref{fig:classification}) and then proceed to be WN and WO after that, a completely new sequence of classes: 
\begin{center}
	O $\rightarrow$ WN $\rightarrow$ WO.
\end{center}
According to Fig.\ref{fig:131evol}, if a WC-phase arises at all between WN and WO, it will not last longer than about $\sim$1-2\% of the total stellar lifetime. 

Finding our stars to be WO instead of WC is in accordance with \citet{Tramper:2013} who concluded that WO stars may be the high-temperature and high-luminosity extensions of WC stars. The WO spectral type does not necessarily mean more oxygen than carbon: in our mid-pMS model, the carbon abundance is actually higher than the oxygen one (as is seen in Fig.~\ref{fig:131evol}). The same was found for the LMC WO stars analysed in \citet{Aadland:2022}. So it is not about abundance but about temperature: carbon is ionized into \ion{C}{V} and thus the otherwise prominent 1550 and 5808 features are weak or absent. 

While WC and WO may not necessarily represent different evolutionary stages from a theorist's point of view, for an observer the distinction could be important. The label ``WO'' is a spectral classification, hence we keep using it here. As is seen in Sect.~\ref{sec:literature}, the detected features of I~Zw~18 are consistent with WO stars. 

Nevertheless, the three phases do not differ much when it comes to the SED, yielding log\,Q$_{\ion{He}{ii}}$ values of 49.80 (early-pMS), 49.79 (mid-pMS) and 49.79 (late-pMS). Note how the surface temperatures of the evolutionary models are similar: the T$_{\rm eff}$ of the early-pMS model is 138.9~kK, mid-pMS 139.2~kK, and late-pMS 138.0~kK
(see Fig.~\ref{fig:classification} where the burning is colour-coded).
In terms of atmosphere models, T$_{\rm eff}$ roughly reflects the wind onset region where $\tau$~$=$~20; but even T$_{2/3}$ (i.e. T at $\tau$~$=$~2/3) is close for the three cases (with early-pMS 56.6~kK, mid-pMS 63.3~kK, late-pMS~70.3 kK), explaining why the predicted \ion{He}{II} flux is not changing considerably during the core-helium burning phase. {Thus, using late-pMS models in our population synthesis to stand for the entirety of the post-main-sequence is well justified when it comes to the ionizing photon count.} 

\begin{figure*}[t]
	\hspace{20pt}\centering
	\includegraphics[height=0.6\columnwidth,page=6,angle=270]{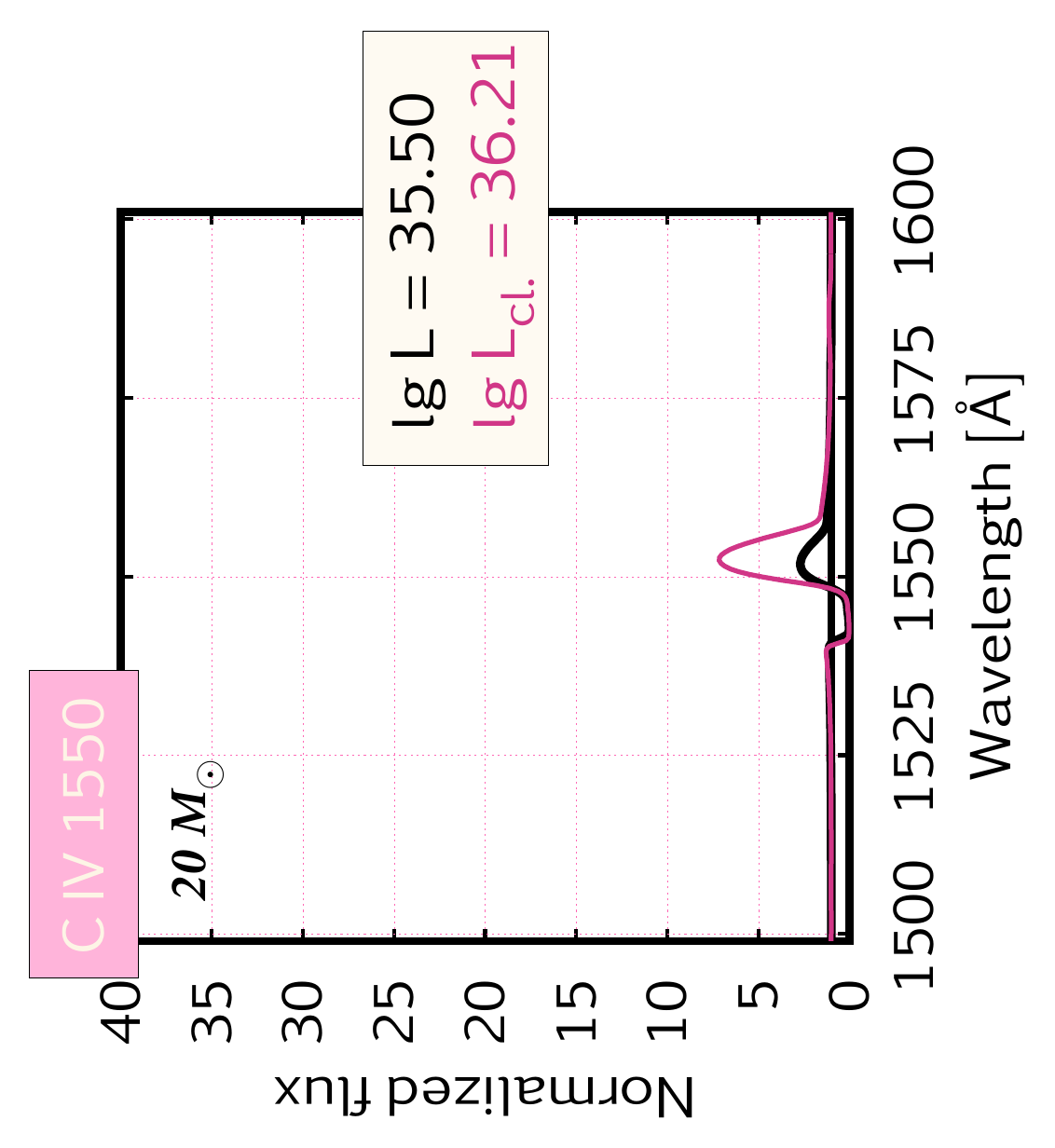}
	\hspace{10pt}
	\includegraphics[height=0.6\columnwidth,page=6,angle=270]{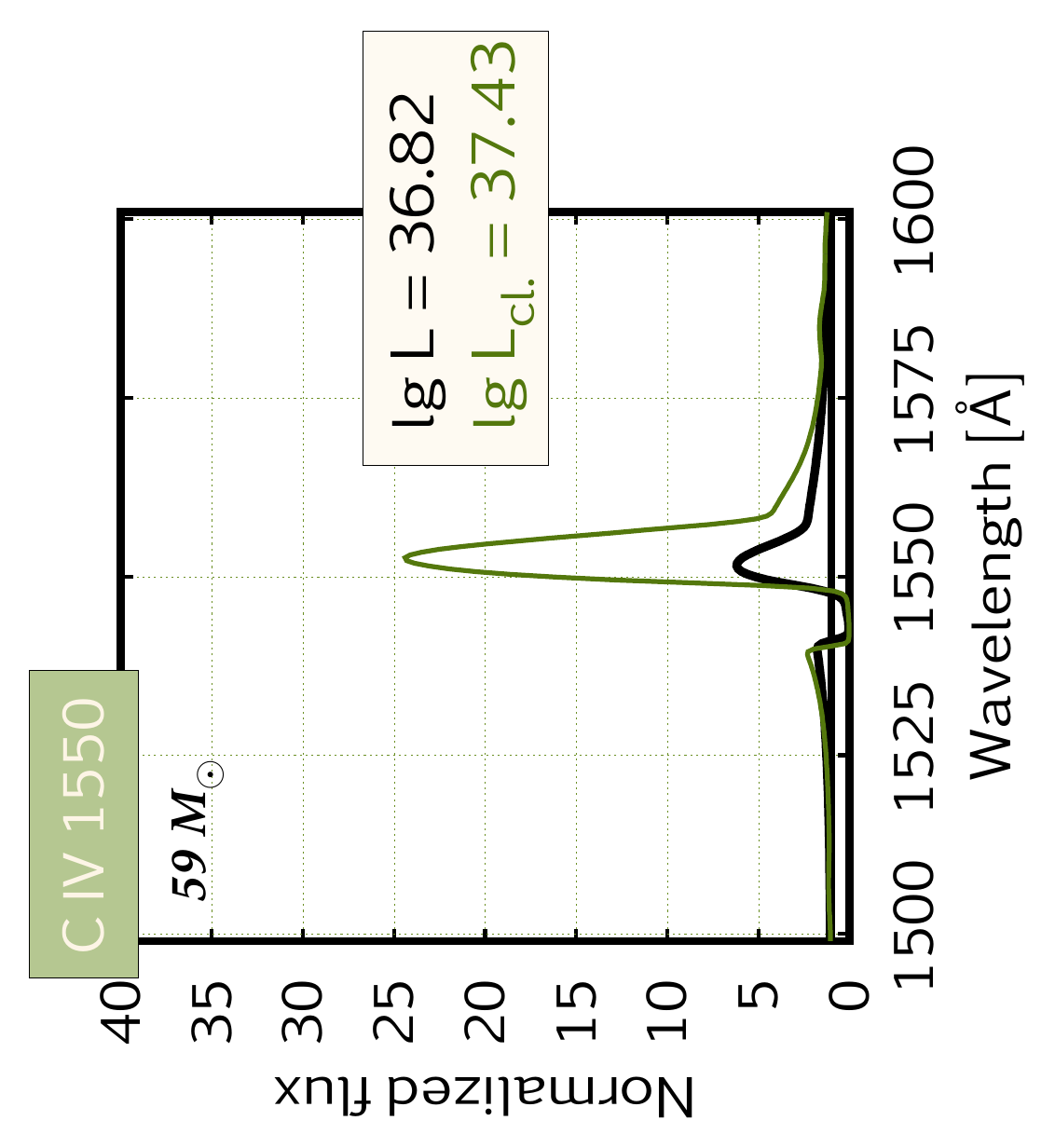}
	\hspace{10pt}
	\includegraphics[height=0.6\columnwidth,page=6,angle=270]{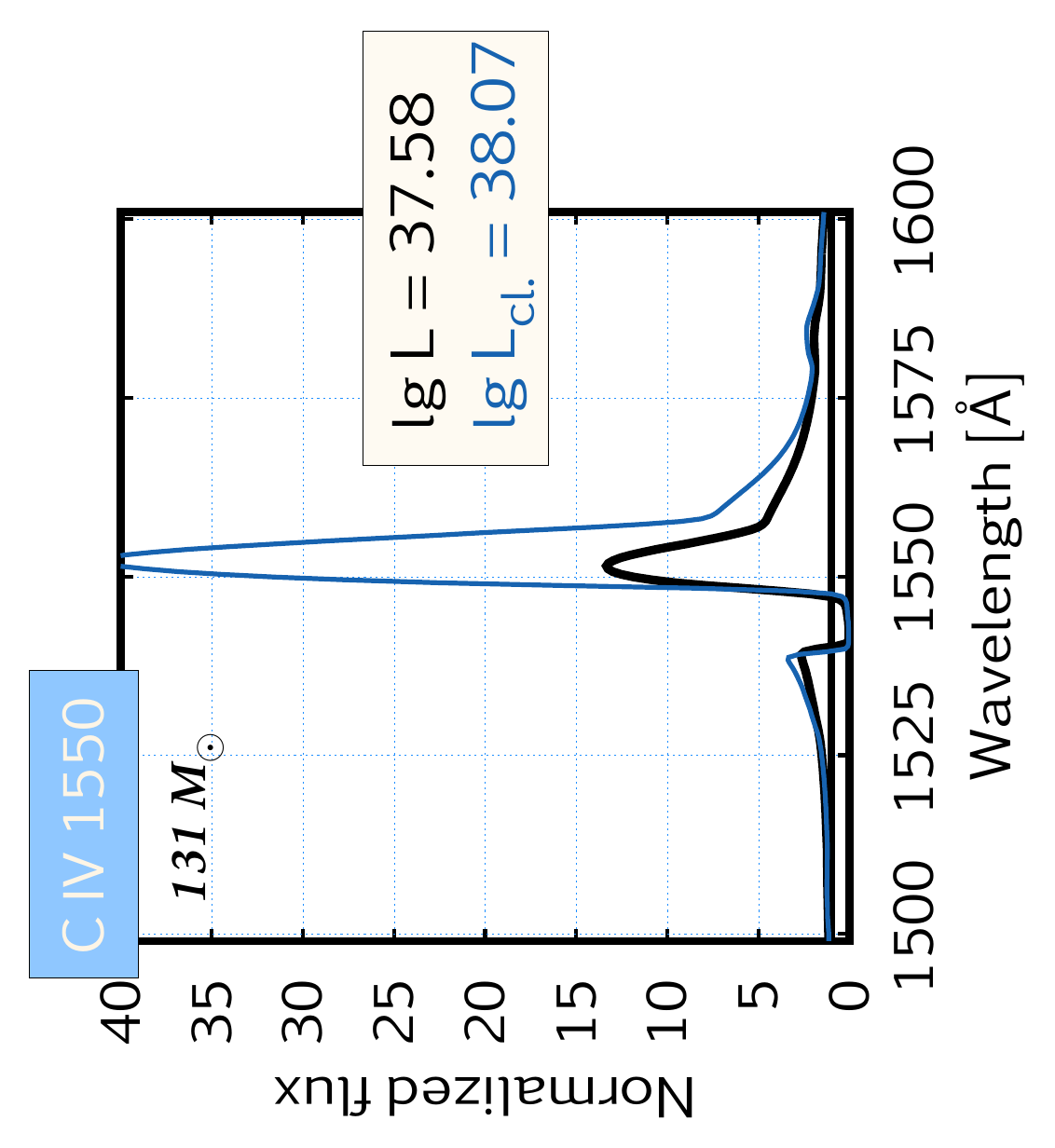}
	\caption{Normalized flux around the UV line \li{C}{IV}{1550} in three post-main-sequence (l-pMS) spectral models with initial masses as indicated in the top left corner of the panels. Line luminosity (computed following Sect.~\ref{sec:luminosity}) is given in the framed boxes in units of log(erg~s$^{-1}$); lg~L stands for unclumped (clumping factor D~$=$~1) while lg~L$_{\rm cl}$ for clumped (D~$=$~10) wind, see details in~\citetalias{Kubatova:2019}. 
	When creating the synthetic population here (Sect.~\ref{sec:population} and Sect.~\ref{sec:main}), we always apply the unclumped models' predictions (i.e. black line). Other emission lines are presented in Figs.~\ref{fig:allopt}-\ref{fig:allUV}. Based on this, we predict a $\sim$40-50~\AA-wide emission bump from our population between 1550-1600~\AA, which is indeed present in the observations in Fig.~\ref{fig:Brown}.
	}\label{fig:Civ}
\end{figure*}

{As for emission-line intensities, the mid- and late-pMS models (both WO) predict L$_{\ion{C}{IV}}$~$=$~2.13$\cdot$10$^{37}$~erg~s$^{-1}$ and 3.77$\cdot$10$^{37}$~erg~s$^{-1}$, respectively. Since clumping introduces a factor of 3 uncertainty (as is shown in Fig.~\ref{fig:Civ}), our use of the late-stage value to represent the entire post-main-sequence is justified with the caveat that winds are probably not unclumped (see \citealt{Roy:2025}) and that the observational range we attempt to match (L$^{obs}_{1550}$~$=$~2.2--5.5$\cdot$10$^{37}$~erg~s$^{-1}$) itself includes such an error margin.}

	\begin{table*}[t!]
    \caption{{Summary of our population synthesis runs, best match shaded with colour.}		}\label{tab:popsyn}
		\centering
		\begin{tabular}{c| c c | c c c c c} 
			\hline
			\rowcolor{lightgray}& SFR  & SF for  & Q$^{}_{\ion{He}{II}}$ & L$^{}_{1550}$ & source of \ion{C}{iv} emission & Q$_{\ion{H}{I}}$ & I(4686) \\
			\rowcolor{lightgray}& [M$_{\odot}$\,yr$^{-1}$] & [Myr] & [ph.~s$^{-1}$] & [erg~s$^{-1}$] &  & [ph.~s$^{-1}$] & /I(\ion{H}{$\beta$})\\ [0.5ex] 
			\hline\hline
			fiducial & 0.1 & 3 & 1.80$\cdot$10$^{50}$ & 3.71$\cdot$10$^{37}$ & one/two WOs of $>$90~M$_{\odot}$ & 7.02$\cdot$10$^{52}$ & 0.004 \\ [0.5ex]
			 & \multicolumn{2}{c|}{\fbox{M$_{\rm top}$\,$=$\,150~M$_{\odot}$}} & \small $\hookrightarrow$ \textit{NE: 0.69$\cdot$10$^{\mathit{50}}$} & & & \small $\hookrightarrow$ \textit{NE: 6.21$\cdot$10$^{\mathit{52}}$} &  \\ 
			& & & \small $\hookrightarrow$ \textit{CHE: 1.11$\cdot$10$^{\mathit{50}}$} & & & \small $\hookrightarrow$ \textit{CHE: 0.81$\cdot$10$^{\mathit{52}}$} &  \\
			[0.5ex] 
			\hline
			{SF-10} & 0.03 & 10 & 1.96$\cdot$10$^{50}$ & 3.43$\cdot$10$^{37}$ & six WOs of 40~M$_{\odot}$ 
			& 3.15$\cdot$10$^{52}$ & 0.011 \\ [0.5ex] 
			 & \multicolumn{2}{c|}{\fbox{M$_{\rm top}$\,$=$\,120~M$_{\odot}$}} & \small $\hookrightarrow$ \textit{NE: 0.14$\cdot$10$^{\mathit{50}}$} & & \& one of 80~M$_{\odot}$ & \small $\hookrightarrow$ \textit{NE: 2.62$\cdot$10$^{\mathit{52}}$} &  \\ 
			& & & \small $\hookrightarrow$ \textit{CHE: 1.82$\cdot$10$^{\mathit{50}}$} & & & \small $\hookrightarrow$ \textit{CHE: 0.53$\cdot$10$^{\mathit{52}}$} &  \\
			[0.5ex] 
			\hline
			\rowcolor{bestmatch} {SF-15} & 0.02 & 15 & 2.19$\cdot$10$^{50}$ & 2.83$\cdot$10$^{37}$ & $\sim$twenty WOs of 25~M$_{\odot}$ & 2.23$\cdot$10$^{52}$ & 0.017 \\ [0.5ex] \rowcolor{bestmatch}
			& \multicolumn{2}{c|}{\fcolorbox{black}{gold!50}{M$_{\rm top}$\,$=$\,120~M$_{\odot}$}} & \small $\hookrightarrow$ \textit{NE: 0.09$\cdot$10$^{\mathit{50}}$} & & \& two of 50~M$_{\odot}$ & \small $\hookrightarrow$ \textit{NE: 1.75$\cdot$10$^{\mathit{52}}$} &  \\ 
			\rowcolor{bestmatch}& & & \small $\hookrightarrow$ \textit{CHE: 2.10$\cdot$10$^{\mathit{50}}$} & & & \small $\hookrightarrow$ \textit{CHE: 0.48$\cdot$10$^{\mathit{52}}$} &  \\
            [0.5ex] 
			\hline
			{SF-15-80} & 0.02 & 15 & 1.94$\cdot$10$^{50}$ & 2.15$\cdot$10$^{37}$ & seventeen WOs of 20~M$_{\odot}$ & 1.96$\cdot$10$^{52}$ & 0.017 \\ [0.5ex] 
			& \multicolumn{2}{c|}{\fbox{M$_{\rm top}$\,$=$\,80~M$_{\odot}$}} & \small $\hookrightarrow$ \textit{NE: 0.06$\cdot$10$^{\mathit{50}}$} & & \& $\sim$four of 40~M$_{\odot}$ & \small $\hookrightarrow$ \textit{NE: 1.53$\cdot$10$^{\mathit{52}}$} &  \\ 
			& & & \small $\hookrightarrow$ \textit{CHE: 1.88$\cdot$10$^{\mathit{50}}$} & & & \small $\hookrightarrow$ \textit{CHE: 0.43$\cdot$10$^{\mathit{52}}$} &  \\
            [0.5ex] 
			\hline
			{SF-15-40} & 0.02 & 15 & 1.31$\cdot$10$^{50}$ & 0.91$\cdot$10$^{37}$ & seventeen WOs of 20~M$_{\odot}$ & 1.38$\cdot$10$^{52}$ & 0.016 \\ [0.5ex] 
			& \multicolumn{2}{c|}{\fbox{M$_{\rm top}$\,$=$\,40~M$_{\odot}$}} & \small $\hookrightarrow$ \textit{NE: 0.01$\cdot$10$^{\mathit{50}}$} & & \& $\sim$three of 30~M$_{\odot}$ & \small $\hookrightarrow$ \textit{NE: 1.07$\cdot$10$^{\mathit{52}}$} &  \\ 
			& & & \small $\hookrightarrow$ \textit{CHE: 1.30$\cdot$10$^{\mathit{50}}$} & & & \small $\hookrightarrow$ \textit{CHE: 0.31$\cdot$10$^{\mathit{52}}$} &  \\
			[1ex] 
			\hline
		\end{tabular}
\tablefoot{All runs assumed a Salpeter IMF {with M$_{\rm top}$~$\leq$~150~M$_{\odot}$ (see label)} and 10\% of stars rotating fast enough for chemically homogeneous evolution. The total size of the region formed is 3$\cdot$10$^5$\,M$_{\odot}$ in all cases.
		For reference, the observed values are Q$^{obs}_{\ion{He}{II}}$~$=$~1.33$\cdot$10$^{50}$~photons~s$^{-1}$ and L$^{obs}_{1550}$~$=$~2.2--5.5$\cdot$10$^{37}$~erg~s$^{-1}$. Our results from {most runs are able to} account for these, {regardless of the various assumptions about the star-formation history and thus, the stellar content}. {The fiducial case mimics the scenario in \citetalias{Szecsi:2015} (details are given in Sect.~\ref{sec:population}), while the others  relax the SFR, lower the M$_{\rm top}$, and lengthen the age of the population, meaning later phases of lower-mass models are captured. This is why the source of the stellar \ion{C}{IV} emission gets less and less extreme: after the third case, no very massive star (VMS) is needed at all.
		As for} the observed hardness, i.e. the relative intensity of I(4686)/I(\ion{H}{$\beta$})~$=$~A$\cdot$Q(\ion{He}{II})/Q(\ion{H}{I}), it should be between 0.02$-$0.04 observationally, {which renders our third population most consistent with measurements (including the \ion{C}{iv} line luminosity, which becomes lower with smaller M$_{\rm top}$) and thus most favoured,} as explained in Sect.~\ref{sec:hardness} {(also see Fig.~\ref{fig:LymanHeII})}.}
	\end{table*}

\subsection{Population synthesis}\label{sec:population}

Equipped with all these model spectra, we now revisit the question of photoionization in I~Zw~18 by building a synthetic population of metal-poor massive stars. Since chemically homogeneous stars 
(TWUIN or O/WN/WO)
are a straightforward outcome of rotating stellar evolution at the metallicity of I~Zw~18, these stars are automatically included.

It was shown in \citetalias{Szecsi:2015} that a population of single metal-poor massive stars \textit{on the main sequence} is able to explain I~Zw~18's \ion{He}{II} ionizing flux 
(\citetalias{Szecsi:2015}, Sect.~10.4). {No other observable was investigated in that simple population calculation.}
Further assumptions were
(i) a Salpeter-type initial mass function (IMF); 
(ii) ongoing star formation for the past 3~Myr with a rate of 0.1~M$_{\odot}$~yr$^{-1}$ \citep[this rate was observed by][]{Lebouteiller:2013} resulting a 300~000~M$_{\odot}$ total mass in the region;
(iii) an upper mass limit set to 500~M${_{\odot}}$;
(iv) a rather high fraction, $\sim$20\% of stars rotating fast enough to evolve chemically homogeneously and therefore becoming TWUIN~stars (instead of cool, non-ionizing supergiants); and
(v) {black body emission}\footnote{{The question was followed up in a thesis \citep{Szecsi:2016}, where the black body emission was corrected for the wind optical depth according to \citet{Langer:1989a}.}}.

We revise these assumptions here. Most importantly, we included the full evolution, i.e. both the main- and the post-main-sequences of our stellar models.
We still used a regular Salpeter IMF \citep{Salpeter:1955} for a continuous star formation episode lasting for 3~Myr forming 0.1~M$_{\odot}$~yr$^{-1}$, same as before. The upper mass limit was, however, set to a more realistic 150~M$_{\odot}$ (as opposed to 500~M$_{\odot}$; though note that even 200~M$_{\odot}$ might be acceptable in light of \citealt{Bestenlehner:2020}). The ratio of chemically homogeneously evolving stars was also relaxed: 10\% now (as opposed to~20\%), which is better supported by the rotational velocity distribution measured in the Small Magellanic Cloud \citep[1/5~Z$_{\odot}$,][]{Mokiem:2006},
where about~10-15\% of massive stars rotate faster than what is needed for CHE at the ten times lower metallicity of I~Zw~18. 
\textit{This means that, in the population we built here, 90\% of the massive stars are in fact `normal', slow-rotating stars, i.e. not chemically homogeneous, but classical single stellar evolution leading to O/B stars and supergiants.} No classical WR stars though, as at this metallicity the supergiants do not lose their complete envelopes in the rather weak stellar winds, as is shown by \citet{Szecsi:2016,Szecsi:2019,Szecsi:2022ok}. 
These `normally' evolving O/B stars are taken into account {assuming black body emission}: 
their contribution to the ionizing flux is {only significant near the zero-age main sequence (ZAMS) when they are the hottest, while they do not contribute to emission line luminosities, for even in their early lives they are less hot than their chemically homogeneously evolving counterparts (as is seen in Figure~\ref{fig:classification}, right panel); thus, they surely experience transparent winds and absorption-type spectra.}
We include low-mass stars ($<$~9~M$_{\odot}$) only as mass holders down to 0.5~M$_{\odot}$, the same way as in \citetalias{Szecsi:2015}. 

To account for the radiation field emitted by the chemically homogeneously evolving TWUIN/WO stars, we used the SED predicted by our synthetic spectra from \citetalias{Kubatova:2019} with a nominal mass-loss rate and smooth wind (Fig.~\ref{fig:BB}). 
This way there is no need to correct for wind optical depth. Such a correction is a common practice for evolutionary models, done in for example \citet{Groh:2019}, \citet{Schootemeijer:2018}, {and \citet{Szecsi:2016}}. They followed the estimates of \citet{Langer:1989a}, which took only electron-scattering opacity into account: 
a fast but simplistic method. Since our spectra were created with state-of-the-art physical treatment of all elements in \PoWR, including the high ionization stages of iron group elements (albeit with reduced abundances corresponding to the low metallicity), all sources of opacity are taken into account. 

As the spectra in \citetalias{Kubatova:2019} are only provided for three initial masses, we linearly interpolated between the predicted flux values to get a smooth distribution. This is shown in Figure~\ref{fig:compare}. {The most massive stars' post-main-sequence phase (spectral type WO) provides the largest ionizing photon count.} So even if such stars are rare according to a typical IMF, their contribution to the \ion{He}{II} ionization {can be} significant. Yet, the population synthesis includes the contribution of all the hot O-stars, as explained above. 
{It depends on the star formation history (see Sect.~\ref{sec:results}) whether very massive stars would form and dominate over the population or not.}
{In either case, the ratio of WR/O stars in our synthetic population is around 0.001, which is consistent with the lowest-Z observations (upper limits) of \citet[][their Fig.\,5]{LopezSanchez:2010a}.}

To sum up, all our assumptions about the synthetic population are either less extreme than those in \citetalias[][]{Szecsi:2015} or identical, while the models we apply are more elaborate. Our results, {including further parameter studies,} are reported and compared to observational data in Sect.~\ref{sec:results}.

\subsection{Luminosity in emission lines}\label{sec:luminosity}

We performed spectral synthesis for emission lines by computing the line luminosity from our spectra. 
Figure~\ref{fig:Civ} shows an example: the flux around 1550~$\AA$ is plotted for the models where the \ion{C}{iv} line is in emission (that is, post-main-sequence models with a nominal mass-loss rate). 

The \li{C}{iv}{1550} emission line luminosity depends strongly on the mass-loss, meaning it depends on the mass of the model: for the M$_{ini}$~$=$~20~M$_{\odot}$ model (the mass of which is only 17~M$_{\odot}$ in the post-main-sequence phase; there is mass-loss over the evolution) it is in the order of 10$^{35}$~erg~s$^{-1}$, while for our very massive model of M$_{ini}$~$=$~131~M$_{\odot}$ (post-main-sequence phase mass only 95~M$_{\odot}$) it is in the order of 10$^{37}$~erg~s$^{-1}$. Although we focus on the models with smooth wind (i.e. without wind clumping) in this work, Fig.~\ref{fig:Civ} attests that clumped winds at the same mass-loss can increase the line luminosity by almost an order of magnitude. 

Interpolation between the \li{C}{iv}{1550} line luminosities is presented in Fig.~\ref{fig:compare}. Population synthesis was done on these values according to the assumptions made in Sect.~\ref{sec:population}.
Other emission line luminosities, both UV and optical, were treated the same way.


\section{Comparing theory to observations}\label{sec:results}

\subsection{Ionizing \ion{He}{II} emission and \ion{C}iv line luminosity}\label{sec:main}

Table~\ref{tab:popsyn} summarizes our results.
From our fiducial population built in Sects.~\ref{sec:population}--\ref{sec:luminosity}, we predict a total \ion{He}{II} ionizing flux of 
Q$_{\ion{He}{II}}$~$=$~1.80$\cdot$10$^{50}$~photons~s$^{-1}$
and a total \li{C}{IV}{1550} emission line luminosity of
L$^{}_{1550}$~$=$~3.71$\cdot$10$^{37}$~erg~s$^{-1}$. These are {not far from what is} measured in I~Zw~18: Q$^{obs}_{\ion{He}{II}}$~$=$~1.33$\cdot$10$^{50}$~photons~s$^{-1}$ and L$^{obs}_{1550}$~$=$~{2.2--5.5}$\cdot$10$^{37}$~erg~s$^{-1}$ (see Sect.~\ref{sec:intro})\footnote{The range 2.2$-$5.5$\cdot$10$^{37}$~erg~s$^{-1}$ can be derived from the data presented in \citet{Brown:2002} by taking the reported L$_{\ion{He}{II}}$=2.4$\cdot$10$^{37}$~erg~s$^{-1}$ for their Fig.1\textit{c} and L$_{\ion{He}{II}}$=3.2$\cdot$10$^{37}$~erg~s$^{-1}$ for their Fig.1\textit{d} (reproduced here in Figs.~\ref{fig:Brown} \&~\ref{fig:COSBrown}) and their statement ``The ratio of \ion{C}{IV} to \ion{He}{II} emission in Figure~1\textit{c} is $\sim$2.3 [while it is] only $\sim$0.7 in Figure~1\textit{d}'', from which one gets L$_{\ion{C}{IV}}$~$\sim$~5.5$\cdot$10$^{37}$~erg~s$^{-1}$ and L$_{\ion{C}{IV}}$~$\sim$~2.2$\cdot$10$^{37}$~erg~s$^{-1}$for their Figs.1\textit{c} and \textit{d}, respectively.}. 
Translating our statistics-based predictions into number of stars, we get about one or two post-main-sequence {chemically homogeneously evolving} objects -- that is, WO stars (see Sect.~\ref{sec:models}-\ref{sec:newmodel}) -- of mass $>$90~M$_{\odot}$ in the population. The contribution of these one or two very massive WR~type stars is responsible for the \ion{C}{IV} line luminosity, while the \ion{He}{II} photon emission is coming from them \textit{and} the rest of the hot-star population {(including normally evolving models)}.

\begin{figure*}[!t]\centering
	\includegraphics[width=0.97\columnwidth]{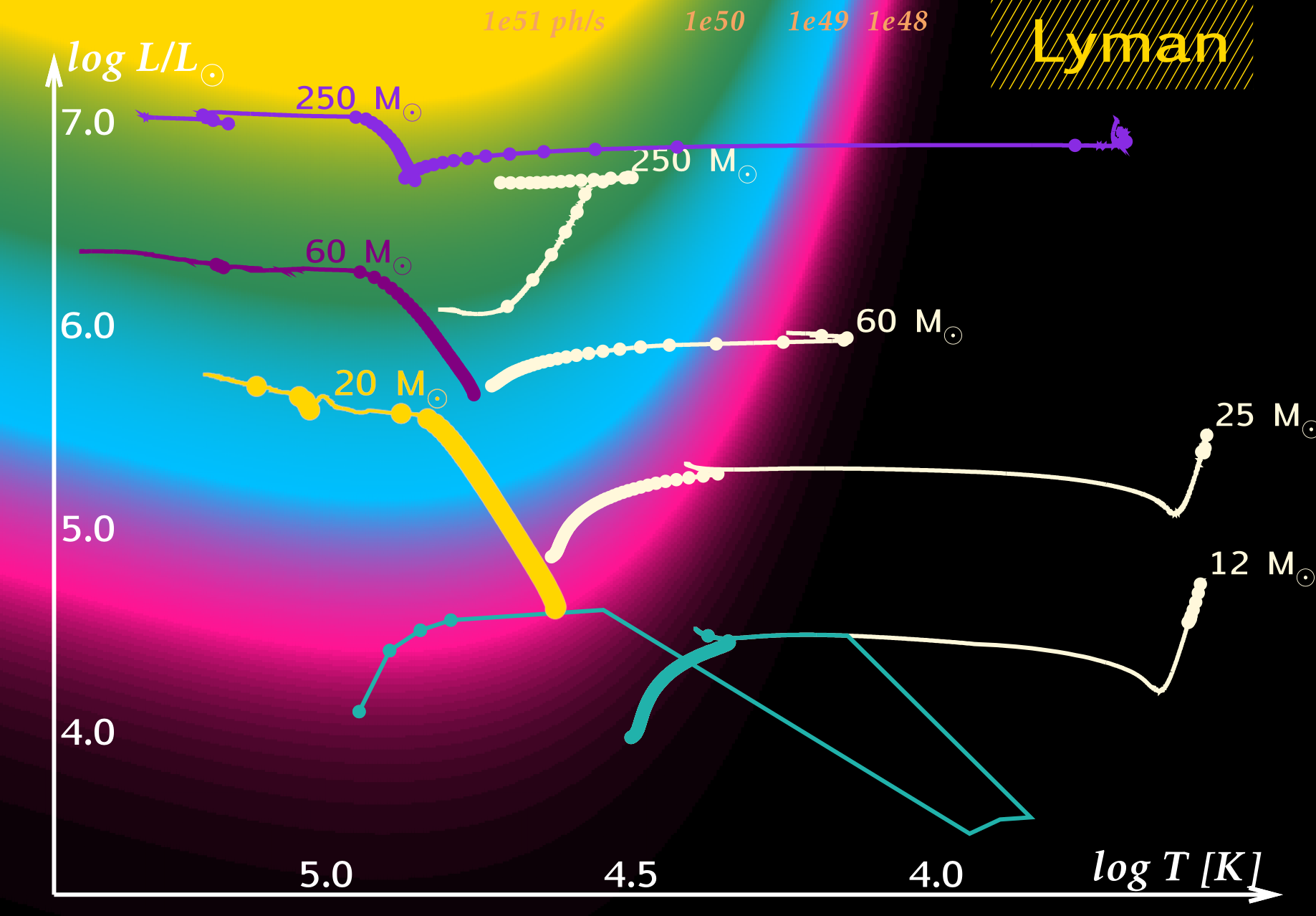}
	\hspace{14pt}\includegraphics[width=0.97\columnwidth]{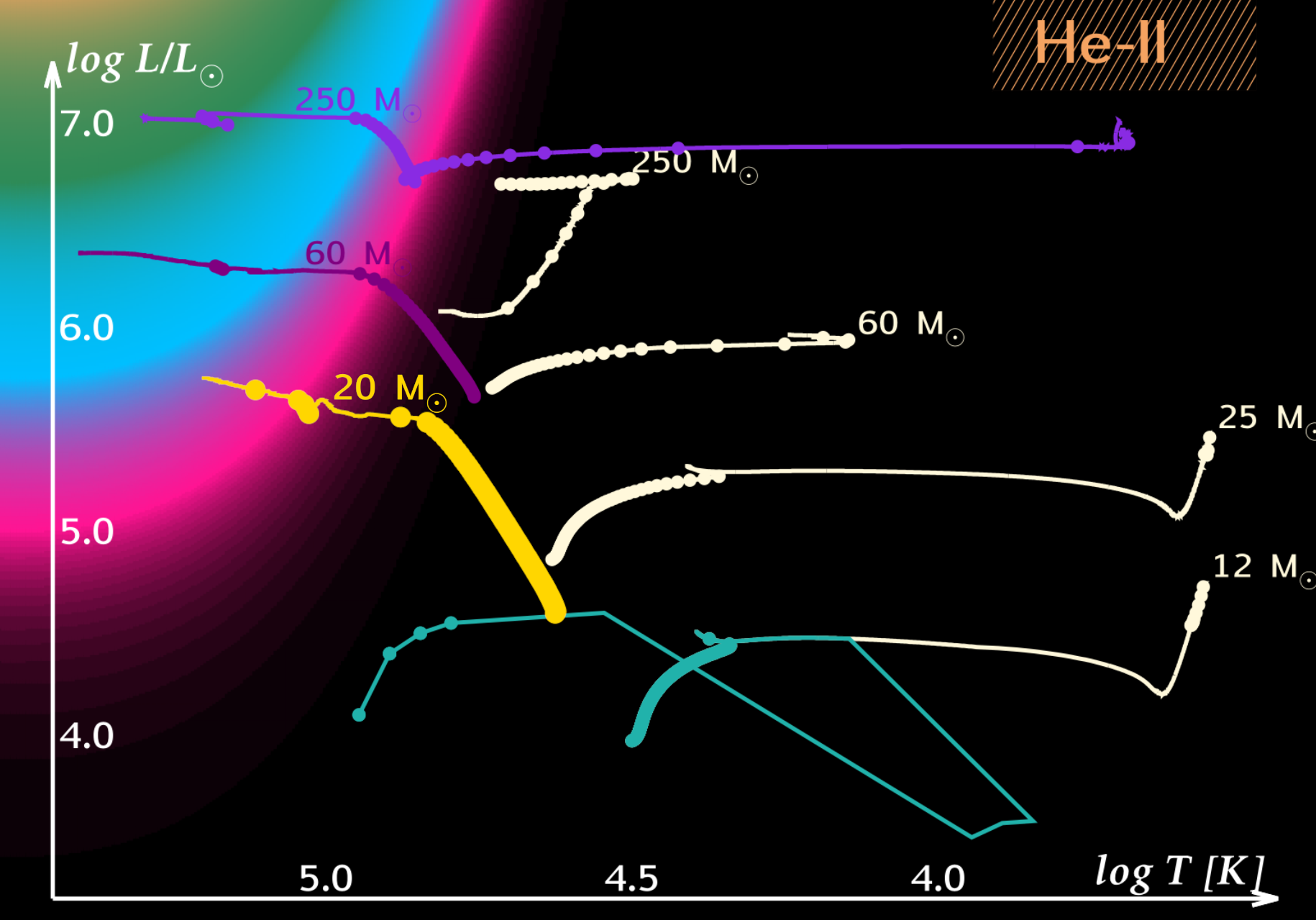}
	\caption{{Hertzsprung--Russell diagrams with background colours representing the number of Lyman (\textit{left}) and He-II (\textit{right}) continuum photons emitted by a black body with a given temperature and luminosity. Units are associated with colours (pink: 10$^{48}$~ph~s$^{-1}$, blue: 10$^{49}$~ph~s$^{-1}$, green: 10$^{50}$~ph~s$^{-1}$, yellow: 10$^{51}$~ph~s$^{-1}$; a black background means $<$\,10$^{48}$~ph~s$^{-1}$). Stellar evolutionary model sequences are labelled with their initial masses. They are taken from the BoOST~project \citep{Szecsi:2022ok}, except for the turquoise line, which is a stripped binary model from \citet{Goetberg:2017}. White lines mean Milky Way composition, where all models evolve the `normal' way (although the most massive ones turn to the left eventually: this is classical Wolf--Rayet evolution). Purple and golden lines mean I~Zw~18 composition, with one of the very massive models evolving normally (initial rotational velocity of 100~km~s$^{-1}$) and the others chemically homogeneously due to fast rotation (500~km~s$^{-1}$). Every 10$^5$~yr of evolution is marked with dot on the tracks. Including the 20~M$_{\odot}$ model into our population synthesis runs (with a sufficiently long star formation episode, see Table~\ref{tab:popsyn}) hardens the combined spectra considerably (reaching close to the observed I(4686)/I($\beta$)~$\sim$~0.02), as the late phases of this model enter the pink zone in terms of He-II photons while not leaving the blue zone in terms of Lyman photons. For more details on the population synthesis and spectral hardness, see Sects.~\ref{sec:population} and~\ref{sec:hardness}, respectively.}
	}
	\label{fig:LymanHeII}
\end{figure*}

There is no need for very massive stars, however, since alternative star formation histories are within reasonable margins of theoretical and observational uncertainties. For example, a continuous episode lasting for 10~Myr with a much more moderate rate of only 0.03~M$_{\odot}$~yr$^{-1}$ also results in a 300~000~M$_{\odot}$ region; this population gives us Q$^{}_{\ion{He}{II}}$~$=$~1.96$\cdot$10$^{50}$~photons~s$^{-1}$ and L$^{}_{1550}$~$=$~3.43$\cdot$10$^{37}$~erg~s$^{-1}$.{We have lowered the top mass value M$_{\rm top}$ of the IMF from 150 to 120~M$_{\odot}$ for this run, because already this more moderate value gives us the right predictions.} In this case, the main source of the UV carbon line is made up by six chemically homogeneously evolving WO stars of mass $\sim$40~M$_{\odot}$ and another one of {mass $\sim$80~M$_{\odot}$;} while the He$^+$-emission is again explained by the combined contribution of these and the rest of the hot-star population (including those classified as type~O in Fig.~\ref{fig:classification}). 
So instead of a couple of very massive WO~stars, a select number of more or less average ones can do the same job. 

{We can relax our assumptions even further, by applying a 0.02~M$_{\odot}$~yr$^{-1}$ rate for a 15~Myr long star formation episode (again with M$_{\rm top}$~$=$~120~M$_{\odot}$; see Table~\ref{tab:popsyn}). This way our 20~M$_{\odot}$ model's post-main-sequence phase (which hardens the resulting ionizing emission considerably; see Sect.~\ref{sec:hardness}) would be captured (the model's total lifetime is 12.3~Myr). In such a scenario, about twenty-three WO stars of 20--25~M$_{\odot}$ and two of around 50~M$_{\odot}$ are responsible for a predicted carbon emission of L$^{}_{1550}$~$=$~2.83$\cdot$10$^{37}$~erg~s$^{-1}$ and an ionizing flux of Q$^{}_{\ion{He}{II}}$~$=$~2.19$\cdot$10$^{50}$~photons~s$^{-1}$. Sect.~\ref{sec:hardness} shows that this third scenario is the most preferred in terms of the spectral hardness, a further observational constraint.}
{Table~\ref{tab:popsyn} presents a fourth and a fifth scenarios, which are even more moderate in terms of M$_{\rm top}$ (80 an 40~M$_{\odot}$), but they fail to account for the observed \ion{C}{iv} luminosity.}

As for correcting for extinction, \citet{Brown:2002} reported that E(B-V)~$=$~0.06 for an SMC-type extinction law fits the data well. From this one obtains a factor $\sim$2.3 attenuation
at $\sim$1500$\,\AA$. 
The recent study of \citet{Berg:2022} quotes E(B-V)~$=$~0.0753$\pm$0.007, which gives a higher correction ($\sim$2.7).
With these numbers, one can estimate
L$_{\ion{C}{IV}}$~$\sim$~(5$-$15)$\cdot$10$^{37}$~erg~s$^{-1}$, which is around the same or up to 3 times higher than the L$^{obs}_{1550}$ {range} we attempt to match here ({2.2--5.5}$\cdot$10$^{37}$~erg~s$^{-1}$). However, as Fig.~\ref{fig:Civ} shows, introducing wind-clumping into our models increases our predicted line luminosity significantly -- up to 3 times higher values with an extreme clumping factor. As is explained in Sect.~\ref{sec:caveats} among caveats, wind-clumping is one of the most underconstrained parameters of massive-star atmospheres, and supposing that the winds are completely unclumped (i.e. smooth) is a rather conservative assumption \citep[see also][]{Roy:2025}. 

We conclude therefore that a population of metal-poor massive single star models -- including those with fast rotation that evolve to be TWUIN stars -- can successfully explain the puzzling properties of the very dwarf galaxy, I~Zw~18, which they were created for. The emission is dominated by WO~stars, {with a total \ion{C}{iv} emission line intensity around 2$-$3$\cdot$10$^{37}$~erg~s$^{-1}$.}

\subsection{Nebular line ratios versus continuum photon ratios}\label{sec:hardness}

The number of ionizing photons in I~Zw~18 was obtained from measuring the strength of the line \li{He}{II}{4686} which originates in the ionized interstellar nebula. \citet{Kehrig:2015} report a line luminosity of 1.12$\cdot$10$^{38}$~erg~s$^{-1}$ from which they derive Q$^{obs}_{\ion{He}{II}}$ using common assumptions about the interstellar medium's properties. 
An important observable to investigate is how the ratio of \ion{He}{II} photons to Lyman continuum photons relate to the observed line intensities of the corresponding, excited nebular lines, \li{He}{II}{4686} and \li{H}{$\beta$}{4861}. Observationally, this hardness ratio is measured to be in the range of I(4686)/I(\ion{H}{$\beta$})~$=$~0.02$-$0.04 (\citealt{Izotov:1998,Lebouteiller:2017}, although \citealt{Kehrig:2015} report values as high as 0.08 in individual spaxels). 
From a theoretical population of massive stars, one can predict this ratio as I(4686)/I(\ion{H}{$\beta$})~$=$~A$\cdot$Q(\ion{He}{II})/Q(\ion{H}{I}), where A is a factor taken to be 1.74 for typical nebular conditions \citep{Stasinska:2015}.

\begin{table*}[t!]
	\centering
	\caption{Summary of the various measurements from the literature reproduced and discussed here.}
	\begin{tabular}{lcccl}\hline\small
		\rule[0mm]{0mm}{5.5mm}\vspace{4pt}
		& Band & Range & Instrument & Find here \\\hline\hline
		\citet{Izotov:1997} & optical & $\sim$3700$-$6800~\AA & MMTO 6.5~m & Fig.~\ref{fig:Izotov} \\\hline
		\citet{Thuan:2005} & optical & $\sim$3200$-$5200~\AA & MMTO 6.5~m & Fig.~\ref{fig:Thuan} \\\hline
		\citet{Kehrig:2015} & optical & $\sim$4440$-$5200~\AA & PMAS 3.5~m (integral field) & Fig.~\ref{fig:Kehrig} \\\hline
		\citet{Brown:2002} & UV & $\sim$1160$-$1710~\AA & HST \textit{STIS} & Fig.~\ref{fig:Brown} \\\hline
		\citet{LecavelierdesEtangs:2004} & UV & $\sim$910$-$1185~\AA & FUSE & Fig.~\ref{fig:Lecavelier} \\\hline
		\citet{Heap:2015}\,/\,\citet{Berg:2022} & UV & $\sim$1160$-$1645~\AA & HST \textit{COS} & Fig.~\ref{fig:Heap} \\\hline
	\end{tabular}
	\label{tab:review}
\end{table*}

In our {fiducial} population
(containing not just WO stars but all hot stars, including O-type stars {from `normal' evolution}; see Sect.~\ref{sec:population}), the predicted photon count is Q(\ion{H}{I})~$=$~7.02$\cdot$10$^{52}$~s$^{-1}$ in the \ion{H}{I} continuum (as reported in Table~\ref{tab:popsyn}) and Q(\ion{He}{II})~$=$~1.80$\cdot$10$^{50}$~s$^{-1}$ in the \ion{He}{II} continuum. From these, we derive the hardness ratio to be {0.004} (Table~\ref{tab:popsyn}), which is {an order of magnitude below what is observed (0.02-0.04).}

{However, for the alternative populations, when lower-mass stars are captured by relaxing the length of the starburst (Sect.~\ref{sec:main}), the value rises to 0.011 and 0.017, approaching the expected range. From this we deduce that the extremely short and intense scenario suggested by \citetalias{Szecsi:2015} (our fiducial population) where a couple of very massive stars dominate the emission is rendered unlikely. Instead, a population where all massive stars' post-main-sequence phases are contributing, not just the youngest (very massive) ones, should be considered realistic. In our second scenario (SF-10 in Table~\ref{tab:popsyn}) the age of the starburst excludes post-main-sequence objects with initial masses below $\sim$\,30~M$_{\odot}$, while in our third scenario (SF-15 in Table~\ref{tab:popsyn}), everything down to $\sim$\,20~M$_{\odot}$ is incorporated. The contribution of this evolutionary model to the hardness of the population is demonstrated by Figure~\ref{fig:LymanHeII}, where evolutionary sequences are compared in terms of their (black-body-estimated) ionizing emissions.}
{We suggest that in the future, \PoWR\ models down to 12~M$_{\odot}$ could be computed and tested for their effect of increasing the hardness -- an apparent trend -- fitting the observed range even better.} 

{Since lowering M$_{\rm top}$ from 150 to 120~M$_{\odot}$ helped us match the measured hardness, we checked even lower values, down to 80 and even 40~M$_{\odot}$. They give less good matches: the general trend is that lowering M$_{\rm top}$ does not influence our \ion{He}{ii} and hardness predictions at all (at least as long as the 15~Myr star formation history is applied), but the \ion{C}{iv} line luminosity decreases below the observed range.}

{We conclude that our theory is consistent with the measured hardness in I~Zw~18, and by matching this value as an additional constraint we can draw important conclusions on the length/intensity of the starbursts, as well as the stellar content, favouring less violent and more relaxed scenarios over those dominated by very-massive stars.} 

\section{Available observations in the literature}\label{sec:literature}

Apart from the \li{C}{iv}{1550} line and the continuum photons, there are further observational features our results should be consistent with. To make such a comparison, we review here the published spectral data of I~Zw~18 focusing on individual emission lines/bumps. The figures below (Figs. \ref{fig:Izotov}--\ref{fig:Heap})
are reproduced from the literature for the sake of this discussion. Table~\ref{tab:review} lists the papers, instruments and spectral ranges (optical or UV) mentioned here. As we shall see, paying attention to the observing strategy of any given survey is key before drawing conclusions.

{We remind the reader that our best fit population predicts $\sim$L$_{\ion{C}{iv}}$~$=$~2.83$\cdot$10$^{37}$~erg~s$^{-1}$ (coming from twenty-something WO stars between 20-50~M$_{\odot}$). This is actually close to what our newly computed \PoWR\ model (with 131~M$_{\odot}$ in the mid-pMS) predicts (2.13$\cdot$10$^{37}$~erg~s$^{-1}$). Therefore, to ease the following discussion, we simply consider this mid-post-main-sequence model
(blue line in Fig.~\ref{fig:newmodel}) representative of the combined spectrum of the population.}

\subsection{Optical emission lines/bumps}\label{sec:opt}
\begin{figure}[h!]\centering
	\includegraphics[width=0.9\columnwidth]{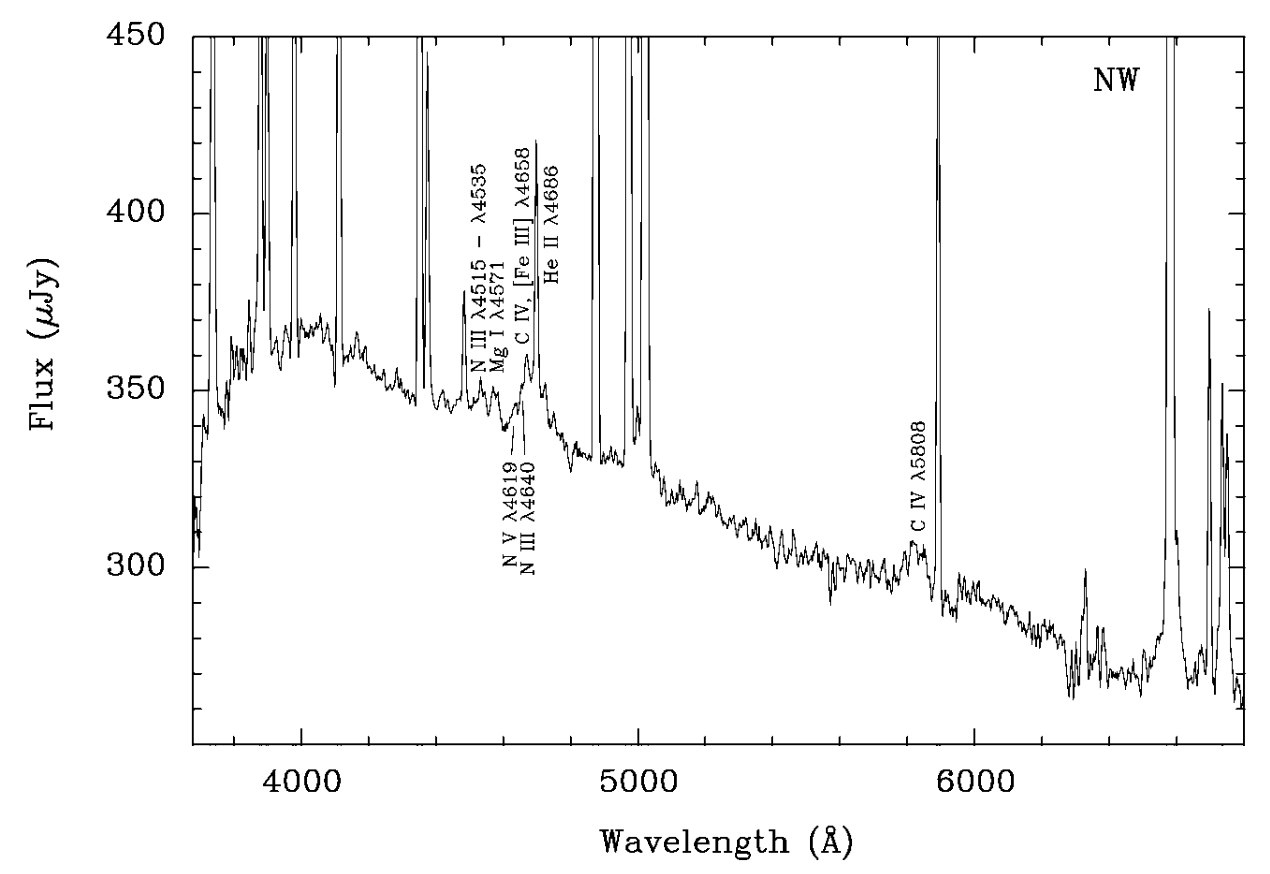}
	\caption{Optical spectra reproduced from \citet[][their Fig.~1]{Izotov:1997}. Spectrophotometric observations were obtained by the MMT Observatory using a 1''.5 $\times$ 180'' slit 
		(blue channel) and a spatial scale of 0''3 per pixel along the slit. Note the prominent broad \li{He}{II}{4686} and \li{C}{IV}{5808} bumps, both explained by our WO stars (as well as the classical WC-star scenario). As for the \li{O}{VI}{3818}, which would prove that the source is WO (and not WC), it is impossible to say if a broad component is there or not, as the flux calibration seems off at this position. 
        For more discussion, see Sect.~\ref{sec:opt}.
	}\label{fig:Izotov}
\end{figure}

\begin{figure}[h!]\centering
	\includegraphics[width=0.9\columnwidth]{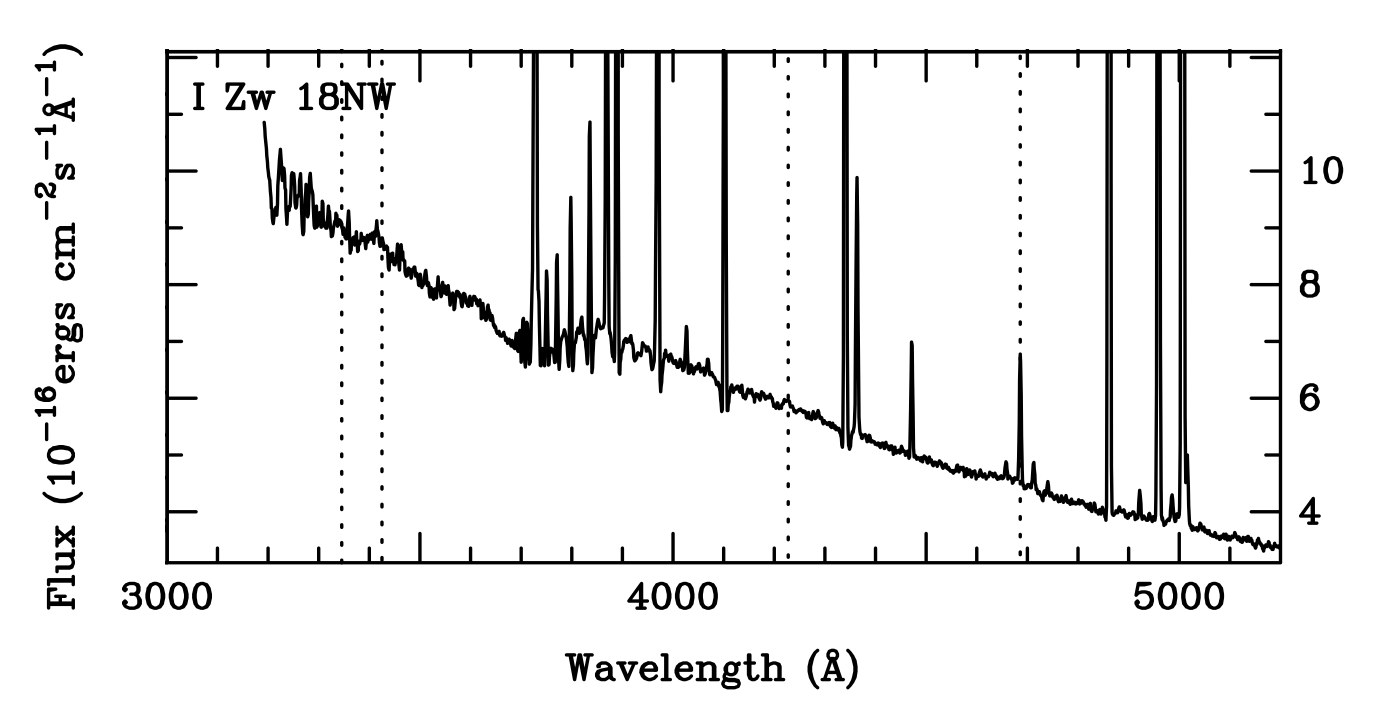}
	\caption{Optical spectra reproduced from \citet[][their Fig.~2]{Thuan:2005}. Observation taken with the MMT spectrograph (blue channel) at the 6.5~m MMTO. A 2''$\times$300'' slit was used, and the brightest part of the north-western \ion{H}{II} region was extracted using a 6''$\times$2'' extraction aperture. For our discussion, see Sect.~\ref{sec:opt}.
	}\label{fig:Thuan}
\end{figure}

\liu{O}{VI}{3818}. \ 
There are two observations available in the literature for this spectral region, \citet{Izotov:1997} and \citet{Thuan:2005}. The first is reproduced here in Fig.~\ref{fig:Izotov} and the second, in Fig.~\ref{fig:Thuan}.
According to our model spectra in Fig.~\ref{fig:newmodel}, this emission is predicted to be not a line but a huge double bump, ranging from 3780$-$3850~\AA\ (see also Fig.~\ref{fig:allopt} showing our line predictions). 
In the \citet{Izotov:1997} spectra  (Fig.~\ref{fig:Izotov}), the flux calibration seems to be off at the position in question (it suddenly drops at the edge) making it hard to tell whether a broad component is there or not.
The second paper, \citet{Thuan:2005}, 
did not report detection of the \li{O}{VI}{3818} line (see their Table~3), which is consistent with what one sees looking at Fig.~\ref{fig:Thuan}.
However, the Thuan spectra also do not show the 
blue bump centred at $\lambda$~$\approx$~4650~$\AA$
that is so prominent in the Izotov spectra. Thus, seeing other wide bumps of stellar origin should not be expected in the Thuan data, either. 
We conclude that the currently available literature does not exclude our model predictions about this emission bump. 

\begin{figure}[h!]\centering
	\includegraphics[width=0.7\columnwidth]{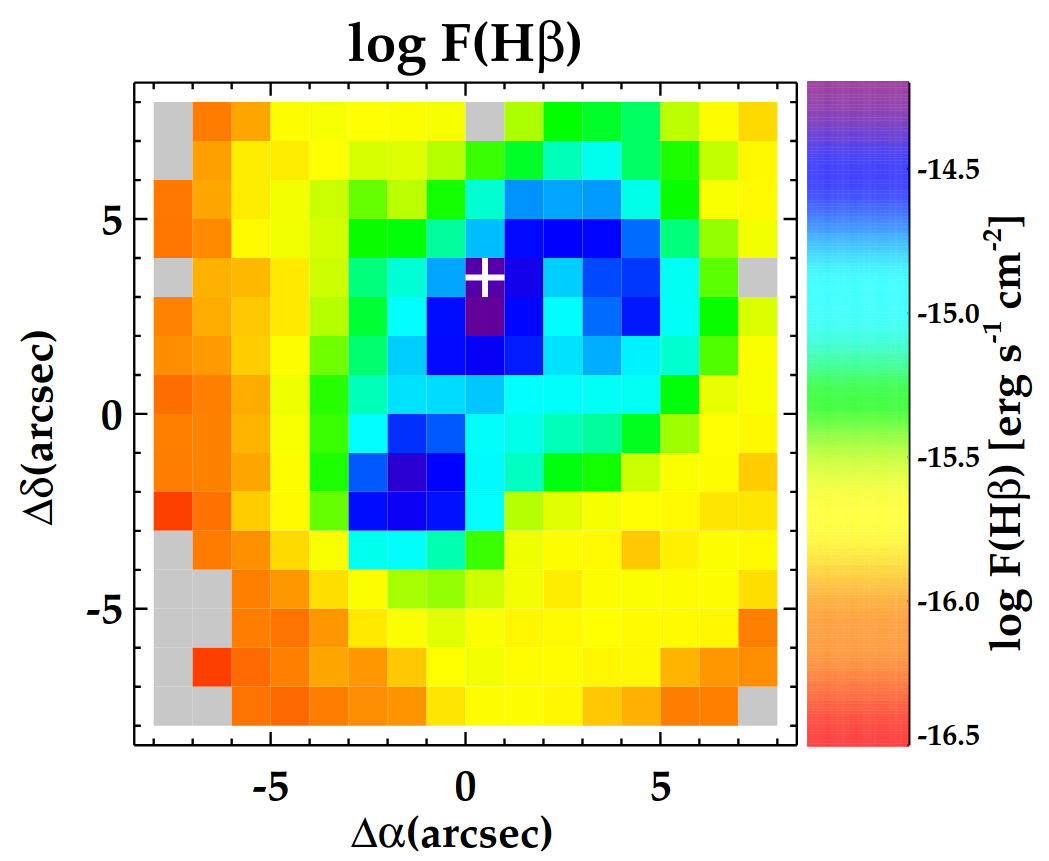}
    \includegraphics[width=0.7\columnwidth]{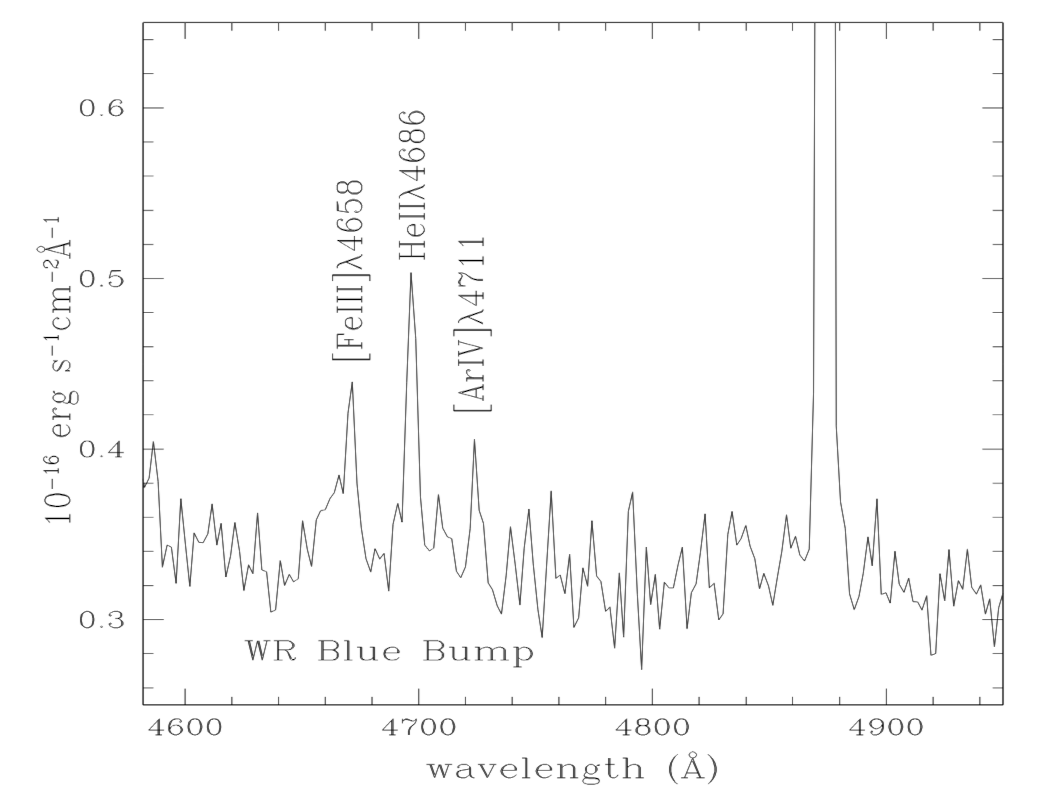}
	\includegraphics[width=\columnwidth]{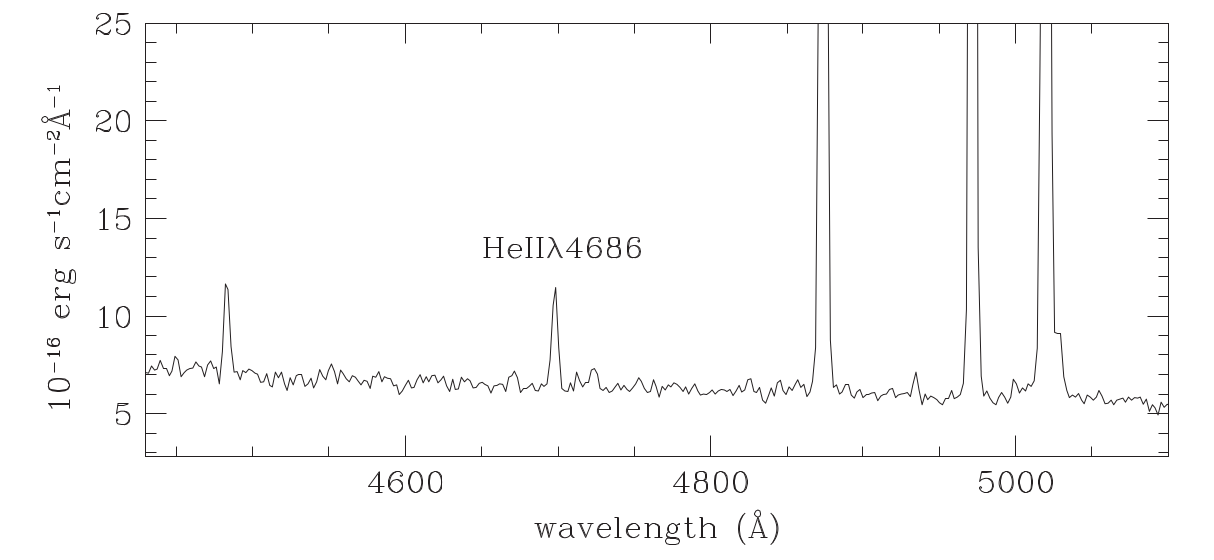}
	\caption{Optical spectra reproduced from \citet[their Figs.~2 \& 6, here \textit{top} \& \textit{middle}, respectively]{Kehrig:2016} and \citet[their Fig.~3, \textit{bottom}]{Kehrig:2015}. Integral field spectroscopy was obtained with the Potsdam Multi-Aperture Spectrophotometer (PMAS) on the 3.5 m telescope at the Calar Alto Observatory (Almeria, Spain). \textit{Top:} H$\beta$ map with a cross marking where the spectrum in the middle was extracted from. 
    \textit{Middle:} Spectrum showing a prominent WR-bump associated with stellar emission underneath the narrow, nebular \li{He}{II}{4686} line, discussed in Sect.~\ref{sec:opt}.
    \textit{Bottom:} Integrated spectrum of the north-western region of I~Zw~18, demonstrating how stellar emission bumps clear out completely when the full galaxy is integrated over, leaving only the narrow nebular component. Compare this to the \textit{FUSE} data in Fig.~\ref{fig:Lecavelier}, which we discuss in Sect.~\ref{sec:uv}.
	}\label{fig:Kehrig}
\end{figure}

\liu{He}{II}{4686}. \ 
This optical line is peculiar because it may have a stellar and a nebular component. The nebular one is narrow, the stellar (if there) broad, bump-like: our theory predicts a wide double bump between 4640-4700~\AA \ (Fig.~\ref{fig:newmodel}) that should be about as prominent as the \ion{C}{iv} bump around 5808~\AA.
The \citet{Izotov:1997} spectra therefore supports our theory: a bump is clearly there in Fig.~\ref{fig:Izotov} under the narrow nebular component. The same is true for the spectra presented in \citet{Kehrig:2016}, shown here in Fig.~\ref{fig:Kehrig}. As for \citet{Thuan:2005}, they measure an equivalent width of 2.5~\AA \ (their Table~3) for the nebular line; 
whether there is a 60~\AA-wide, almost completely cleared-out bump of stellar origin underneath it (Fig.~\ref{fig:Thuan}), it is hard to say.
We speculate that the lack of stellar emission bumps in the Thuan data may be attributed to the extraction aperture (6''$\times$2'') not crossing over the star clusters of interest, while the aperture of the Izotov data (5'', centred on the north-western region) includes these regions.
Indeed, the position and size of the extraction apertures may play a key role in the detection of (or lack of) spectral features, as demonstrated by \citet[][see their Fig.~11]{Kehrig:2013}.
We conclude that the existing literature does not exclude our model predictions about this line. 

\liu{C}{IV}{5808}. \ 
Only the \citet{Izotov:1997} observation covers this region (Fig.~\ref{fig:Izotov}). This measurement does show a bump at 5808, thus supporting our theory. Even the relative strength of it compared to the above discussed \li{He}{II}{4686} bump is tentatively consistent with our predictions in Fig.~\ref{fig:newmodel}.

\liu{C}{IV}{7724}. \ 
We did not found any spectra in the literature covering this wavelength.

\subsection{UV emission lines/bumps}\label{sec:uv}

\begin{figure}[h!]\centering
	\includegraphics[width=0.5\columnwidth]{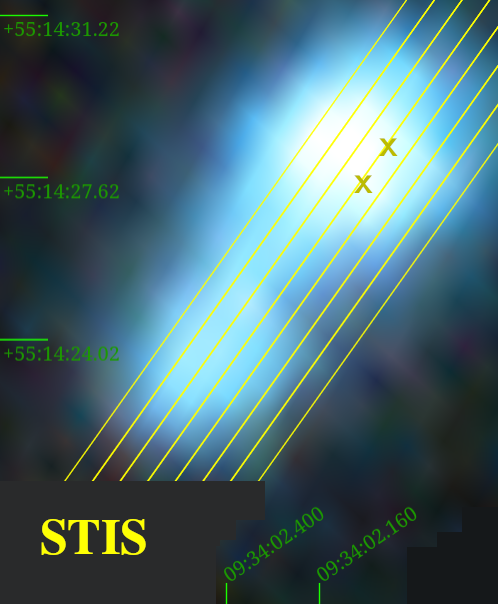}
	\includegraphics[width=0.7\columnwidth]{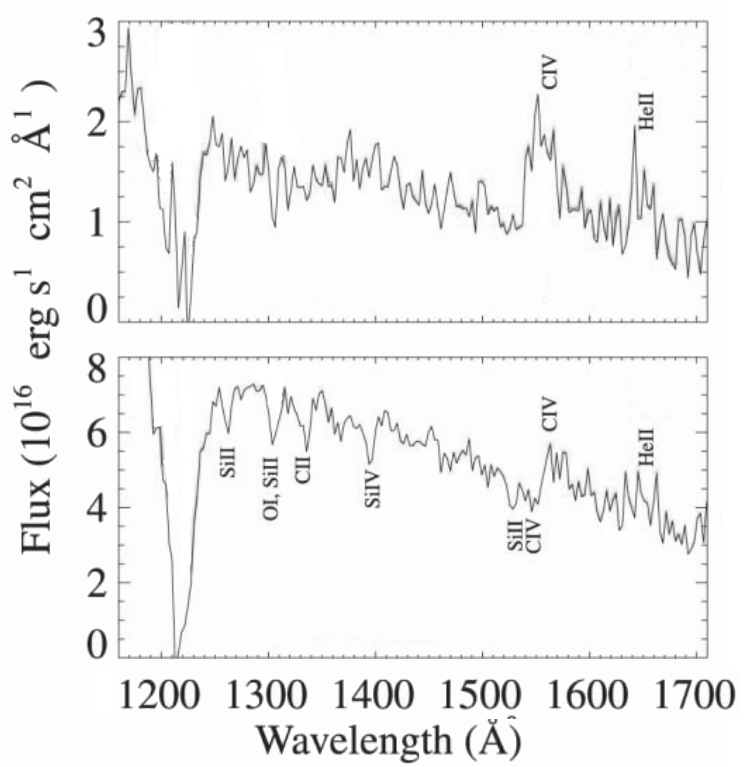}
	\caption{UV spectra reproduced from \citet[][their Fig.~1]{Brown:2002}. Observations taken with Hubble's Space Telescope Imaging Spectrograph (STIS) using the G140L grating and the 52'' $\times$ 0''.5 slit. The two spectra (\textit{bottom}) were taken from two clusters in the north-western region that contain \ion{C}{iv} bumps (\textit{top}, where {Xs mark the approximate positions from which those spectra were extracted; see our Fig.~\ref{fig:COSBrown} where these positions are more reliably pinpointed by the authors themselves}). Note the presence of \ion{He}{ii} bumps, which are similar in size to the \ion{C}{iv} bumps, {consistently with out models} (see Sect.~\ref{sec:uv}). 
	}\label{fig:Brown}
\end{figure}

\liu{He}{II}{1640}. \ 
The appearance of this UV emission line (bump) is crucial, as it is typically used as a proxy for first stars (Pop-III). However, there are other ways to produce this line, as our models (massive Pop-II stars) attest.
The bump is visible in the HST \textit{STIS} spectra of \citet{Brown:2002}, as is seen in Fig.~\ref{fig:Brown}, and its size is comparable to that of the \li{C}{IV}{1550} bump.  
Our theoretical populations nicely account for this. 
We predict line luminosities of lg\,L~$=$~37.58 for \ion{C}{iv} (see Fig.~\ref{fig:Civ}) and lg\,L~$=$~37.35 for \ion{He}{ii} (see Fig.~\ref{fig:allUV}) in the case of our mid-post-main-sequence WO~model 
that dominates the population. As for the alternative population where a select number of more moderate-mass models dominate (see Sect.~\ref{sec:main}), the situation is similar: the values the WO~star with M$_{ini}$~$=$~59~M$_{\odot}$ gives are lg\,L~$=$~36.82 for \ion{C}{iv} and lg\,L~$=$~36.63 for \ion{He}{ii}. Thus, this UV emission line supports our theory. We note however that when attempting to repeat the data reduction (the observations are publicly available in the MAST archive\footnote{\url{http://mast.stsci.edu/portal/Mashup/Clients/Mast/Portal.html}}) we found it challenging to identify the prominent bumps reported by \citet{Brown:2002}. While reanalysing archival HST data is outside the scope of the current work, we encourage ongoing projects (e.g. S.~Heap, private comm.) in this direction. 

\begin{figure}[h!]\centering
	\includegraphics[width=0.5\columnwidth]{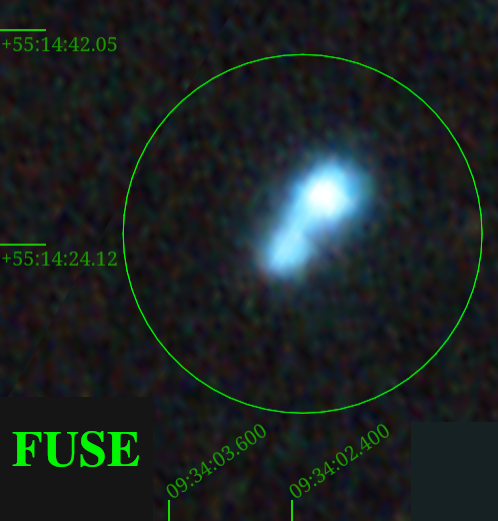}
	\includegraphics[width=\columnwidth]{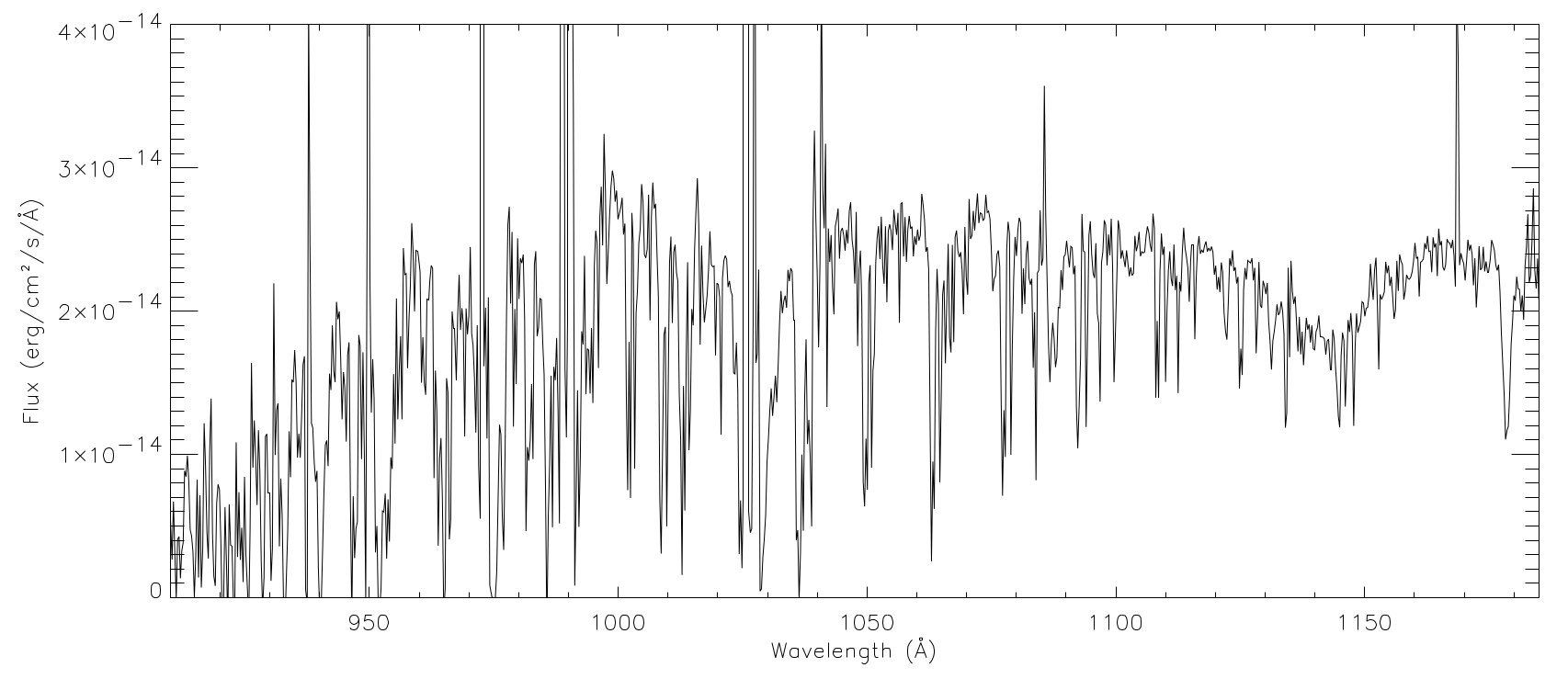}
	\caption{UV spectra reproduced from \citet[][their Fig.~1]{LecavelierdesEtangs:2004}. Observation taken with \textit{FUSE}, with an aperture (30”x20”) covering the whole galaxy. \textit{Top:} Position of the observation; picture is taken from the MAST archives {with background image from Pan-STARRS}. \textit{Bottom:} Spectra presented in \citet{LecavelierdesEtangs:2004}. Since the full galaxy is in the aperture, the features of the localized stellar populations are not expected to be seen (see Fig.~\ref{fig:Kehrig}, as well as our discussion in Sect.~\ref{sec:uv}).
	}\label{fig:Lecavelier}
\end{figure}

\liu{O}{VI}{1037}. \ 
This UV line is even stronger in our model spectra 
than \li{C}{IV}{1550} (see Figs.~\ref{fig:newmodel} and \ref{fig:allUV}). Hence, we should look for it in the observations. \citet[][Fig.~\ref{fig:Brown}]{Brown:2002} 
do not cover the position of \li{O}{VI}{1037}, unfortunately. Other works, such as \citet{LecavelierdesEtangs:2004} do, although without covering \li{C}{IV}{1550}, as is seen in Fig.~\ref{fig:Lecavelier}. However, the observational strategy is again an important factor. \citet{Brown:2002} took their UV spectrum as a slit crossing through the star-forming region in the north-western cluster (see their Fig.~1 where they show their slits on the detailed map of I~Zw~18, as well as our less detailed Fig.~\ref{fig:Brown} where we show the same). \citet{LecavelierdesEtangs:2004} applied another strategy: the FUSE aperture they used 
covers 
the full galaxy (30”x20”), as is shown in Fig.~\ref{fig:Lecavelier}.
In this integrated spectrum, a stellar emission in \li{O}{VI}{1037} is not necessarily expected to be seen -- the same way stellar emissions in the optical are not seen in the integrated spectrum presented by \citealt{Kehrig:2015}: the 4650 bump, otherwise identifiable in targeted observations (e.g. \citealt{Izotov:1997}, see Fig~\ref{fig:Izotov})
is completely cleared out when the data is summed over,
as is shown by Fig.~\ref{fig:Kehrig} (reproduced from Fig.~3 of \citealt{Kehrig:2015}).
Still, one might identify a P-Cygni profile around \ion{O}{VI}{1032} in the FUSE spectrum, just as we predict (see Fig.~\ref{fig:allUV}).
We conclude that the available data in the literature does not exclude our model predictions about this line.

\begin{figure}[h!]\centering
	\includegraphics[width=0.5\columnwidth]{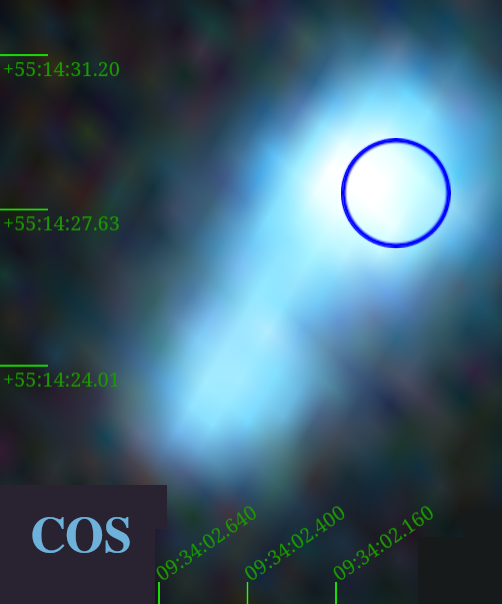}
	\includegraphics[width=\columnwidth]{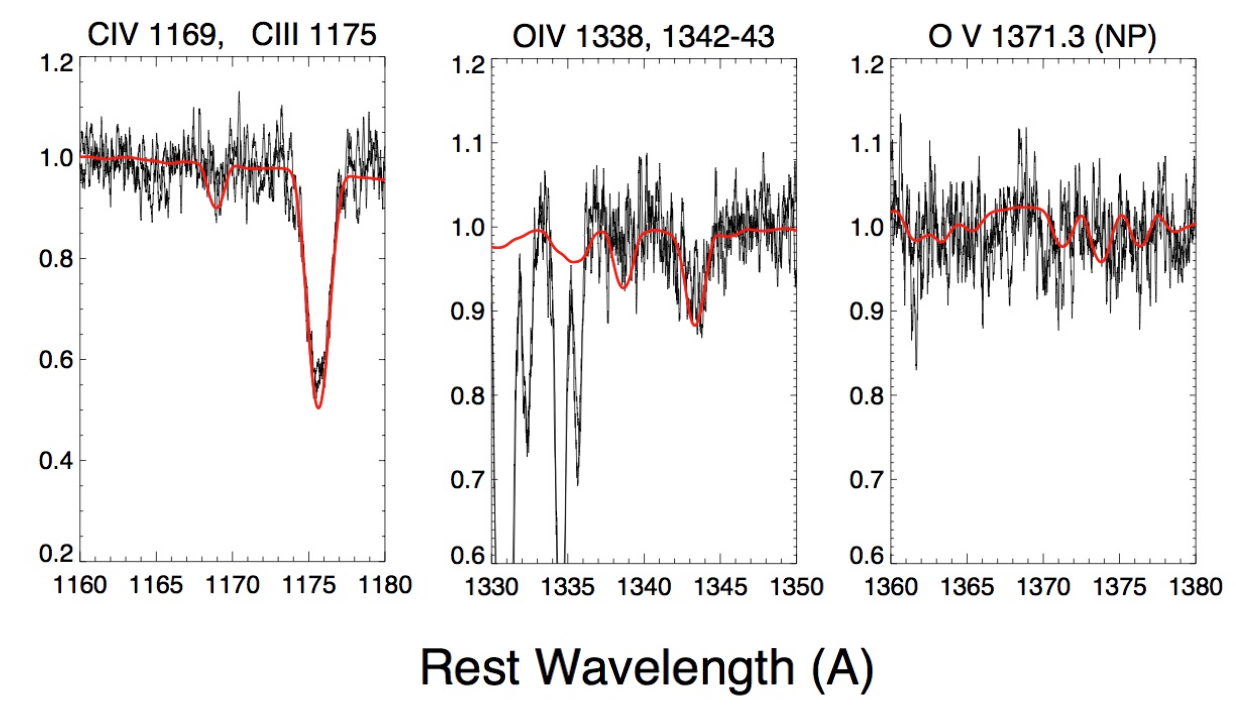}
	\caption{UV spectra reproduced from \citet[][their Fig.~5]{Heap:2015}. \textit{Top:} Regions of the \textit{Hubble COS} observations according to the MAST archive (PI: J.C.~Grean, year: 2010, aperture: 2''.5), {background image taken from Pan-STARRS}. \textit{Bottom:} Observation taken with \textit{COS} (black); a model spectrum (red) of a 5~Myr stellar population using the \textsc{Tlusty} atmosphere code is overplotted. For details, see Sect.~\ref{sec:uv} and Fig.~\ref{fig:COSBrown}.
	}\label{fig:Heap}
\end{figure}

\begin{figure*}[h!]
	\centering\includegraphics[width=1.6\columnwidth]{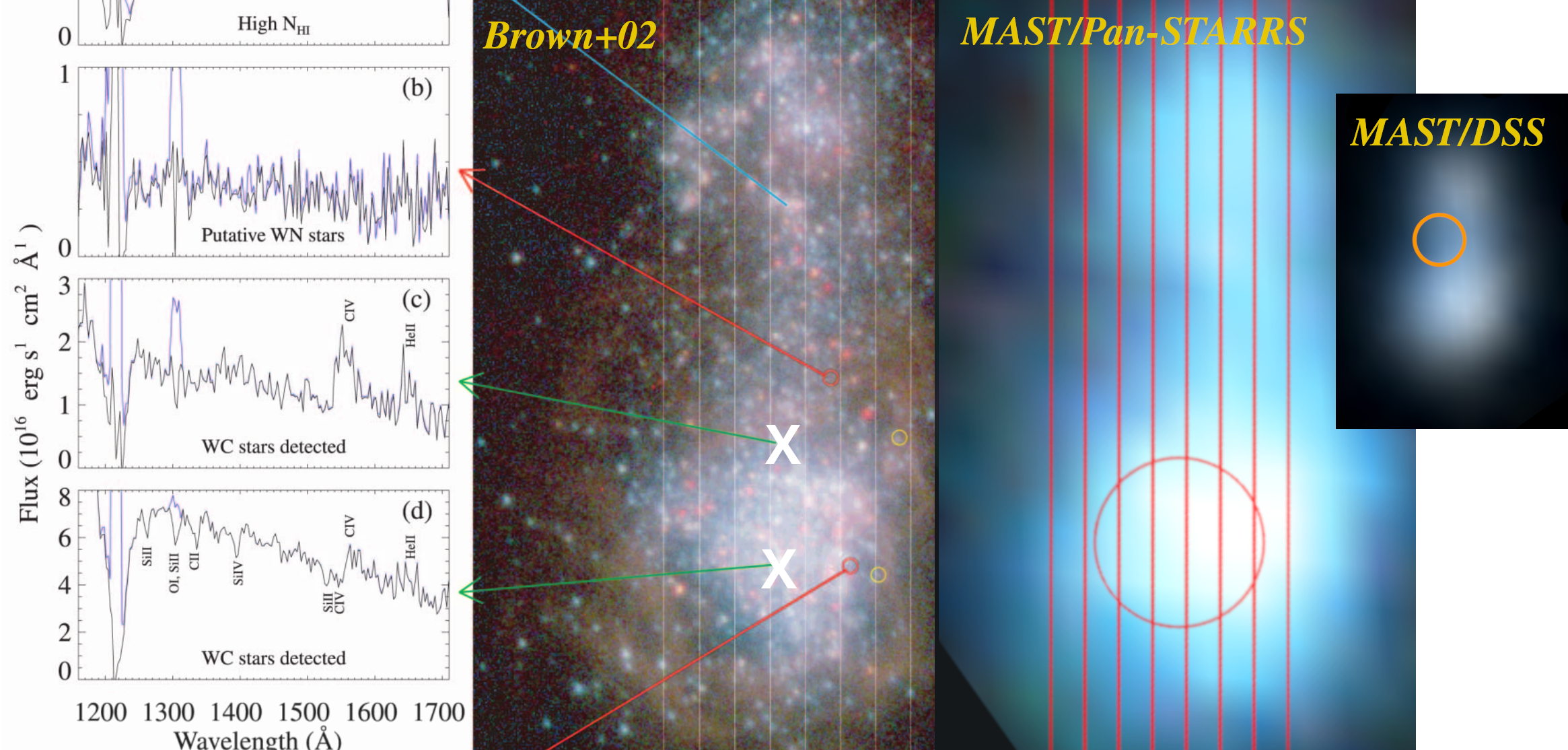}
	\caption{{HST measurements of I~Zw~18.} The panel labelled \textit{MAST/Pan-STARRS} shows the circular aperture used for obtaining the highest signal-to-noise-ratio UV spectrum of I~Zw~18 available to date, taken by COS (\citealt{Berg:2022}). {Circle and background image are the same as in Fig.~\ref{fig:Heap}} (taken from the MAST archive applying background image from the Pan-STARRS survey). The panels to the left correspond to STIS copied from \citet[][mentioned also in Sect.~\ref{sec:uv} and Fig.~\ref{fig:Brown}, see details there]{Brown:2002}. The positions where \citet{Brown:2002} identified WR-like emission bumps (\textit{c} and \textit{d}) are marked by Xs for convenience. {The panel labelled \textit{MAST/DSS} demonstrates a discrepancy we found in the MAST Archive's AstroView tool (see Sect.~\ref{sec:MAST} for details).}
    {Our explanation for why the COS spectrum analysed by \citet[][see their Fig.\,12]{Berg:2022} does not display \ion{C}{iv} emission -- despite at least one of the Brown-sources seemingly being within the aperture -- is that the field of view includes too much contaminating light: as demonstrated by \citet{Kehrig:2016} using integral field spectroscopy of I~Zw~18 (see our Fig.~\ref{fig:Kehrig}), stellar emission bumps clear out completely unless the spectra is taken from the right physical position of the galaxy. --- \textit{After carefully reviewing these and other measurements in the literature, we conclude that a population of chemically-homogeneously-evolving O and WO stars existing in I~Zw~18 and being responsible for the high observed photoionization is fully supported by the currently available observational evidence.}} 
	}\label{fig:COSBrown}
\end{figure*}

\liu{C}{iv}{1169}, \liu{C}{iii}{1175}, \liu{O}{IV}{1340}, \liu{O}{V}{1371}. \ 
These lines/bumps are mentioned in \citet{Heap:2015} who presents 
\textit{Hubble}’s Cosmic Origins Spectrograph (COS) observations
for them (their Fig.~5, reproduced here as Fig.~\ref{fig:Heap}). Since emission seems to be absent in their standard fit, they conclude that ,,the ultraviolet spectrum of [the north-western region] shows no evidence of extremely hot, luminous stars''. {The same conclusion was arrived at by \citet{Berg:2022}, after analysing the highest signal-to-noise-ratio UV spectrum of I~Zw~18 available to date, taken by COS/HST (and incorporated in CLASSY).} However, the observational strategy is again important here. As Fig.~\ref{fig:Heap} shows, the COS {field of view extends to a substantially large fraction of the north-western region, as opposed to using slits crossing through the sources of interest (similarly to the STIS data, see Fig.~\ref{fig:COSBrown} for a direct comparison). As is demonstrated by the integral-field-spectroscopy results of \citet[][which we discussed in Fig.~\ref{fig:Kehrig}]{Kehrig:2016}, emission bumps of stellar origin are only visible when the spectrum is extracted from the right spatial position where those massive stars (star clusters) are physically located. When the entirety of the light is integrated over, the stellar emission clears out (Fig.~\ref{fig:Kehrig}, bottom). Therefore, we the fact that the COS data show no stellar emission bumps does not necessarily mean that those stars are not there. It may simply mean that the COS aperture is too large to detect them, and more targeted campaigns are needed.
}
We conclude that these observations do not disprove our theory either.

\liu{O}{VI}{2070}. \ We did not found any spectra in the literature covering this wavelength. 

\subsection{On using MAST AstroView}\label{sec:MAST}

{While conducting this literature review, we have found a discrepancy in the AstroView tool of the MAST Archive.
As is shown by Figure~\ref{fig:COSBrown}, the two available backgrounds give strikingly different visuals at the position of I~Zw~18 (RA\,$\sim$\,09:34:02, DEC\,$\sim$\,+55:14:31). While the COS aperture's position is the same, the DSS background (default option) is off by $\sim$\,4 arcsec compared to the Pan-STARRS background (alternative option in the AstroView settings). This can easily lead to confusion when interpreting the COS data (or any other measurement).}

{According to the MAST Team (Y.~Li, private comm.), the Pan-STARRS version is more accurate, although they have not confirmed this to be always true. They hope that these image products will be improved in the future, and they do not recommend using them for scientific analysis, only for preview. While for our present purposes, such previews are quite enough, this is a warning to the community that discrepancies of a few arcseconds can be expected when it comes to I~Zw~18 and these images. For proper studies, coordinates from recent reference catalogues such as Gaia and Pan-STARRS1 should be used to confirm the accuracy of the galaxy's -- and especially its various regions' -- celestial positions.}

\subsection{Summary of the literature review}\label{sec:sum}

As is seen from this careful analysis of the literature -- paying close attention to the observational strategies such as which apertures were used for which measurement -- the currently available data either support, or at least never exclude, our theoretical predictions.
Even the Brown data, which claims the presence of WC stars, do not exclude the possibility of these being, in fact, WO stars: the \li{O}{iv}{1340} and \li{O}{V}{1371} bumps are actually visible in their STIS spectra in Fig.~\ref{fig:Brown}. Our models predict these lines to be nowhere near as relevant as for example \li{C}{IV}{1550} or \li{He}{II}{4686} (Figs.~\ref{fig:newmodel} and \ref{fig:allUV}), which is exactly what we see in the Brown spectra.
We conclude therefore that our suggestion about chemically homogeneously evolved WO~stars being responsible for the emission bumps so far associated with WC~stars is consistent with all available observations of I~Zw~18.


\section{Discussion}\label{sec:disc}

\subsection{Other possible explanations}

The suggestion that nearly metal-free, hot massive stars (Pop~III-like) are responsible for the \ion{He}{II} emission in I~Zw~18 \citep{Kehrig:2015,Kehrig:2015b,Heap:2015} is certainly appealing, but relies on the galaxy
keeping (or obtaining) pockets of primordial gas that have just recently started to form massive stars for some {unclear} reason {(see the review of \citealt{Klessen:2023} claiming that Pop III star formation ended roughly at z~$\sim$~5)}. 
Our scenario follows more straightforwardly from the present day conditions in I~Zw~18. 
Our stellar evolutionary models were computed with the measured composition of the very dwarf galaxy we study (following the gas composition reported in \citealt{Lebouteiller:2013}), while our synthetic spectra were created based on these evolutionary models with the most state-of-the-art physics. The population synthesis is a rather general one, without extreme assumptions. So if metal-poor massive stars behave the way the models predict, then both Q$^{obs}_{\ion{He}{II}}$ and L$^{obs}_{1550}$ are properly accounted for -- without needing to postulate the presence of Pop~III-like stars. Note that according to the seminal work of \citet{Yoon:2012}, Pop~III stars also evolve chemically homogeneously when rotation is included, the same way as our metal-poor (i.e. massive Pop~II) stars do.

As for the scenario proposed by \citet{Pequignot:2008,Lebouteiller:2017} and \citet{Schaerer:2019} on X-ray binaries, we note that chemically homogeneously evolving stars have been shown to lead to high-mass X-ray binaries in case they are born in a close binary system \citep{Marchant:2017}. While \citet{Kehrig:2021} refutes that high-mass X-rays binaries and/or diffuse soft X-ray photons would be the explanation for the high-ionizing features in I~Zw~18, it may still be true in other galaxies. Hence, we suggest our scenario to be complementary to that of \citet{Schaerer:2019}, and welcome further research into the direction of combining these two possibilities (e.g. applying the methods developed in \citealt{Sen:2024} for predicting faint X-ray emissions). 

Indeed, while we only use single star models here, massive binaries are able to evolve chemically homogeneously too \citep[e.g.][]{deMink:2009b,deMink:2016,Marchant:2016}. Alternatively, they may become analogous to our single, chemically homogeneous stars after losing their envelopes in a mass transfer phase \citep[][]{Goetberg:2017}. 
Since double systems may be common in massive stellar populations \citep{Sana:2012,Hainich:2018,Spencer:2018}, a more comprehensive picture of the contribution of chemically homogeneously evolving stars to the photoionization and carbon line emission in metal-poor galaxies will need to involve not only our fast rotating single stars \citep[especially since there is ongoing debate on the effectiveness of rotationally induced mixing, e.g.][]{Higgins:2019}, but binaries as well. 

As for the model of \citet{Oskinova:2022} suggesting cluster-winds emitting X-rays, we note that this again can be nicely reconciled with our models. Indeed, our very evolutionary models have been applied in cluster-wind studies accounting for X-ray emission in Green Pea galaxies \citep{Franeck:2022}. A similar study for I~Zw~18 would be, therefore, quite possible, and we foresee a follow-up project combining the ionizing emission from massive stars \textit{and} their shocked winds.

{\citet{Roy:2025} recently suggested that classical WR stars with clumpy winds can produce hard photons in local and high-z galaxies. To test this scenario's validity in I~Zw~18, the \ion{C}{IV} and other observed emission line luminosities that their models predict would need to be compared to observations. The same is true for the seminal work of \citet{Lecroq:2024}, in which they build a complex spectral synthesis tool including binary processes and investigate various emission lines (both nebular and stellar) in metal-poor stellar populations. We hope that by compiling existing UV and optical spectra from the literature on I~Zw~18 -- one of the best-studied dwarf galaxies out there -- and by critically discussing this set of data in terms of its capacity to guide modelling efforts (Sect.~\ref{sec:literature}), our work will assist future endeavours like these when testing their predictions directly against I~Zw~18.} 

\subsection{Caveats}\label{sec:caveats}

Caveats in our theory include uncertainties inherent in the simulations (see Sect.~6.2 of \citetalias{Kubatova:2019} for a detailed discussion on this, as well as \citealt{Agrawal:2022}). Some of the main sources of uncertainty are mass loss, wind clumping and rotational mixing efficiency. Especially mass loss and wind clumping can directly influence the strength of the spectral lines -- but not the ionizing flux.
For example, reducing the mass-loss rate by two orders of magnitude erases any trace of the carbon emission even in the post-main-sequence phase of our models, as is seen in Fig.~B4 of \citetalias[][]{Kubatova:2019}; however, this has practically no effect on the total number of ionizing photons (see Fig.~B1 of the same paper). On the other hand, supposing that winds are clumped has an \textit{opposite} effect on the emission lines: making them more prominent, as is shown by Figs.~\ref{fig:Civ} and~\ref{fig:allopt}-\ref{fig:allUV} while not influencing the SED. This means that if mass-loss rates turn out to be lower then the rather high value assumed here, increasing wind clumping -- another unconstrained property \citep[see][]{Roy:2025} -- may still lead to the same conclusion, making our results quite robust.

As for the efficiency of rotational mixing, it is a widely studied yet unconstrained parameter in stellar evolution theory \citep{Maeder:2000,Meynet:2000,Brott:2011a,deMink:2013,Burssens:2023}. However, our results are rather robust in this context as well. As is explained in Sect.~\ref{sec:population}, we make the simple assumption that 10\% of stars rotate fast enough for CHE (guided by the observed initial rotational-velocity distribution of massive single stars in the SMC). So even if rotational mixing is found to be weaker than in our models, our results would be unchanged {as long as 10\% of the population follows the chemically homogeneous path}. 

The same is true for other modelling ingredients that influence stellar evolution (e.g. mass-loss rate prescriptions, convective overshooting, semi-convection, core-envelope coupling): with $\sim$10\% of massive stars in the population evolving chemically homogeneously, either as single stars or binaries\footnote{This ratio is in line with \citet{Dorozsmai:2024} who investigated populations of hierarchical triples with a chemically homogeneously evolving inner binary (based on the rapid binary population synthesis method developed by \citealt{Riley:2021}): in their Table~1, they report 10\% of their systems following the chemically homogeneous path. (Whether such triples survive until core collapse is another question, as is shown by \citealt{Vigna-Gomez:2025}.)}, their spectra will look like how the \PoWR\ models predict: so they will help us explain the peculiar features of I~Zw~18. 

As was shown recently by the seminal work of \citet[][see their Sect.~6.1.3]{Abdellaoui:2023}, fast rotation has a strong effect on stellar winds, as well as on the emergent ionizing fluxes. Including these effects (by recomputing our \PoWR\ models using the mass-loss rates obtained by \citealt{Abdellaoui:2023}) will be part of a future work. To repeat, we do not have any ``classically'' occurring WO~stars in our population (i.e. WOs that got stripped by stellar-wind mass loss) because the evolutionary models we rely on do not predict such strong stellar winds at this low metal content \citep{Szecsi:2015,Szecsi:2022ok}. All our WOs evolved via the chemically homogeneous route, as explained in Sect.~\ref{sec:population}.

The star formation history of I~Zw~18 is not conclusively established \citep{Kunth:2000,Papaderos:2002,Izotov:2004,Papaderos:2012}, and indeed it has been debated whether it is a typical dwarf galaxy at all \citep{Aloisi:1999,Guseva:2000,Annibali:2013}. For example, the IMF may be top-heavy in starbursts; this would help our case. 
Still, our result that massive stars of I~Zw~18-composition are consistently able to account for two, previously unreconcilable observational characteristics while fulfilling all observational constraints available in the literature, implies that their existence should be supposed -- and, therefore, further studied -- in other galaxies of this type too. For example, some dwarf galaxies may be different from I~Zw~18 \textit{not} because of the typical evolutionary behaviours of their massive stars, but because their star formation history progressed another way due to, for example, different dynamical interactions with other galaxies \citep[see][]{Christensen:2016}, or to different thermodynamic/magneto-hydrodynamic conditions in their interstellar gas \citep[see][]{Hopkins:2011,Seifried:2017,Girichidis:2018,Haid:2019}, leading to different rotational velocities (or different binary fraction) in the current massive stellar population. Indeed, to obtain a comprehensive theory of low-metallicity massive stars, one will need to correct for such environmental characteristics when accounting for a larger sample of dwarf galaxy observations \citep[see also the conclusions in][]{Stanway:2019}, including any possible contribution of the interstellar medium to the emission lines commonly attributed to stellar origin.


\section{Conclusions}\label{sec:conc}

We have shown that a population of massive Pop-II stars can consistently reproduce the observationally derived \ion{He}{II} ionizing flux and the observed \li{C}{IV}{1550} line luminosity in the dwarf galaxy I~Zw~18. 
The source of the emission bump is a small number of \textit{chemically homogeneously evolved} WO~stars: these, combined with the rest of the hot O~star population ({both from normal evolution and CHE, the latter} dubbed `TWUIN' in \citetalias{Szecsi:2015} and \citetalias{Kubatova:2019} in order to raise attention to their predicted low wind-opacity and thus the lack of emission features) explain the high photoionization. The evolutionary and atmospheric models were created with the measured composition of the gas in I~Zw~18, thereby making our results self-consistent. Additionally, we carefully evaluated the literature and showed that neither the existing UV spectra nor the optical spectra are able to contradict our scenario. 

When performing population and spectral synthesis, we assumed a normal Salpeter mass function {up to 120~M$_{\odot}$} and a rather small fraction (10\%) of single stars rotating fast enough for CHE. {This makes our result quite robust, as it means that the same conclusions would be drawn by a binary-population study too, as long as 10\% of the systems evolve chemically homogeneously}. By varying our assumptions about the star formation history, {the best match we found predicts $\sim$\,20 WO stars that have a mass of around 25~M$_{\odot}$ and two of 50~M$_{\odot}$} 
to account for the observations, {including the spectral hardness}.

Furthermore, we have established a new sequence of evolutionary phases, typical for chemically homogeneously evolving stars at Z=0.0002: O $\rightarrow$ WN $\rightarrow$ WO. The WC~phase, if it arises at all between WN and WO, lasts no longer than $\sim$1\% of the total lifetime. {While we have based this study on single stars, we expect the same sequence for binaries in the chemically homogeneously evolving channel \citep{Hainich:2018}.}

From our careful review of the literature, we conclude that while a certain kind of signal may dominate in one measurement (similarly to how stellar emission dominates in some of the spectra we discuss in Sect.~\ref{sec:literature}), it does not follow that it will dominate all other measurements from the same, spatially extended galaxy. The telescopes may point at different regions, apply differing extraction apertures, or have lower sensitivity in a given spectral range. Paying attention to the observational strategy (e.g. slit vs integrated spectrum, pointing, size of aperture) is crucial when comparing various pieces of data from various instruments, especially when it comes to spatially resolved objects such as I~Zw~18. {In this context, detecting the emission bump around \li{O}{VI}{3818} (where the existing data is problematic, Sect.~\ref{sec:opt}) is a worthwhile future goal that could conclusively prove our theory about these WO stars -- and thus, the existence of CHE. At the faintness of I~Zw~18, such a measurement can only be done by a 6-10m class telescope with a low- or medium-resolution spectrograph.}

Our stars are expected to end their lives with certain explosions such as long-duration gamma-ray bursts \citep{Yoon:2005,Woosley:2006,Yoon:2006}, supernovae of type~I~b/c, or even superluminous supernovae of the hydrogen-poor type~I \citep[][which of these would happen depends mainly on their mass]{Szecsi:2016,Szecsi:2017short,Szecsi:2017long,AguileraDena:2018}. In a close binary system, they may even form a double compact object that leads to gravitational wave emission upon merging \citep{deMink:2009b,deMink:2016,Mandel:2016,duBuisson:2020G,Riley:2021,Sharpe:2024G}.
Finding evidence of the existence of this evolutionary pathway in I~Zw~18 is therefore of upmost importance, and serves as motivation to study this dwarf galaxy further in the future. 

Our research on metal-poor massive stars in I~Zw~18 has implications not only for local dwarf galaxies but for other types of galaxies as well. For example, it has been suggested \citep{Micheva:2017} that the so-called Green Pea galaxies \citep{Izotov:2011,Jaskot:2013,Yang:2016,Orlitova:2018,Franeck:2022,ArroyoPolonio:2023,Smith:2023} are dwarf-galaxy analogues at intermediate cosmological distances; if so, their properties may only be understood by including chemically homogeneously evolving massive stars into the picture. Similarly, our knowledge of high-redshift galaxies, for example in the epoch of cosmic reionization, may need to be re-evaluated in light of what chemically homogeneous stars can bring to the table in terms of ionizing radiation \citep{Eldridge:2012,Sobral:2015,Stanway:2016,Visbal:2016,Sobral:2019} and other types of stellar feedback \citep{Bowler:2017}. Given the recent result of \citet{Liu:2025} showing -- with a completely independent method from ours -- that chemically homogeneously evolving stars may indeed exist in early cosmic epochs, our conclusions have far-reaching implications for future surveys with JWST and other campaigns targeting the dawn of our Universe.

\begin{acknowledgements}
{We thank both our anonymous referees, the first for making us add the entirety of Sect.~4 on comparing our results with all available observations from the literature, which made this work a truly unique contribution to the field, and the second for spotting mistakes and raising our attention to various omissions, such as the importance of including NE models when computing the spectral hardness.}
    This research was funded in part by the National Science
	Center (NCN), Poland under grant number OPUS 2021/41/B/ST9/00757. For
	the purpose of Open Access, the author has applied a CC-BY public copyright
	license to any Author Accepted Manuscript (AAM) version arising from this
	submission. 
	D.Sz. was supported by the Alexander von Humboldt Foundation.
        AACS is funded by the German \emph{Deut\-sche For\-schungs\-ge\-mein\-schaft, DFG\/} in the form of an Emmy Noether Research Group -- Project-ID 445674056 (SA4064/1-1, PI Sander). AACS further acknowledges financial support by the Federal Ministry for Economic Affairs and Climate Action (BMWK) via the German Aerospace Center (Deutsches Zentrum f\"ur Luft- und Raumfahrt, DLR) grant 50 OR 2503 (PI: Sander). This project was co-funded by the European Union (Project 101183150 - OCEANS).
	 This research is partially done in the frame of the European Union’s Framework Programme for Research and Innovation Horizon 2020 (2014-2020) under the Marie Skłodowska-Curie Grant Agreement No. 823734. B.K. and J.K. acknowledge the support of the Grant Agency of the Czech Republic (GA\v{C}R 25-15910S) and RVO:67985815. CK acknowledges financial support from the State Agency for Research of the Spanish MCIU through Center of Excellence Severo Ochoa award to the Instituto de Astrofísica de Andalucía CEX2021- 001131-S funded by MCIN/AEI/10.13039/501100011033, and from the grant PID2022-136598NB-C32 “Estallidos8”.
M. Garcia gratefully acknowledges support by grants PID2019-105552RB-C41 and PID2022-137779OB-C41, funded by the Spanish Ministry of Science, Innovation and Universities/State Agency of Research MICIU/AEI/10.13039/501100011033 and by “ERDF A way of making Europe”. 
D.~Sz. is grateful for the relevant discussions with S.~Heap, R.~Sarwar, H.~Stinshoff, K.~Sen, G.~Telford, N.~Langer, P.~Crowther, C.~Leitherer, G.~Gr\"afener, I.~Mandel, S.~Justham, C.~Neijssel, A.~Vigna-G\'omez, and S.~Vinciguerra, and Á. Szabó. 
\end{acknowledgements}


\bibliographystyle{aa} 
\bibliography{References4}

@Article{Surlan:2013,
  Title                    = {{Macroclumping as solution of the discrepancy between H{$\alpha$} and P v mass loss diagnostics for O-type stars}},
  Author                   = {{{\v S}urlan}, B. and {Hamann}, W.-R. and {Aret}, A. and {Kub{\'a}t}, J. and {Oskinova}, L.~M. and {Torres}, A.~F.},
  Journal                  = {\aap},
  Year                     = {2013},

  Month                    = nov,
  Pages                    = {A130},
  Volume                   = {559},
  Adsurl                   = {http://adsabs.harvard.edu/abs/2013A%26A...559A.130S},
  Archiveprefix            = {arXiv},
  Doi                      = {10.1051/0004-6361/201322390},
  Eid                      = {A130},
  Eprint                   = {1310.0449},
  Primaryclass             = {astro-ph.SR}
}

@Article{AguileraDena:2018,
  Title                    = {{Related Progenitor Models for Long-duration Gamma-Ray Bursts and Type Ic Superluminous Supernovae}},
  Author                   = {{Aguilera-Dena}, D.~R. and {Langer}, N. and {Moriya}, T.~J. and {Schootemeijer}, A.},
  Journal                  = {\apj},
  Year                     = {2018},

  Month                    = may,
  Pages                    = {115},
  Volume                   = {858},
  Adsurl                   = {http://adsabs.harvard.edu/abs/2018ApJ...858..115A},
  Archiveprefix            = {arXiv},
  Doi                      = {10.3847/1538-4357/aabfc1},
  Eid                      = {115},
  Eprint                   = {1804.07317},
  Primaryclass             = {astro-ph.SR}
}

@Article{Aloisi:1999,
  Author                   = {Aloisi, A. and Tosi, M. and Greggio, L.},
  Journal                  = {ApJ},
  Year                     = {1999},
  Pages                    = {302-322},
  Volume                   = {118}
}

@Article{Annibali:2013,
  Title                    = {{The Star Formation History of the Very Metal-poor Blue Compact Dwarf I Zw 18 from HST/ACS Data}},
  Author                   = {{Annibali}, F. and {Cignoni}, M. and {Tosi}, M. and {van der Marel}, R.~P. and {Aloisi}, A. and {Clementini}, G. and {Contreras Ramos}, R. and {Fiorentino}, G. and {Marconi}, M. and {Musella}, I.},
  Journal                  = {\aj},
  Year                     = {2013},

  Month                    = dec,
  Pages                    = {144},
  Volume                   = {146},

  Eid                      = {144}
}

@Article{Bowler:2017,
  Title                    = {{No evidence for Population III stars or a Direct Collapse Black Hole in the z = 6.6 Lyman-{$\alpha$} emitter 'CR7'}},
  Author                   = {{Bowler}, R.~A.~A. and {McLure}, R.~J. and {Dunlop}, J.~S. and {McLeod}, D.~J. and {Stanway}, E.~R. and {Eldridge}, J.~J. and {Jarvis}, M.~J.},
  Journal                  = {\mnras},
  Year                     = {2017},

  Month                    = jul,
  Pages                    = {448-458},
  Volume                   = {469},
  Adsurl                   = {http://adsabs.harvard.edu/abs/2017MNRAS.469..448B},
  Archiveprefix            = {arXiv},
  Doi                      = {10.1093/mnras/stx839},
  Eprint                   = {1609.00727}
}

@Article{Brott:2011a,
  Title                    = {{Rotating massive main-sequence stars. I. Grids of evolutionary models and isochrones}},
  Author                   = {{Brott}, I. and {de Mink}, S.~E. and {Cantiello}, M. and {Langer}, N. and {de Koter}, A. and {Evans}, C.~J. and {Hunter}, I. and {Trundle}, C. and {Vink}, J.S.},
  Journal                  = {\aap},
  Year                     = {2011},

  Month                    = jun,
  Pages                    = {A115},
  Volume                   = {530},

  Eid                      = {A115}
}

@Article{Brown:2002,
  Title                    = {{Isolating Clusters with Wolf-Rayet Stars in I Zw 18}},
  Author                   = {{Brown}, T.~M. and {Heap}, S.~R. and {Hubeny}, I. and {Lanz}, T. and {Lindler}, D.},
  Journal                  = {\apjl},
  Year                     = {2002},

  Month                    = nov,
  Pages                    = {L75-L78},
  Volume                   = {579},
  Adsurl                   = {http://adsabs.harvard.edu/abs/2002ApJ...579L..75B},
  Doi                      = {10.1086/345336},
  Eprint                   = {astro-ph/0210089}
}

@Article{Christensen:2016,
  Title                    = {{In-N-Out: The Gas Cycle from Dwarfs to Spiral Galaxies}},
  Author                   = {{Christensen}, C.~R. and {Dav{\'e}}, R. and {Governato}, F. and {Pontzen}, A. and {Brooks}, A. and {Munshi}, F. and {Quinn}, T. and {Wadsley}, J.},
  Journal                  = {\apj},
  Year                     = {2016},

  Month                    = jun,
  Pages                    = {57},
  Volume                   = {824},
  Adsurl                   = {http://adsabs.harvard.edu/abs/2016ApJ...824...57C},
  Archiveprefix            = {arXiv},
  Doi                      = {10.3847/0004-637X/824/1/57},
  Eid                      = {57},
  Eprint                   = {1508.00007}
}

@Article{Crowther:1998,
  Title                    = {{Quantitative classification of WC and WO stars}},
  Author                   = {{Crowther}, P.~A. and {De Marco}, O. and {Barlow}, M.~J.},
  Journal                  = {\mnras},
  Year                     = {1998},

  Month                    = may,
  Pages                    = {367-378},
  Volume                   = {296},
  Adsurl                   = {http://adsabs.harvard.edu/abs/1998MNRAS.296..367C},
  Doi                      = {10.1046/j.1365-8711.1998.01360.x}
}

@Article{Crowther:2006,
  Title                    = {{Reduced Wolf-Rayet line luminosities at low metallicity}},
  Author                   = {{Crowther}, P.~A. and {Hadfield}, L.~J.},
  Journal                  = {\aap},
  Year                     = {2006},

  Month                    = apr,
  Pages                    = {711-722},
  Volume                   = {449}
}

@Article{Crowther:1995,
  Title                    = {{Fundamental parameters of Wolf-Rayet stars. I. Ofpe/WN9 stars.}},
  Author                   = {{Crowther}, P.~A. and {Hillier}, D.~J. and {Smith}, L.~J.},
  Journal                  = {\aap},
  Year                     = {1995},

  Month                    = jan,
  Pages                    = {172-197},
  Volume                   = {293},
  Adsurl                   = {http://adsabs.harvard.edu/abs/1995A%26A...293..172C}
}

@Article{Crowther:2011,
  Title                    = {{Spectral classification of O2-3.5 If*/WN5-7 stars}},
  Author                   = {{Crowther}, P.~A. and {Walborn}, N.~R.},
  Journal                  = {\mnras},
  Year                     = {2011},

  Month                    = sep,
  Pages                    = {1311-1323},
  Volume                   = {416},
  Adsurl                   = {http://adsabs.harvard.edu/abs/2011MNRAS.416.1311C},
  Archiveprefix            = {arXiv},
  Doi                      = {10.1111/j.1365-2966.2011.19129.x},
  Eprint                   = {1105.4757},
  Primaryclass             = {astro-ph.SR}
}

@Article{deMink:2009b,
  Title                    = {{Rotational mixing in massive binaries. Detached short-period systems}},
  Author                   = {{de Mink}, S.~E. and {Cantiello}, M. and {Langer}, N. and {Pols}, O.~R. and {Brott}, I. and {Yoon}, S.-C.},
  Journal                  = {\aap},
  Year                     = {2009},

  Month                    = apr,
  Pages                    = {243-253},
  Volume                   = {497},
  Adsurl                   = {http://adsabs.harvard.edu/abs/2009A%26A...497..243D},
  Archiveprefix            = {arXiv},
  Doi                      = {10.1051/0004-6361/200811439},
  Eprint                   = {0902.1751},
  Primaryclass             = {astro-ph.SR}
}

@Article{deMink:2013,
  Title                    = {{The Rotation Rates of Massive Stars: The Role of Binary Interaction through Tides, Mass Transfer, and Mergers}},
  Author                   = {{de Mink}, S.~E. and {Langer}, N. and {Izzard}, R.~G. and {Sana}, H. and {de Koter}, A.},
  Journal                  = {\apj},
  Year                     = {2013},

  Month                    = feb,
  Pages                    = {166},
  Volume                   = {764},

  Eid                      = {166}
}

@Article{deMink:2016,
  Title                    = {{The chemically homogeneous evolutionary channel for binary black hole mergers: rates and properties of gravitational-wave events detectable by advanced LIGO}},
  Author                   = {{de Mink}, S.~E. and {Mandel}, I.},
  Journal                  = {\mnras},
  Year                     = {2016},

  Month                    = aug,
  Pages                    = {3545-3553},
  Volume                   = {460},
  Adsurl                   = {http://adsabs.harvard.edu/abs/2016MNRAS.460.3545D},
  Archiveprefix            = {arXiv},
  Doi                      = {10.1093/mnras/stw1219},
  Eprint                   = {1603.02291},
  Primaryclass             = {astro-ph.HE}
}

@Article{Eldridge:2016,
  Title                    = {{BPASS predictions for binary black hole mergers}},
  Author                   = {{Eldridge}, J.~J. and {Stanway}, E.~R.},
  Journal                  = {\mnras},
  Year                     = {2016},

  Month                    = nov,
  Pages                    = {3302-3313},
  Volume                   = {462},
  Adsurl                   = {http://adsabs.harvard.edu/abs/2016MNRAS.462.3302E},
  Archiveprefix            = {arXiv},
  Doi                      = {10.1093/mnras/stw1772},
  Eprint                   = {1602.03790},
  Primaryclass             = {astro-ph.HE}
}

@Article{Eldridge:2012,
  Title                    = {{The effect of stellar evolution uncertainties on the rest-frame ultraviolet stellar lines of C IV and He II in high-redshift Lyman-break galaxies}},
  Author                   = {{Eldridge}, J.~J. and {Stanway}, E.~R.},
  Journal                  = {\mnras},
  Year                     = {2012},

  Month                    = jan,
  Pages                    = {479-489},
  Volume                   = {419},
  Adsurl                   = {http://adsabs.harvard.edu/abs/2012MNRAS.419..479E},
  Archiveprefix            = {arXiv},
  Doi                      = {10.1111/j.1365-2966.2011.19713.x},
  Eprint                   = {1109.0288}
}

@Article{Eldridge:2019,
  Title                    = {{A consistent estimate for gravitational wave and electromagnetic transient rates}},
  Author                   = {{Eldridge}, J.~J. and {Stanway}, E.~R. and {Tang}, P.~N.},
  Journal                  = {\mnras},
  Year                     = {2019},

  Month                    = jan,
  Pages                    = {870-880},
  Volume                   = {482},
  Adsurl                   = {http://adsabs.harvard.edu/abs/2019MNRAS.482..870E},
  Archiveprefix            = {arXiv},
  Doi                      = {10.1093/mnras/sty2714},
  Eprint                   = {1807.07659},
  Primaryclass             = {astro-ph.HE}
}

@Article{Evans:2019,
  Title                    = {{First stellar spectroscopy in Leo P}},
  Author                   = {{Evans}, C.~J. and {Castro}, N. and {Gonzalez}, O.~A. and {Garcia}, M. and {Bastian}, N. and {Cioni}, M.-R.~L. and {Clark}, J.~S. and {Davies}, B. and {Ferguson}, A.~M.~N. and {Kamann}, S. and {Lennon}, D.~J. and {Patrick}, L.~R. and {Vink}, J.~S. and {Weisz}, D.~R.},
  Journal                  = {\aap},
  Year                     = {2019},

  Month                    = feb,
  Pages                    = {A129},
  Volume                   = {622},
  Adsurl                   = {http://adsabs.harvard.edu/abs/2019A%26A...622A.129E},
  Archiveprefix            = {arXiv},
  Doi                      = {10.1051/0004-6361/201834145},
  Eid                      = {A129},
  Eprint                   = {1901.01295},
  Primaryclass             = {astro-ph.SR}
}

@Article{Goetberg:2017,
  Title                    = {{Ionizing spectra of stars that lose their envelope through interaction with a binary companion: role of metallicity}},
  Author                   = {{G{\"o}tberg}, Y. and {de Mink}, S.~E. and {Groh}, J.~H.},
  Journal                  = {\aap},
  Year                     = {2017},

  Month                    = nov,
  Pages                    = {A11},
  Volume                   = {608},
  Adsurl                   = {http://adsabs.harvard.edu/abs/2017A%26A...608A..11G},
  Archiveprefix            = {arXiv},
  Doi                      = {10.1051/0004-6361/201730472},
  Eid                      = {A11},
  Eprint                   = {1701.07439},
  Primaryclass             = {astro-ph.SR}
}

@Article{Garcia:2019,
  Title                    = {{Ongoing star formation at the outskirts of Sextans A: spectroscopic detection of early O-type stars}},
  Author                   = {{Garcia}, M. and {Herrero}, A. and {Najarro}, F. and {Camacho}, I. and {Lorenzo}, M.},
  Journal                  = {\mnras},
  Year                     = {2019},

  Month                    = mar,
  Pages                    = {422-430},
  Volume                   = {484},
  Adsurl                   = {http://adsabs.harvard.edu/abs/2019MNRAS.484..422G},
  Archiveprefix            = {arXiv},
  Doi                      = {10.1093/mnras/sty3503},
  Eprint                   = {1901.02466}
}

@Article{Georgy:2012,
  Title                    = {{Grids of stellar models with rotation. II. WR populations and supernovae/GRB progenitors at Z = 0.014}},
  Author                   = {{Georgy}, C. and {Ekstr{\"o}m}, S. and {Meynet}, G. and {Massey}, P. and {Levesque}, E.~M. and {Hirschi}, R. and {Eggenberger}, P. and {Maeder}, A.},
  Journal                  = {\aap},
  Year                     = {2012},

  Month                    = jun,
  Pages                    = {A29},
  Volume                   = {542},

  Eid                      = {A29}
}

@Article{Girichidis:2018,
  Title                    = {{The SILCC project - V. The impact of magnetic fields on the chemistry and the formation of molecular clouds}},
  Author                   = {{Girichidis}, P. and {Seifried}, D. and {Naab}, T. and {Peters}, T. and {Walch}, S. and {W{\"u}nsch}, R. and {Glover}, S.~C.~O. and {Klessen}, R.~S.},
  Journal                  = {\mnras},
  Year                     = {2018},

  Month                    = nov,
  Pages                    = {3511-3540},
  Volume                   = {480},
  Adsurl                   = {http://adsabs.harvard.edu/abs/2018MNRAS.480.3511G},
  Archiveprefix            = {arXiv},
  Doi                      = {10.1093/mnras/sty2016},
  Eprint                   = {1808.05222}
}

@Article{Grafener:2002,
  Title                    = {{Line-blanketed model atmospheres for WR stars}},
  Author                   = {{Gr{\"a}fener}, G. and {Koesterke}, L. and {Hamann}, W.-R.},
  Journal                  = {\aap},
  Year                     = {2002},

  Month                    = may,
  Pages                    = {244-257},
  Volume                   = {387},
  Adsurl                   = {http://adsabs.harvard.edu/abs/2002A%26A...387..244G},
  Doi                      = {10.1051/0004-6361:20020269}
}

@Article{Groh:2019,
       author = {{Groh}, J.~H. and {Ekstr{\"o}m}, S. and {Georgy}, C. and {Meynet}, G. and {Choplin}, A. and {Eggenberger}, P. and {Hirschi}, R. and {Maeder}, A. and {Murphy}, L.~J. and {Boian}, I. and {Farrell}, E.~J.},
        title = "{Grids of stellar models with rotation. IV. Models from 1.7 to 120 M$_{{\ensuremath{\odot}}}$ at a metallicity Z = 0.0004}",
      journal = {\aap},
         year = 2019,
        month = jul,
       volume = {627},
          eid = {A24},
        pages = {A24},
          doi = {10.1051/0004-6361/201833720},
archivePrefix = {arXiv},
       eprint = {1904.04009},
 primaryClass = {astro-ph.SR},
       adsurl = {https://ui.adsabs.harvard.edu/abs/2019A&A...627A..24G}
}

@Article{Groh:2014,
  Title                    = {{The evolution of massive stars and their spectra. I. A non-rotating 60 Ms star from the zero-age main sequence to the pre-supernova stage}},
  Author                   = {{Groh}, J.~H. and {Meynet}, G. and {Ekstr{\"o}m}, S. and {Georgy}, C. },
  Journal                  = {\aap},
  Year                     = {2014},

  Month                    = apr,
  Pages                    = {A30},
  Volume                   = {564},

  Eid                      = {A30}
}

@Article{Guseva:2000,
  Title                    = {{A Spectroscopic Study of a Large Sample Of Wolf-Rayet Galaxies}},
  Author                   = {{Guseva}, N.~G. and {Izotov}, Y.~I. and {Thuan}, T.~X.},
  Journal                  = {\apj},
  Year                     = {2000},

  Month                    = mar,
  Pages                    = {776-803},
  Volume                   = {531},
  Adsurl                   = {http://adsabs.harvard.edu/abs/2000ApJ...531..776G},
  Doi                      = {10.1086/308489},
  Eprint                   = {astro-ph/9910432}
}

@Article{Haid:2019,
  Title                    = {{SILCC-Zoom: The early impact of ionizing radiation on forming molecular clouds}},
  Author                   = {{Haid}, S. and {Walch}, S. and {Seifried}, D. and {W{\"u}nsch}, R. and {Dinnbier}, F. and {Naab}, T.},
  Journal                  = {\mnras},
  Year                     = {2019},

  Month                    = jan,
  Pages                    = {4062-4083},
  Volume                   = {482},
  Adsurl                   = {http://adsabs.harvard.edu/abs/2019MNRAS.482.4062H},
  Archiveprefix            = {arXiv},
  Doi                      = {10.1093/mnras/sty2938},
  Eprint                   = {1810.08210}
}

@Article{Hainich:2018,
  Title                    = {{Observational properties of massive black hole binary progenitors}},
  Author                   = {{Hainich}, R. and {Oskinova}, L.~M. and {Shenar}, T. and {Marchant}, P. and {Eldridge}, J.~J. and {Sander}, A.~A.~C. and {Hamann}, W.-R. and {Langer}, N. and {Todt}, H.},
  Journal                  = {\aap},
  Year                     = {2018},

  Month                    = jan,
  Pages                    = {A94},
  Volume                   = {609},
  Adsurl                   = {http://adsabs.harvard.edu/abs/2018A%26A...609A..94H},
  Archiveprefix            = {arXiv},
  Doi                      = {10.1051/0004-6361/201731449},
  Eid                      = {A94},
  Eprint                   = {1707.01912},
  Primaryclass             = {astro-ph.SR}
}

@Article{Hainich:2015,
  Title                    = {{Wolf-Rayet stars in the Small Magellanic Cloud. I. Analysis of the single WN stars}},
  Author                   = {{Hainich}, R. and {Pasemann}, D. and {Todt}, H. and {Shenar}, T. and {Sander}, A. and {Hamann}, W.-R.},
  Journal                  = {\aap},
  Year                     = {2015},

  Month                    = sep,
  Pages                    = {A21},
  Volume                   = {581},
  Adsurl                   = {http://adsabs.harvard.edu/abs/2015A%26A...581A..21H},
  Archiveprefix            = {arXiv},
  Doi                      = {10.1051/0004-6361/201526241},
  Eid                      = {A21},
  Eprint                   = {1507.04000},
  Primaryclass             = {astro-ph.SR}
}

@Article{Hamann:2004,
  Title                    = {{Grids of model spectra for WN stars, ready for use}},
  Author                   = {{Hamann}, W.-R. and {Gr{\"a}fener}, G.},
  Journal                  = {\aap},
  Year                     = {2004},

  Month                    = nov,
  Pages                    = {697-704},
  Volume                   = {427},
  Adsurl                   = {http://adsabs.harvard.edu/abs/2004A%26A...427..697H},
  Doi                      = {10.1051/0004-6361:20040506}
}

@Article{Hamann:2003,
  Title                    = {{A temperature correction method for expanding atmospheres}},
  Author                   = {{Hamann}, W.-R. and {Gr{\"a}fener}, G.},
  Journal                  = {\aap},
  Year                     = {2003},

  Month                    = nov,
  Pages                    = {993-1000},
  Volume                   = {410},
  Adsurl                   = {http://adsabs.harvard.edu/abs/2003A%26A...410..993H},
  Doi                      = {10.1051/0004-6361:20031308}
}

@Article{Hamann:1995,
  Author                   = {Hamann, W.-R. and Koesterke, L. and Wessolowski, U.},
  Journal                  = {A\&A},
  Year                     = {1995},
  Pages                    = {151-162},
  Volume                   = {299}
}

@Article{Heap:2015,
  Title                    = {{Population III Stars in I Zw 18}},
  Author                   = {{Heap}, S. and {Bouret}, J.-C. and {Hubeny}, I.},
  Journal                  = {ArXiv e-print 1504.02742},
  Year                     = {2015},

  Month                    = apr
}

@Article{Higgins:2019,
  Title                    = {{Massive star evolution: rotation, winds, and overshooting vectors in the mass-luminosity plane. I. A calibrated grid of rotating single star models}},
  Author                   = {{Higgins}, E.~R. and {Vink}, J.~S.},
  Journal                  = {\aap},
  Year                     = {2019},

  Month                    = feb,
  Pages                    = {A50},
  Volume                   = {622},
  Adsurl                   = {http://adsabs.harvard.edu/abs/2019A%26A...622A..50H},
  Archiveprefix            = {arXiv},
  Doi                      = {10.1051/0004-6361/201834123},
  Eid                      = {A50},
  Eprint                   = {1811.12190},
  Primaryclass             = {astro-ph.SR}
}

@Article{Hopkins:2011,
  Title                    = {{Self-regulated star formation in galaxies via momentum input from massive stars}},
  Author                   = {{Hopkins}, P.~F. and {Quataert}, E. and {Murray}, N.},
  Journal                  = {\mnras},
  Year                     = {2011},

  Month                    = oct,
  Pages                    = {950-973},
  Volume                   = {417},
  Adsurl                   = {http://adsabs.harvard.edu/abs/2011MNRAS.417..950H},
  Archiveprefix            = {arXiv},
  Doi                      = {10.1111/j.1365-2966.2011.19306.x},
  Eprint                   = {1101.4940},
  Primaryclass             = {astro-ph.CO}
}

@Article{Hunter:1995,
  Author                   = {Hunter, D.A. and Thronson, H.A.},
  Journal                  = {ApJ},
  Year                     = {1995},
  Pages                    = {238-252},
  Volume                   = {452}
}

@Article{Izotov:2004,
  Author                   = {Izotov, Y.I. and Thuan, T.X.},
  Journal                  = {ApJ},
  Year                     = {2004},
  Pages                    = {768-782},
  Volume                   = {616}
}

@Article{Izotov:2002,
  Author                   = {Izotov, Y.I. and Thuan, T.X.},
  Journal                  = {ApJ},
  Year                     = {2002},
  Number                   = {875},
  Volume                   = {567}
}

@Article{Izotov:1997,
  Title                    = {{I Zw 18: A New Wolf-Rayet Galaxy}},
  Author                   = {{Izotov}, Y.~I. and {Foltz}, C.~B. and {Green}, R.~F. and {Guseva}, N.~G. and {Thuan}, T.~X.},
  Journal                  = {\apjl},
  Year                     = {1997},

  Month                    = sep,
  Pages                    = {L37-L40},
  Volume                   = {487}
}

@Article{Izotov:2011,
  Title                    = {{Green Pea Galaxies and Cohorts: Luminous Compact Emission-line Galaxies in the Sloan Digital Sky Survey}},
  Author                   = {{Izotov}, Y.~I. and {Guseva}, N.~G. and {Thuan}, T.~X.},
  Journal                  = {\apj},
  Year                     = {2011},

  Month                    = feb,
  Pages                    = {161},
  Volume                   = {728},
  Adsurl                   = {http://adsabs.harvard.edu/abs/2011ApJ...728..161I},
  Archiveprefix            = {arXiv},
  Doi                      = {10.1088/0004-637X/728/2/161},
  Eid                      = {161},
  Eprint                   = {1012.5639}
}

@Article{Izotov:2016,
  Title                    = {{Detection of high Lyman continuum leakage from four low-redshift compact star-forming galaxies}},
  Author                   = {{Izotov}, Y.~I. and {Schaerer}, D. and {Thuan}, T.~X. and {Worseck}, G. and {Guseva}, N.~G. and {Orlitov{\'a}}, I. and {Verhamme}, A.},
  Journal                  = {\mnras},
  Year                     = {2016},

  Month                    = oct,
  Pages                    = {3683-3701},
  Volume                   = {461},
  Adsurl                   = {http://adsabs.harvard.edu/abs/2016MNRAS.461.3683I},
  Archiveprefix            = {arXiv},
  Doi                      = {10.1093/mnras/stw1205},
  Eprint                   = {1605.05160}
}

@Article{Izotov:1998,
  Title                    = {{Reexamining the Helium Abundance of I ZW 18}},
  Author                   = {{Izotov}, Y.~I. and {Thuan}, T.~X.},
  Journal                  = {\apj},
  Year                     = {1998},

  Month                    = apr,
  Pages                    = {227-237},
  Volume                   = {497},
  Adsurl                   = {http://adsabs.harvard.edu/abs/1998ApJ...497..227I},
  Doi                      = {10.1086/305440}
}

@Article{Jaskot:2013,
  Title                    = {{The Origin and Optical Depth of Ionizing Radiation in the ``Green Pea'' Galaxies}},
  Author                   = {{Jaskot}, A.~E. and {Oey}, M.~S.},
  Journal                  = {\apj},
  Year                     = {2013},

  Month                    = apr,
  Pages                    = {91},
  Volume                   = {766},
  Adsurl                   = {http://adsabs.harvard.edu/abs/2013ApJ...766...91J},
  Archiveprefix            = {arXiv},
  Doi                      = {10.1088/0004-637X/766/2/91},
  Eid                      = {91},
  Eprint                   = {1301.0530}
}

@Article{Kehrig:2013,
  Title                    = {{Uncovering multiple Wolf-Rayet star clusters and the ionized ISM in Mrk 178: the closest metal-poor Wolf-Rayet H II galaxy}},
  Author                   = {{Kehrig}, C. and {P{\'e}rez-Montero}, E. and {V{\'{\i}}lchez}, J.~M. and {Brinchmann}, J. and {Kunth}, D. and {Garc{\'{\i}}a-Benito}, R. and {Crowther}, P.~A. and {Hern{\'a}ndez-Fern{\'a}ndez}, J. and {Durret}, F. and {Contini}, T. and {Fern{\'a}ndez-Mart{\'{\i}}n}, A. and {James}, B.~L.},
  Journal                  = {\mnras},
  Year                     = {2013},

  Month                    = jul,
  Pages                    = {2731-2745},
  Volume                   = {432}
}

@Article{Kehrig:2018,
  Title                    = {{The extended He II {$\lambda$}4686 emission in the extremely metal-poor galaxy SBS 0335 - 052E seen with MUSE}},
  Author                   = {{Kehrig}, C. and {V{\'{\i}}lchez}, J.~M. and {Guerrero}, M.~A. and {Iglesias-P{\'a}ramo}, J. and {Hunt}, L.~K. and {Duarte-Puertas}, S. and {Ramos-Larios}, G.},
  Journal                  = {\mnras},
  Year                     = {2018},

  Month                    = oct,
  Pages                    = {1081-1095},
  Volume                   = {480},
  Adsurl                   = {http://adsabs.harvard.edu/abs/2018MNRAS.480.1081K},
  Archiveprefix            = {arXiv},
  Doi                      = {10.1093/mnras/sty1920},
  Eprint                   = {1807.09307}
}

@InProceedings{Kehrig:2015b,
  Title                    = {{PopIII-star siblings in IZw18 and metal-poor WR galaxies unveiled from integral field spectroscopy}},
  Author                   = {{Kehrig}, C. and {V{\'{\i}}lchez}, J.~M. and {P{\'e}rez-Montero}, E. and {Iglesias-P{\'a}ramo}, J. and {Brinchmann}, J. and {Crowther}, P.~A. and {Durret}, F. and {Kunth}, D.},
  Booktitle                = {Wolf-Rayet Stars: Proceedings of an International Workshop held in Potsdam, Germany, 1-5 June 2015. Edited by Wolf-Rainer Hamann, Andreas Sander, Helge Todt. Universit{\"a}tsverlag Potsdam, 2015., p.55-58},
  Year                     = {2015},
  Editor                   = {{Hamann}, W.-R. and {Sander}, A. and {Todt}, H.},
  Pages                    = {55-58},
  Adsurl                   = {http://adsabs.harvard.edu/abs/2015wrs..conf...55K},
  Archiveprefix            = {arXiv},
  Eprint                   = {1509.06285}
}

@Article{Kehrig:2015,
  Title                    = {{The Extended He II {$\lambda$}4686-emitting Region in IZw 18 Unveiled: Clues for Peculiar Ionizing Sources}},
  Author                   = {{Kehrig}, C. and {V{\'{\i}}lchez}, J.~M. and {P{\'e}rez-Montero}, E. and {Iglesias-P{\'a}ramo}, J. and {Brinchmann}, J. and {Kunth}, D. and {Durret}, F. and {Bayo}, F.~M.},
  Journal                  = {\apjl},
  Year                     = {2015},

  Month                    = mar,
  Pages                    = {L28},
  Volume                   = {801},
  Adsurl                   = {http://adsabs.harvard.edu/abs/2015ApJ...801L..28K},
  Archiveprefix            = {arXiv},
  Doi                      = {10.1088/2041-8205/801/2/L28},
  Eid                      = {L28},
  Eprint                   = {1502.00522}
}

@Article{Kehrig:2016,
  Title                    = {{Spatially resolved integral field spectroscopy of the ionized gas in IZw18}},
  Author                   = {{Kehrig}, C. and {V{\'{\i}}lchez}, J.~M. and {P{\'e}rez-Montero}, E. and {Iglesias-P{\'a}ramo}, J. and {Hern{\'a}ndez-Fern{\'a}ndez}, J.~D. and {Duarte Puertas}, S. and {Brinchmann}, J. and {Durret}, F. and {Kunth}, D.},
  Journal                  = {\mnras},
  Year                     = {2016},

  Month                    = jul,
  Pages                    = {2992-3004},
  Volume                   = {459},
  Adsurl                   = {http://adsabs.harvard.edu/abs/2016MNRAS.459.2992K},
  Archiveprefix            = {arXiv},
  Doi                      = {10.1093/mnras/stw806},
  Eprint                   = {1604.08555}
}

@Article{Kubatova:2019,
  Title                    = {{Low-metallicity massive single stars with rotation. II. Predicting spectra and spectral classes of chemically homogeneously evolving stars}},
  Author                   = {{Kub{\'a}tov{\'a}}, B. and {Sz{\'e}csi}, D. and {Sander}, A.~A.~C. and {Kub{\'a}t}, J. and {Tramper}, F. and {Krti{\v c}ka}, J. and {Kehrig}, C. and {Hamann}, W.-R. and {Hainich}, R. and {Shenar}, T. },
  Journal                  = {\aap},
  Year                     = {2019},

  Month                    = mar,
  Pages                    = {A8},
  Volume                   = {623},
  Adsurl                   = {http://adsabs.harvard.edu/abs/2019A%26A...623A...8K},
  Archiveprefix            = {arXiv},
  Doi                      = {10.1051/0004-6361/201834360},
  Eid                      = {A8},
  Eprint                   = {1810.01267},
  Primaryclass             = {astro-ph.SR}
}

@Article{Kunth:2000,
  Title                    = {{The most metal-poor galaxies}},
  Author                   = {{Kunth}, D. and {{\"O}stlin}, G.},
  Journal                  = {\aapr},
  Year                     = {2000},
  Pages                    = {1-79},
  Volume                   = {10},
  Adsurl                   = {http://adsabs.harvard.edu/abs/2000A%26ARv..10....1K},
  Doi                      = {10.1007/s001590000005},
  Eprint                   = {astro-ph/9911094}
}

@Article{Langer:1989a,
  Title                    = {{Standard models of Wolf-Rayet stars}},
  Author                   = {{Langer}, N.},
  Journal                  = {\aap},
  Year                     = {1989},

  Month                    = feb,
  Pages                    = {93-113},
  Volume                   = {210}
}

@Article{Lebouteiller:2013,
  Title                    = {{Chemical enrichment and physical conditions in I Zw 18}},
  Author                   = {{Lebouteiller}, V. and {Heap}, S. and {Hubeny}, I. and {Kunth}, D. },
  Journal                  = {\aap},
  Year                     = {2013},

  Month                    = may,
  Pages                    = {A16},
  Volume                   = {553},

  Eid                      = {A16}
}

@Article{Lebouteiller:2017,
  Title                    = {{Neutral gas heating by X-rays in primitive galaxies: Infrared observations of the blue compact dwarf I Zw 18 with Herschel}},
  Author                   = {{Lebouteiller}, V. and {P{\'e}quignot}, D. and {Cormier}, D. and {Madden}, S. and {Pakull}, M.~W. and {Kunth}, D. and {Galliano}, F. and {Chevance}, M. and {Heap}, S.~R. and {Lee}, M.-Y. and {Polles}, F.~L. },
  Journal                  = {\aap},
  Year                     = {2017},

  Month                    = jun,
  Pages                    = {A45},
  Volume                   = {602},
  Adsurl                   = {http://adsabs.harvard.edu/abs/2017A%26A...602A..45L},
  Archiveprefix            = {arXiv},
  Doi                      = {10.1051/0004-6361/201629675},
  Eid                      = {A45},
  Eprint                   = {1702.07377}
}

@Article{LecavelierdesEtangs:2004,
  Title                    = {{FUSE observations of the H I interstellar gas of I Zw 18}},
  Author                   = {{Lecavelier des Etangs}, A. and {D{\'e}sert}, J.-M. and {Kunth}, D. and {Vidal-Madjar}, A. and {Callejo}, G. and {Ferlet}, R. and {H{\'e}brard}, G. and {Lebouteiller}, V.},
  Journal                  = {\aap},
  Year                     = {2004},

  Month                    = jan,
  Pages                    = {131-137},
  Volume                   = {413}
}

@Article{Maeder:2000,
  Title                    = {{The Evolution of Rotating Stars}},
  Author                   = {{Maeder}, A. and {Meynet}, G.},
  Journal                  = {\araa},
  Year                     = {2000},
  Pages                    = {143-190},
  Volume                   = {38}
}

@Article{Mandel:2016,
  Title                    = {{Merging binary black holes formed through chemically homogeneous evolution in short-period stellar binaries}},
  Author                   = {{Mandel}, I. and {de Mink}, S.~E.},
  Journal                  = {\mnras},
  Year                     = {2016},

  Month                    = may,
  Pages                    = {2634-2647},
  Volume                   = {458},
  Adsurl                   = {http://adsabs.harvard.edu/abs/2016MNRAS.458.2634M},
  Archiveprefix            = {arXiv},
  Doi                      = {10.1093/mnras/stw379},
  Eprint                   = {1601.00007},
  Primaryclass             = {astro-ph.HE}
}

@Article{Marchant:2017,
  Title                    = {{Ultra-luminous X-ray sources and neutron-star-black-hole mergers from very massive close binaries at low metallicity}},
  Author                   = {{Marchant}, P. and {Langer}, N. and {Podsiadlowski}, P. and {Tauris}, T.~M. and {de Mink}, S. and {Mandel}, I. and {Moriya}, T.~J. },
  Journal                  = {\aap},
  Year                     = {2017},

  Month                    = aug,
  Pages                    = {A55},
  Volume                   = {604},
  Adsurl                   = {http://adsabs.harvard.edu/abs/2017A%26A...604A..55M},
  Archiveprefix            = {arXiv},
  Doi                      = {10.1051/0004-6361/201630188},
  Eid                      = {A55},
  Eprint                   = {1705.04734},
  Primaryclass             = {astro-ph.HE}
}

@Article{Marchant:2016,
  Title                    = {{A new route towards merging massive black holes}},
  Author                   = {{Marchant}, P. and {Langer}, N. and {Podsiadlowski}, P. and {Tauris}, T.~M. and {Moriya}, T.~J.},
  Journal                  = {Astronomy \& Astrophysics},
  Year                     = {2016},

  Month                    = apr,
  Pages                    = {A50},
  Volume                   = {588},
  Adsurl                   = {http://adsabs.harvard.edu/abs/2016A%26A...588A..50M},
  Archiveprefix            = {arXiv},
  Doi                      = {10.1051/0004-6361/201628133},
  Eid                      = {A50},
  Eprint                   = {1601.03718},
  Primaryclass             = {astro-ph.SR}
}

@Article{Meynet:2000,
  Title                    = {{Stellar evolution with rotation. V. Changes in all the outputs of massive star models}},
  Author                   = {{Meynet}, G. and {Maeder}, A.},
  Journal                  = {\aap},
  Year                     = {2000},

  Month                    = sep,
  Pages                    = {101-120},
  Volume                   = {361}
}

@Article{Micheva:2017,
  Title                    = {{Mrk 71/NGC 2366: The Nearest Green Pea Analog}},
  Author                   = {{Micheva}, G. and {Oey}, M.~S. and {Jaskot}, A.~E. and {James}, B.~L. },
  Journal                  = {\apj},
  Year                     = {2017},

  Month                    = aug,
  Pages                    = {165},
  Volume                   = {845},
  Adsurl                   = {http://adsabs.harvard.edu/abs/2017ApJ...845..165M},
  Archiveprefix            = {arXiv},
  Doi                      = {10.3847/1538-4357/aa830b},
  Eid                      = {165},
  Eprint                   = {1704.01678}
}

@Article{Mokiem:2006,
  Title                    = {{The VLT-FLAMES survey of massive stars: mass loss and rotation of early-type stars in the SMC}},
  Author                   = {{Mokiem}, M.~R. and {de Koter}, A. and {Evans}, C.~J. and {Puls}, J. and {Smartt}, S.~J. and {Crowther}, P.~A. and {Herrero}, A. and {Langer}, N. and {Lennon}, D.~J. and {Najarro}, F. and {Villamariz}, M.~R. and {Yoon}, S.-C.},
  Journal                  = {\aap},
  Year                     = {2006},

  Month                    = sep,
  Pages                    = {1131-1151},
  Volume                   = {456}
}

@Article{Nugis:2000,
  Title                    = {{Mass-loss rates of Wolf-Rayet stars as a function of stellar parameters}},
  Author                   = {{Nugis}, T. and {Lamers}, H.J.G.L.M.},
  Journal                  = {\aap},
  Year                     = {2000},

  Month                    = aug,
  Pages                    = {227-244},
  Volume                   = {360}
}

@Article{Orlitova:2018,
  Title                    = {{Puzzling Lyman-alpha line profiles in green pea galaxies}},
  Author                   = {{Orlitov{\'a}}, I. and {Verhamme}, A. and {Henry}, A. and {Scarlata}, C. and {Jaskot}, A. and {Oey}, M.~S. and {Schaerer}, D.},
  Journal                  = {\aap},
  Year                     = {2018},

  Month                    = aug,
  Pages                    = {A60},
  Volume                   = {616},
  Adsurl                   = {http://adsabs.harvard.edu/abs/2018A%26A...616A..60O},
  Archiveprefix            = {arXiv},
  Doi                      = {10.1051/0004-6361/201732478},
  Eid                      = {A60},
  Eprint                   = {1806.01027}
}

@Article{Pequignot:2008,
  Title                    = {{Heating of blue compact dwarf galaxies: gas distribution and photoionization by stars in I Zw 18}},
  Author                   = {{P{\'e}quignot}, D.},
  Journal                  = {\aap},
  Year                     = {2008},

  Month                    = {Feb},
  Number                   = {2},
  Pages                    = {371-385},
  Volume                   = {478},
  Adsurl                   = {https://ui.adsabs.harvard.edu/abs/2008A&A...478..371P},
  Archiveprefix            = {arXiv},
  Doi                      = {10.1051/0004-6361:20078344},
  Eprint                   = {0710.5082},
  Primaryclass             = {astro-ph}
}

@Article{Papaderos:2012,
  Title                    = {{I Zw 18 as morphological paradigm for rapidly assembling high-z galaxies}},
  Author                   = {{Papaderos}, P. and {{\"O}stlin}, G.},
  Journal                  = {\aap},
  Year                     = {2012},

  Month                    = jan,
  Pages                    = {A126},
  Volume                   = {537},

  Eid                      = {A126}
}

@Article{Papaderos:2002,
  Title                    = {{The blue compact dwarf galaxy I Zw 18: A comparative study of its low-surface-brightness component}},
  Author                   = {{Papaderos}, P. and {Izotov}, Y.~I. and {Thuan}, T.~X. and {Noeske}, K.~G. and {Fricke}, K.~J. and {Guseva}, N.~G. and {Green}, R.~F.},
  Journal                  = {\aap},
  Year                     = {2002},

  Month                    = oct,
  Pages                    = {461-483},
  Volume                   = {393}
}

@Article{Perley:2016,
  Title                    = {{The Swift GRB Host Galaxy Legacy Survey. II. Rest-frame Near-IR Luminosity Distribution and Evidence for a Near-solar Metallicity Threshold}},
  Author                   = {{Perley}, D.~A. and {Tanvir}, N.~R. and {Hjorth}, J. and {Laskar}, T. and {Berger}, E. and {Chary}, R. and {de Ugarte Postigo}, A. and {Fynbo}, J.~P.~U. and {Kr{\"u}hler}, T. and {Levan}, A.~J. and {Micha{\l}owski}, M.~J. and {Schulze}, S.},
  Journal                  = {\apj},
  Year                     = {2016},

  Month                    = jan,
  Pages                    = {8},
  Volume                   = {817},
  Adsurl                   = {http://adsabs.harvard.edu/abs/2016ApJ...817....8P},
  Archiveprefix            = {arXiv},
  Doi                      = {10.3847/0004-637X/817/1/8},
  Eid                      = {8},
  Eprint                   = {1504.02479}
}

@Article{Salpeter:1955,
  Author                   = {Salpeter, E.E.},
  Journal                  = {ApJ},
  Year                     = {1955},
  Pages                    = {161},
  Volume                   = {121}
}

@Article{Sana:2012,
  Title                    = {{Binary Interaction Dominates the Evolution of Massive Stars}},
  Author                   = {{Sana}, H. and {de Mink}, S.~E. and {de Koter}, A. and {Langer}, N. and {Evans}, C.~J. and {Gieles}, M. and {Gosset}, E. and {Izzard}, R.~G. and {Le Bouquin}, J.-B. and {Schneider}, F.~R.~N.},
  Journal                  = {Science},
  Year                     = {2012},

  Month                    = jul,
  Pages                    = {444},
  Volume                   = {337},
  Adsurl                   = {http://adsabs.harvard.edu/abs/2012Sci...337..444S},
  Archiveprefix            = {arXiv},
  Doi                      = {10.1126/science.1223344},
  Eprint                   = {1207.6397},
  Primaryclass             = {astro-ph.SR}
}

@Article{Sander:2015,
  Title                    = {{On the consistent treatment of the quasi-hydrostatic layers in hot star atmospheres}},
  Author                   = {{Sander}, A. and {Shenar}, T. and {Hainich}, R. and {G{\'{\i}}menez-Garc{\'{\i}}a}, A. and {Todt}, H. and {Hamann}, W.-R.},
  Journal                  = {\aap},
  Year                     = {2015},

  Month                    = may,
  Pages                    = {A13},
  Volume                   = {577},
  Adsurl                   = {http://adsabs.harvard.edu/abs/2015A%26A...577A..13S},
  Archiveprefix            = {arXiv},
  Doi                      = {10.1051/0004-6361/201425356},
  Eid                      = {A13},
  Eprint                   = {1503.01338},
  Primaryclass             = {astro-ph.SR}
}

@Article{Schaerer:2019,
  Title                    = {{X-ray binaries as the origin of nebular He II emission in low-metallicity star-forming galaxies}},
  Author                   = {{Schaerer}, D. and {Fragos}, T. and {Izotov}, Y.~I.},
  Journal                  = {\aap},
  Year                     = {2019},

  Month                    = feb,
  Pages                    = {L10},
  Volume                   = {622},
  Adsurl                   = {http://adsabs.harvard.edu/abs/2019A%26A...622L..10S},
  Archiveprefix            = {arXiv},
  Doi                      = {10.1051/0004-6361/201935005},
  Eid                      = {L10},
  Eprint                   = {1902.10496}
}

@Article{Schootemeijer:2018,
  Title                    = {{Wolf-Rayet stars in the Small Magellanic Cloud as testbed for massive star evolution}},
  Author                   = {{Schootemeijer}, A. and {Langer}, N.},
  Journal                  = {\aap},
  Year                     = {2018},

  Month                    = mar,
  Pages                    = {A75},
  Volume                   = {611},
  Adsurl                   = {http://adsabs.harvard.edu/abs/2018A%26A...611A..75S},
  Archiveprefix            = {arXiv},
  Doi                      = {10.1051/0004-6361/201731895},
  Eid                      = {A75},
  Eprint                   = {1709.08727},
  Primaryclass             = {astro-ph.SR}
}

@Article{Searle:1972,
  Author                   = {Searle, L. and Sargent, W.},
  Journal                  = {ApJ},
  Year                     = {1972},
  Pages                    = {25-33},
  Volume                   = {173}
}

@Article{Seifried:2017,
  Title                    = {{SILCC-Zoom: the dynamic and chemical evolution of molecular clouds}},
  Author                   = {{Seifried}, D. and {Walch}, S. and {Girichidis}, P. and {Naab}, T. and {W{\"u}nsch}, R. and {Klessen}, R.~S. and {Glover}, S.~C.~O. and {Peters}, T. and {Clark}, P.},
  Journal                  = {\mnras},
  Year                     = {2017},

  Month                    = dec,
  Pages                    = {4797-4818},
  Volume                   = {472},
  Adsurl                   = {http://adsabs.harvard.edu/abs/2017MNRAS.472.4797S},
  Archiveprefix            = {arXiv},
  Doi                      = {10.1093/mnras/stx2343},
  Eprint                   = {1704.06487}
}

@Article{Senchyna:2019,
       author = {{Senchyna}, Peter and {Stark}, Daniel P. and {Chevallard}, Jacopo and {Charlot}, St{\'e}phane and {Jones}, Tucker and {Vidal-Garc{\'\i}a}, Alba},
        title = "{Extremely metal-poor galaxies with HST/COS: laboratories for models of low-metallicity massive stars and high-redshift galaxies}",
      journal = {\mnras},
         year = 2019,
        month = sep,
       volume = {488},
       number = {3},
        pages = {3492-3506},
          doi = {10.1093/mnras/stz1907},
archivePrefix = {arXiv},
       eprint = {1904.01615},
 primaryClass = {astro-ph.GA},
       adsurl = {https://ui.adsabs.harvard.edu/abs/2019MNRAS.488.3492S}
}

@Article{Shirazi:2012,
  Title                    = {{Strongly star forming galaxies in the local Universe with nebular He II{$\lambda$}4686 emission}},
  Author                   = {{Shirazi}, M. and {Brinchmann}, J.},
  Journal                  = {\mnras},
  Year                     = {2012},

  Month                    = apr,
  Pages                    = {1043-1063},
  Volume                   = {421}
}

@Article{Smith:1996,
  Title                    = {{A three-dimensional classification for WN stars}},
  Author                   = {{Smith}, L.~F. and {Shara}, M.~M. and {Moffat}, A.~F.~J.},
  Journal                  = {\mnras},
  Year                     = {1996},

  Month                    = jul,
  Pages                    = {163-191},
  Volume                   = {281},
  Adsurl                   = {http://adsabs.harvard.edu/abs/1996MNRAS.281..163S},
  Doi                      = {10.1093/mnras/281.1.163}
}

@Article{Sobral:2019,
  Title                    = {{On the nature and physical conditions of the luminous Ly {$\alpha$} emitter CR7 and its rest-frame UV components}},
  Author                   = {{Sobral}, D. and {Matthee}, J. and {Brammer}, G. and {Ferrara}, A. and {Alegre}, L. and {R{\"o}ttgering}, H. and {Schaerer}, D. and {Mobasher}, B. and {Darvish}, B.},
  Journal                  = {\mnras},
  Year                     = {2019},

  Month                    = jan,
  Pages                    = {2422-2441},
  Volume                   = {482},
  Adsurl                   = {http://adsabs.harvard.edu/abs/2019MNRAS.482.2422S},
  Archiveprefix            = {arXiv},
  Doi                      = {10.1093/mnras/sty2779},
  Eprint                   = {1710.08422}
}

@Article{Sobral:2015,
  Title                    = {{Evidence for PopIII-like Stellar Populations in the Most Luminous Lyman-{$\alpha$} Emitters at the Epoch of Reionization: Spectroscopic Confirmation}},
  Author                   = {{Sobral}, D. and {Matthee}, J. and {Darvish}, B. and {Schaerer}, D. and {Mobasher}, B. and {R{\"o}ttgering}, H.~J.~A. and {Santos}, S. and {Hemmati}, S.},
  Journal                  = {\apj},
  Year                     = {2015},

  Month                    = aug,
  Pages                    = {139},
  Volume                   = {808},
  Adsurl                   = {http://adsabs.harvard.edu/abs/2015ApJ...808..139S},
  Archiveprefix            = {arXiv},
  Doi                      = {10.1088/0004-637X/808/2/139},
  Eid                      = {139},
  Eprint                   = {1504.01734}
}

@Article{Spencer:2018,
  Title                    = {{The Binary Fraction of Stars in Dwarf Galaxies: The Cases of Draco and Ursa Minor}},
  Author                   = {{Spencer}, M.~E. and {Mateo}, M. and {Olszewski}, E.~W. and {Walker}, M.~G. and {McConnachie}, A.~W. and {Kirby}, E.~N.},
  Journal                  = {\aj},
  Year                     = {2018},

  Month                    = dec,
  Pages                    = {257},
  Volume                   = {156},
  Adsurl                   = {http://adsabs.harvard.edu/abs/2018AJ....156..257S},
  Archiveprefix            = {arXiv},
  Doi                      = {10.3847/1538-3881/aae3e4},
  Eid                      = {257},
  Eprint                   = {1811.06597}
}

@Article{Stanway:2019,
  Title                    = {{Initial mass function variations cannot explain the ionizing spectrum of low metallicity starbursts}},
  Author                   = {{Stanway}, E.~R. and {Eldridge}, J.~J.},
  Journal                  = {\aap},
  Year                     = {2019},

  Month                    = jan,
  Pages                    = {A105},
  Volume                   = {621},
  Adsurl                   = {http://adsabs.harvard.edu/abs/2019A%26A...621A.105S},
  Archiveprefix            = {arXiv},
  Doi                      = {10.1051/0004-6361/201834359},
  Eid                      = {A105},
  Eprint                   = {1811.03856}
}

@Article{Stanway:2016,
  Title                    = {{Stellar population effects on the inferred photon density at reionization}},
  Author                   = {{Stanway}, E.~R. and {Eldridge}, J.~J. and {Becker}, G.~D.},
  Journal                  = {\mnras},
  Year                     = {2016},

  Month                    = feb,
  Pages                    = {485-499},
  Volume                   = {456},
  Adsurl                   = {http://adsabs.harvard.edu/abs/2016MNRAS.456..485S},
  Archiveprefix            = {arXiv},
  Doi                      = {10.1093/mnras/stv2661},
  Eprint                   = {1511.03268}
}

@Article{Stasinska:2015,
  Title                    = {{Excitation properties of galaxies with the highest [O iii]/[O ii] ratios. No evidence for massive escape of ionizing photons}},
  Author                   = {{Stasi{\'n}ska}, G. and {Izotov}, Y. and {Morisset}, C. and {Guseva}, N.},
  Journal                  = {\aap},
  Year                     = {2015},

  Month                    = apr,
  Pages                    = {A83},
  Volume                   = {576},
  Adsurl                   = {http://adsabs.harvard.edu/abs/2015A%26A...576A..83S},
  Archiveprefix            = {arXiv},
  Doi                      = {10.1051/0004-6361/201425389},
  Eid                      = {A83},
  Eprint                   = {1503.00320}
}

@Article{Stevenson:2019,
   author = {{Stevenson}, S. and {Sampson}, M. and {Powell}, J. and {Vigna-G{\'o}mez}, A. and 
	{Neijssel}, C.~J. and {Sz{\'e}csi}, D and {Mandel}, I.},
    title = "{The Impact of Pair-instability Mass Loss on the Binary Black Hole Mass Distribution}",
  journal = {\apj},
archivePrefix = "arXiv",
   eprint = {1904.02821},
 primaryClass = "astro-ph.HE",
     year = 2019,
    month = sep,
   volume = 882,
      eid = {121},
    pages = {121},
      doi = {10.3847/1538-4357/ab3981},
   adsurl = {https://ui.adsabs.harvard.edu/abs/2019ApJ...882..121S}
}

@Article{Szecsi:2017long,
  Title                    = {{Single and binary stellar progenitors of long-duration gamma-ray bursts}},
  Author                   = {{Sz{\'e}csi}, D.},
  Journal                  = {Proceedings of Science, PoS(MULTIF2017)065, 2017},
  Year                     = {2017},

  Month                    = oct,
  Adsurl                   = {http://adsabs.harvard.edu/abs/2017arXiv171005655S},
  Archiveprefix            = {arXiv},
  Eprint                   = {1710.05655},
  Primaryclass             = {astro-ph.HE}
}

@Article{Szecsi:2017short,
  Title                    = {{How may short-duration GRBs form? A review of progenitor theories.}},
  Author                   = {{Sz{\'e}csi}, D.},
  Journal                  = {Contributions of the Astronomical Observatory Skalnate Pleso},
  Year                     = {2017},

  Month                    = jul,
  Pages                    = {108, [arXiv:1710.05356]},
  Volume                   = {47},
  Adsurl                   = {http://adsabs.harvard.edu/abs/2017CoSka..47..108S},
  Archiveprefix            = {arXiv},
  Eprint                   = {1710.05356},
  Primaryclass             = {astro-ph.HE}
}

@PhdThesis{Szecsi:2016,
  Title                    = {{The Evolution of Low-Metallicity Massive Stars}},
  Author                   = {{Sz{\'e}csi}, D.},
  School                   = {Bonn University, {\urlstyle{rm}\DeclareUrlCommand\url{\color{black}}
\url{http://bonndoc.ulb.uni-bonn.de/xmlui/handle/20.500.11811/6862}}},
  Year                     = {2016},
  Month                    = {July},
  Adsurl                   = {http://adsabs.harvard.edu/abs/2016PhDT.......375S},
  Doi                      = {10.5281/zenodo.998070}
}

@Article{Szecsi:2015,
  Title                    = {{Low-metallicity massive single stars with rotation. Evolutionary models applicable to I Zwicky 18}},
  Author                   = {{Sz{\'e}csi}, D. and {Langer}, N. and {Yoon}, S.-C. and {Sanyal}, D. and {de Mink}, S. and {Evans}, C.~J. and {Dermine}, T.},
  Journal                  = {\aap},
  Year                     = {2015},

  Month                    = sep,
  Pages                    = {A15},
  Volume                   = {581},

  Eid                      = {A15}
}

@Article{Szecsi:2019,
  Title                    = {{Role of supergiants in the formation of globular clusters}},
  Author                   = {{Sz{\'e}csi}, D. and {W{\"u}nsch}, R.},
  Journal                  = {ApJ, Vol. 871, nr. 1},
  Year                     = {2019},
  Adsurl                   = {http://adsabs.harvard.edu/abs/2018arXiv180901395S},
  Archiveprefix            = {arXiv},
  Eprint                   = {1809.01395}
}

@Article{Tramper:2013,
  Title                    = {{On the nature of WO stars: a quantitative analysis of the WO3 star DR1 in IC 1613}},
  Author                   = {{Tramper}, F. and {Gr{\"a}fener}, G. and {Hartoog}, O.~E. and {Sana}, H. and {de Koter}, A. and {Vink}, J.~S. and {Ellerbroek}, L.~E. and {Langer}, N. and {Garcia}, M. and {Kaper}, L. and {de Mink}, S.~E. },
  Journal                  = {\aap},
  Year                     = {2013},

  Month                    = nov,
  Pages                    = {A72},
  Volume                   = {559},
  Adsurl                   = {http://adsabs.harvard.edu/abs/2013A%26A...559A..72T},
  Archiveprefix            = {arXiv},
  Doi                      = {10.1051/0004-6361/201322155},
  Eid                      = {A72},
  Eprint                   = {1310.2849},
  Primaryclass             = {astro-ph.SR}
}

@Article{Vaduvescu:2007,
  Title                    = {{Chemical Properties of Star-Forming Dwarf Galaxies}},
  Author                   = {{Vaduvescu}, O. and {McCall}, M.~L. and {Richer}, M.~G.},
  Journal                  = {\aj},
  Year                     = {2007},

  Month                    = aug,
  Pages                    = {604-616},
  Volume                   = {134}
}

@Article{VignaGomez:2018,
  Title                    = {{On the formation history of Galactic double neutron stars}},
  Author                   = {{Vigna-G{\'o}mez}, A. and {Neijssel}, C.~J. and {Stevenson}, S. and {Barrett}, J.~W. and {Belczynski}, K. and {Justham}, S. and {de Mink}, S.~E. and {M{\"u}ller}, B. and {Podsiadlowski}, P. and {Renzo}, M. and {Sz{\'e}csi}, D. and {Mandel}, I.},
  Journal                  = {\mnras},
  Year                     = {2018},

  Month                    = dec,
  Pages                    = {4009-4029},
  Volume                   = {481},
  Adsurl                   = {http://adsabs.harvard.edu/abs/2018MNRAS.481.4009V},
  Archiveprefix            = {arXiv},
  Doi                      = {10.1093/mnras/sty2463},
  Eprint                   = {1805.07974},
  Primaryclass             = {astro-ph.SR}
}

@Article{Vink:2001,
  Author                   = {Vink, J.S. and de Koter, A. and Lamers, H.J.G.L.M.},
  Journal                  = {A\&A},
  Year                     = {2001},
  Pages                    = {574-588},
  Volume                   = {369}
}

@Article{Visbal:2016,
  Title                    = {{Formation of massive Population III galaxies through photoionization feedback: a possible explanation for CR 7}},
  Author                   = {{Visbal}, E. and {Haiman}, Z. and {Bryan}, G.~L.},
  Journal                  = {\mnras},
  Year                     = {2016},

  Month                    = jul,
  Pages                    = {L59-L63},
  Volume                   = {460},
  Adsurl                   = {http://adsabs.harvard.edu/abs/2016MNRAS.460L..59V},
  Archiveprefix            = {arXiv},
  Doi                      = {10.1093/mnrasl/slw071},
  Eprint                   = {1602.04843}
}

@Article{Weisz:2014,
  Title                    = {{The Star Formation Histories of Local Group Dwarf Galaxies. I. Hubble Space Telescope/Wide Field Planetary Camera 2 Observations}},
  Author                   = {{Weisz}, D.~R. and {Dolphin}, A.~E. and {Skillman}, E.~D. and {Holtzman}, J. and {Gilbert}, K.~M. and {Dalcanton}, J.~J. and {Williams}, B.~F.},
  Journal                  = {\apj},
  Year                     = {2014},

  Month                    = jul,
  Pages                    = {147},
  Volume                   = {789},
  Adsurl                   = {http://adsabs.harvard.edu/abs/2014ApJ...789..147W},
  Archiveprefix            = {arXiv},
  Doi                      = {10.1088/0004-637X/789/2/147},
  Eid                      = {147},
  Eprint                   = {1404.7144}
}

@Article{Woosley:2006,
  Title                    = {{The Progenitor Stars of Gamma-Ray Bursts}},
  Author                   = {{Woosley}, S.~E. and {Heger}, A.},
  Journal                  = {\apj},
  Year                     = {2006},

  Month                    = feb,
  Pages                    = {914-921},
  Volume                   = {637}
}

@Article{Yang:2016,
  Title                    = {{Green Pea Galaxies Reveal Secrets of Ly{$\alpha$} Escape}},
  Author                   = {{Yang}, H. and {Malhotra}, S. and {Gronke}, M. and {Rhoads}, J.~E. and {Dijkstra}, M. and {Jaskot}, A. and {Zheng}, Z. and {Wang}, J. },
  Journal                  = {\apj},
  Year                     = {2016},

  Month                    = apr,
  Pages                    = {130},
  Volume                   = {820},
  Adsurl                   = {http://adsabs.harvard.edu/abs/2016ApJ...820..130Y},
  Archiveprefix            = {arXiv},
  Doi                      = {10.3847/0004-637X/820/2/130},
  Eid                      = {130},
  Eprint                   = {1506.02885}
}

@Article{Yoon:2012,
  Title                    = {{Evolution of massive Population III stars with rotation and magnetic fields}},
  Author                   = {{Yoon}, S.-C. and {Dierks}, A. and {Langer}, N.},
  Journal                  = {\aap},
  Year                     = {2012},

  Month                    = jun,
  Pages                    = {A113},
  Volume                   = {542},

  Eid                      = {A113}
}

@Article{Yoon:2005,
  Author                   = {Yoon, S.-C. and Langer, N.},
  Journal                  = {A\&A},
  Year                     = {2005},
  Pages                    = {643-648},
  Volume                   = {443}
}

@Article{Yoon:2006,
  Title                    = {{Single star progenitors of long gamma-ray bursts. I. Model grids and redshift dependent GRB rate}},
  Author                   = {{Yoon}, S.-C. and {Langer}, N. and {Norman}, C.},
  Journal                  = {\aap},
  Year                     = {2006},

  Month                    = dec,
  Pages                    = {199-208},
  Volume                   = {460},
  Adsurl                   = {http://adsabs.harvard.edu/abs/2006A%26A...460..199Y},
  Doi                      = {10.1051/0004-6361:20065912},
  Eprint                   = {astro-ph/0606637}
}

@Article{Zhao:2013,
  Title                    = {A Study on the Chemical Properties of Blue Compact Dwarf Galaxies},
  Author                   = {Yinghe Zhao and Yu Gao and Qiusheng Gu},
  Journal                  = {ApJ},
  Year                     = {2013},
  Number                   = {1},
  Pages                    = {44},
  Volume                   = {764}
}

@ARTICLE{Sander:2020,
       author = {{Sander}, Andreas A.~C. and {Vink}, Jorick S.},
        title = "{On the nature of massive helium star winds and Wolf-Rayet-type mass-loss}",
      journal = {\mnras},
         year = 2020,
        month = nov,
       volume = {499},
       number = {1},
        pages = {873-892},
          doi = {10.1093/mnras/staa2712},
archivePrefix = {arXiv},
       eprint = {2009.01849},
 primaryClass = {astro-ph.SR},
       adsurl = {https://ui.adsabs.harvard.edu/abs/2020MNRAS.499..873S}
}

@ARTICLE{Lahen:2023,
       author = {{Lah{\'e}n}, Natalia and {Naab}, Thorsten and {Kauffmann}, Guinevere and {Sz{\'e}csi}, Dorottya and {Hislop}, Jessica May and {Rantala}, Antti and {Kozyreva}, Alexandra and {Walch}, Stefanie and {Hu}, Chia-Yu},
        title = "{Formation of star clusters and enrichment by massive stars in simulations of low-metallicity galaxies with a fully sampled initial stellar mass function}",
      journal = {\mnras},
         year = 2023,
        month = jun,
       volume = {522},
       number = {2},
        pages = {3092-3116},
          doi = {10.1093/mnras/stad1147},
archivePrefix = {arXiv},
       eprint = {2211.15705},
 primaryClass = {astro-ph.GA},
       adsurl = {https://ui.adsabs.harvard.edu/abs/2023MNRAS.522.3092L}
}

@ARTICLE{Romagnolo:2022,
       author = {{Romagnolo}, A. and {Belczynski}, K. and {Klencki}, J. and {Agrawal}, P. and {Shenar}, T. and {Sz{\'e}csi}, D.},
        title = "{The role of stellar expansion on the formation of gravitational wave sources}",
      journal = {\mnras},
         year = 2023,
        month = oct,
       volume = {525},
       number = {1},
        pages = {706-720},
          doi = {10.1093/mnras/stad2366},
archivePrefix = {arXiv},
       eprint = {2211.15800},
 primaryClass = {astro-ph.HE},
       adsurl = {https://ui.adsabs.harvard.edu/abs/2023MNRAS.525..706R}
}

@ARTICLE{Franeck:2022,
       author = {{Franeck}, Annika and {W{\"u}nsch}, Richard and {Mart{\'\i}nez-Gonz{\'a}lez}, Sergio and {Orlitov{\'a}}, Ivana and {Boorman}, Peter and {Svoboda}, Ji{\v{r}}{\'\i} and {Sz{\'e}csi}, Dorottya and {Douna}, Vanesa},
        title = "{X-Ray Emission from Star-cluster Winds in Starburst Galaxies}",
      journal = {\apj},
         year = 2022,
        month = mar,
       volume = {927},
       number = {2},
          eid = {212},
        pages = {212},
          doi = {10.3847/1538-4357/ac4fc2},
archivePrefix = {arXiv},
       eprint = {2201.12339},
 primaryClass = {astro-ph.GA},
       adsurl = {https://ui.adsabs.harvard.edu/abs/2022ApJ...927..212F}
}

@ARTICLE{Chruslinska:2019,
       author = {{Chruslinska}, Martyna and {Nelemans}, Gijs},
        title = "{Metallicity of stars formed throughout the cosmic history based on the observational properties of star-forming galaxies}",
      journal = {\mnras},
         year = 2019,
        month = oct,
       volume = {488},
       number = {4},
        pages = {5300-5326},
          doi = {10.1093/mnras/stz2057},
archivePrefix = {arXiv},
       eprint = {1907.11243},
 primaryClass = {astro-ph.GA},
       adsurl = {https://ui.adsabs.harvard.edu/abs/2019MNRAS.488.5300C}
}

@ARTICLE{Oskinova:2022,
       author = {{Oskinova}, Lidia M. and {Schaerer}, Daniel},
        title = "{Ionization of He II in star-forming galaxies by X-rays from cluster winds and superbubbles}",
      journal = {\aap},
         year = 2022,
        month = may,
       volume = {661},
          eid = {A67},
        pages = {A67},
          doi = {10.1051/0004-6361/202142520},
archivePrefix = {arXiv},
       eprint = {2203.04987},
 primaryClass = {astro-ph.GA},
       adsurl = {https://ui.adsabs.harvard.edu/abs/2022A&A...661A..67O}
}

@ARTICLE{Smith:2023,
       author = {{Smith}, Linda J. and {Oey}, M.~S. and {Hernandez}, Svea and {Ryon}, Jenna and {Leitherer}, Claus and {Charlot}, Stephane and {Bruzual}, Gustavo and {Calzetti}, Daniela and {Chu}, You-Hua and {Hayes}, Matthew J. and {James}, Bethan L. and {Jaskot}, Anne E. and {{\"O}stlin}, G{\"o}ran},
        title = "{HST FUV Spectroscopy of Super Star Cluster A in the Green Pea Analog Mrk 71: Revealing the Presence of Very Massive Stars}",
      journal = {\apj},
         year = 2023,
        month = dec,
       volume = {958},
       number = {2},
          eid = {194},
        pages = {194},
          doi = {10.3847/1538-4357/ad00b4},
archivePrefix = {arXiv},
       eprint = {2310.03413},
 primaryClass = {astro-ph.GA},
       adsurl = {https://ui.adsabs.harvard.edu/abs/2023ApJ...958..194S}
}

@ARTICLE{Burssens:2023,
       author = {{Burssens}, Siemen and {Bowman}, Dominic M. and {Michielsen}, Mathias and {Sim{\'o}n-D{\'\i}az}, Sergio and {Aerts}, Conny and {Vanlaer}, Vincent and {Banyard}, Gareth and {Nardetto}, Nicolas and {Townsend}, Richard H.~D. and {Handler}, Gerald and {Mombarg}, Joey S.~G. and {Vanderspek}, Roland and {Ricker}, George},
        title = "{A calibration point for stellar evolution from massive star asteroseismology}",
      journal = {Nature Astronomy},
         year = 2023,
        month = aug,
       volume = {7},
        pages = {913-930},
          doi = {10.1038/s41550-023-01978-y},
archivePrefix = {arXiv},
       eprint = {2306.11798},
 primaryClass = {astro-ph.SR},
       adsurl = {https://ui.adsabs.harvard.edu/abs/2023NatAs...7..913B}
}

@INPROCEEDINGS{Hamann:2015,
       author = {{Hamann}, W. -R. and {Sander}, A. and {Todt}, H.},
        title = "{Wolf-Rayet Stars}",
    booktitle = {Wolf-Rayet Stars: Proceedings of an International Workshop held in Potsdam, Germany},
         year = 2015,
       editor = {{Hamann}, Wolf-Rainer and {Sander}, Andreas and {Todt}, Helge},
        month = jan,
       adsurl = {https://ui.adsabs.harvard.edu/abs/2015wrs..conf.....H}
}

@ARTICLE{duBuisson:2020G,
       author = {{du Buisson}, L. and {Marchant}, P. and {Podsiadlowski}, Ph and {Kobayashi}, C. and {Abdalla}, F.~B. and {Taylor}, P. and {Mandel}, I. and {de Mink}, S.~E. and {Moriya}, T.~J. and {Langer}, N.},
        title = "{Cosmic rates of black hole mergers and pair-instability supernovae from chemically homogeneous binary evolution}",
      journal = {\mnras},
         year = 2020,
        month = dec,
       volume = {499},
       number = {4},
        pages = {5941-5959},
          doi = {10.1093/mnras/staa3225},
archivePrefix = {arXiv},
       eprint = {2002.11630},
 primaryClass = {astro-ph.HE},
       adsurl = {https://ui.adsabs.harvard.edu/abs/2020MNRAS.499.5941D}
}

@ARTICLE{Sharpe:2024G,
       author = {{Sharpe}, K. and {van Son}, L.~A.~C. and {de Mink}, S.~E. and {Farmer}, R. and {Marchant}, P. and {Koenigsberger}, G.},
        title = "{Investigating the Chemically Homogeneous Evolution Channel and Its Role in the Formation of the Enigmatic Binary Black Hole Progenitor Candidate HD 5980}",
      journal = {\apj},
         year = 2024,
        month = may,
       volume = {966},
       number = {1},
          eid = {9},
        pages = {9},
          doi = {10.3847/1538-4357/ad2f3e},
archivePrefix = {arXiv},
       eprint = {2402.12438},
 primaryClass = {astro-ph.SR},
       adsurl = {https://ui.adsabs.harvard.edu/abs/2024ApJ...966....9S}
}

@ARTICLE{Gull:2022,
       author = {{Gull}, Maude and {Weisz}, Daniel R. and {Senchyna}, Peter and {Sandford}, Nathan R. and {Choi}, Yumi and {McLeod}, Anna F. and {El-Badry}, Kareem and {G{\"o}tberg}, Ylva and {Gilbert}, Karoline M. and {Boyer}, Martha and {Dalcanton}, Julianne J. and {GuhaThakurta}, Puragra and {Goldman}, Steven and {Marigo}, Paola and {McQuinn}, Kristen B.~W. and {Pastorelli}, Giada and {Stark}, Daniel P. and {Skillman}, Evan and {Ting}, Yuan-sen and {Williams}, Benjamin F.},
        title = "{A Panchromatic Study of Massive Stars in the Extremely Metal-poor Local Group Dwarf Galaxy Leo A}",
      journal = {\apj},
         year = 2022,
        month = dec,
       volume = {941},
       number = {2},
          eid = {206},
        pages = {206},
          doi = {10.3847/1538-4357/aca295},
archivePrefix = {arXiv},
       eprint = {2211.14349},
 primaryClass = {astro-ph.SR},
       adsurl = {https://ui.adsabs.harvard.edu/abs/2022ApJ...941..206G}
}

@ARTICLE{Senchyna:2020,
       author = {{Senchyna}, Peter and {Stark}, Daniel P. and {Mirocha}, Jordan and {Reines}, Amy E. and {Charlot}, St{\'e}phane and {Jones}, Tucker and {Mulchaey}, John S.},
        title = "{High-mass X-ray binaries in nearby metal-poor galaxies: on the contribution to nebular He II emission}",
      journal = {\mnras},
         year = 2020,
        month = may,
       volume = {494},
       number = {1},
        pages = {941-957},
          doi = {10.1093/mnras/staa586},
archivePrefix = {arXiv},
       eprint = {1909.10574},
 primaryClass = {astro-ph.GA},
       adsurl = {https://ui.adsabs.harvard.edu/abs/2020MNRAS.494..941S}
}

@ARTICLE{Bestenlehner:2020,
       author = {{Bestenlehner}, Joachim M. and {Crowther}, Paul A. and {Caballero-Nieves}, Saida M. and {Schneider}, Fabian R.~N. and {Sim{\'o}n-D{\'\i}az}, Sergio and {Brands}, Sarah A. and {de Koter}, Alex and {Gr{\"a}fener}, G{\"o}tz and {Herrero}, Artemio and {Langer}, Norbert and {Lennon}, Daniel J. and {Ma{\'\i}z Apell{\'a}niz}, Jesus and {Puls}, Joachim and {Vink}, Jorick S.},
        title = "{The R136 star cluster dissected with Hubble Space Telescope/STIS - II. Physical properties of the most massive stars in R136}",
      journal = {\mnras},
         year = 2020,
        month = dec,
       volume = {499},
       number = {2},
        pages = {1918-1936},
          doi = {10.1093/mnras/staa2801},
archivePrefix = {arXiv},
       eprint = {2009.05136},
 primaryClass = {astro-ph.SR},
       adsurl = {https://ui.adsabs.harvard.edu/abs/2020MNRAS.499.1918B}
}

@ARTICLE{Aadland:2022,
       author = {{Aadland}, Erin and {Massey}, Philip and {Hillier}, D. John and {Morrell}, Nidia I. and {Neugent}, Kathryn F. and {Eldridge}, J.~J.},
        title = "{WO-type Wolf-Rayet Stars: The Last Hurrah of Massive Star Evolution}",
      journal = {\apj},
         year = 2022,
        month = jun,
       volume = {931},
       number = {2},
          eid = {157},
        pages = {157},
          doi = {10.3847/1538-4357/ac66e7},
archivePrefix = {arXiv},
       eprint = {2204.04258},
 primaryClass = {astro-ph.SR},
       adsurl = {https://ui.adsabs.harvard.edu/abs/2022ApJ...931..157A}
}

@ARTICLE{LopezSanchez:2010b,
       author = {{L{\'o}pez-S{\'a}nchez}, {\'A}. R. and {Esteban}, C.},
        title = "{Massive star formation in Wolf-Rayet galaxies. IV. Colours, chemical-composition analysis and metallicity-luminosity relations}",
      journal = {\aap},
         year = 2010,
        month = jul,
       volume = {517},
          eid = {A85},
        pages = {A85},
          doi = {10.1051/0004-6361/201014156},
archivePrefix = {arXiv},
       eprint = {1004.0626},
 primaryClass = {astro-ph.CO},
       adsurl = {https://ui.adsabs.harvard.edu/abs/2010A&A...517A..85L}
}

@ARTICLE{LopezSanchez:2010a,
       author = {{L{\'o}pez-S{\'a}nchez}, {\'A}. R. and {Esteban}, C.},
        title = "{Massive star formation in Wolf-Rayet galaxies. III. Analysis of the O and WR populations}",
      journal = {\aap},
         year = 2010,
        month = jun,
       volume = {516},
          eid = {A104},
        pages = {A104},
          doi = {10.1051/0004-6361/200913434},
archivePrefix = {arXiv},
       eprint = {1004.0051},
 primaryClass = {astro-ph.CO},
       adsurl = {https://ui.adsabs.harvard.edu/abs/2010A&A...516A.104L}
}

@ARTICLE{Sander:2022IAU,
       author = {{Sander}, Andreas A.~C.},
        title = "{Massive Stars in the Far and Extreme Ultraviolet}",
      journal = {IAU GA 2022 Proceedings},
         year = 2022,
        month = nov,
          eid = {arXiv:2211.05424},
        pages = {arXiv:2211.05424},
          doi = {10.48550/arXiv.2211.05424},
archivePrefix = {arXiv},
       eprint = {2211.05424},
 primaryClass = {astro-ph.SR},
       adsurl = {https://ui.adsabs.harvard.edu/abs/2022arXiv221105424S}
}

@ARTICLE{Abdellaoui:2023,
       author = {Abdellaoui, Slah},
        title = "{Hydrodynamic modeling of
stellar winds}",
      journal = {PhD thesis, MASARYK UNIVERSITY, Faculty of Science},
         year = 2023,
        url = {https://is.muni.cz/th/kq4lg/Hydrodynamic_Modeling_of_Hot_Stars.pdf}
}

@ARTICLE{Sen:2024,
       author = {{Sen}, K. and {El Mellah}, I. and {Langer}, N. and {Xu}, X. -T. and {Quast}, M. and {Pauli}, D.},
        title = "{Whispering in the dark: Faint X-ray emission from black holes with OB star companions}",
      journal = {\aap},
         year = 2024,
        month = oct,
       volume = {690},
          eid = {A256},
        pages = {A256},
          doi = {10.1051/0004-6361/202450940},
archivePrefix = {arXiv},
       eprint = {2406.08596},
 primaryClass = {astro-ph.HE},
       adsurl = {https://ui.adsabs.harvard.edu/abs/2024A&A...690A.256S}
}

@ARTICLE{ArroyoPolonio:2023,
       author = {{Arroyo-Polonio}, A. and {Iglesias-P{\'a}ramo}, J. and {Kehrig}, C. and {V{\'\i}lchez}, J.~M. and {Amor{\'\i}n}, R. and {Breda}, I. and {P{\'e}rez-Montero}, E. and {P{\'e}rez-D{\'\i}az}, B. and {Hayes}, M.},
        title = "{A MUSE/VLT spatially resolved study of the emission structure of Green Pea galaxies}",
      journal = {\aap},
         year = 2023,
        month = sep,
       volume = {677},
          eid = {A114},
        pages = {A114},
          doi = {10.1051/0004-6361/202346192},
archivePrefix = {arXiv},
       eprint = {2309.09585},
 primaryClass = {astro-ph.GA},
       adsurl = {https://ui.adsabs.harvard.edu/abs/2023A&A...677A.114A}
}

@ARTICLE{Szecsi:2022ok,
       author = {{Sz{\'e}csi}, Dorottya and {Agrawal}, Poojan and {W{\"u}nsch}, Richard and {Langer}, Norbert},
        title = "{Bonn Optimized Stellar Tracks (BoOST). Simulated populations of massive and very massive stars for astrophysical applications}",
      journal = {\aap},
         year = 2022,
        month = feb,
       volume = {658},
          eid = {A125},
        pages = {A125},
          doi = {10.1051/0004-6361/202141536},
archivePrefix = {arXiv},
       eprint = {2004.08203},
 primaryClass = {astro-ph.SR},
       adsurl = {https://ui.adsabs.harvard.edu/abs/2022A&A...658A.125S}
}

@ARTICLE{Agrawal:2022,
       author = {{Agrawal}, Poojan and {Sz{\'e}csi}, Dorottya and {Stevenson}, Simon and {Eldridge}, Jan J. and {Hurley}, Jarrod},
        title = "{Explaining the differences in massive star models from various simulations}",
      journal = {\mnras},
         year = 2022,
        month = jun,
       volume = {512},
       number = {4},
        pages = {5717-5725},
          doi = {10.1093/mnras/stac930},
archivePrefix = {arXiv},
       eprint = {2112.02800},
       adsurl = {https://ui.adsabs.harvard.edu/abs/2022MNRAS.512.5717A}
}

@ARTICLE{Thuan:2005,
       author = {{Thuan}, Trinh X. and {Izotov}, Yuri I.},
        title = "{High-Ionization Emission in Metal-deficient Blue Compact Dwarf Galaxies}",
      journal = {\apjs},
         year = 2005,
        month = dec,
       volume = {161},
       number = {2},
        pages = {240-270},
          doi = {10.1086/491657},
archivePrefix = {arXiv},
       eprint = {astro-ph/0507209},
 primaryClass = {astro-ph},
       adsurl = {https://ui.adsabs.harvard.edu/abs/2005ApJS..161..240T}
}

@ARTICLE{Kehrig:2021,
       author = {{Kehrig}, C. and {Guerrero}, M.~A. and {V{\'\i}lchez}, J.~M. and {Ramos-Larios}, G.},
        title = "{On the Contribution of the X-Ray Source to the Extended Nebular He II Emission in IZW18}",
      journal = {\apjl},
         year = 2021,
        month = feb,
       volume = {908},
       number = {2},
          eid = {L54},
        pages = {L54},
          doi = {10.3847/2041-8213/abe41b},
archivePrefix = {arXiv},
       eprint = {2103.06599},
 primaryClass = {astro-ph.GA},
       adsurl = {https://ui.adsabs.harvard.edu/abs/2021ApJ...908L..54K}
}

@ARTICLE{ArroyoPolonio:2024,
       author = {{Arroyo-Polonio}, A. and {Kehrig}, C. and {Iglesias-P{\'a}ramo}, J. and {V{\'\i}lchez}, J.~M. and {P{\'e}rez-Montero}, E. and {Duarte Puertas}, S. and {Gallego}, J. and {Reverte}, D. and {Cabrera-Lavers}, A.},
        title = "{Unraveling the kinematics of IZw18: A detailed study of ionized gas with MEGARA/GTC}",
      journal = {\aap},
         year = 2024,
        month = jul,
       volume = {687},
          eid = {A77},
        pages = {A77},
          doi = {10.1051/0004-6361/202449606},
archivePrefix = {arXiv},
       eprint = {2404.13956},
 primaryClass = {astro-ph.GA},
       adsurl = {https://ui.adsabs.harvard.edu/abs/2024A&A...687A..77A}
}

@ARTICLE{Berg:2022,
       author = {{Berg}, Danielle A. and {James}, Bethan L. and {King}, Teagan and {McDonald}, Meaghan and {Chen}, Zuyi and {Chisholm}, John and {Heckman}, Timothy and {Martin}, Crystal L. and {Stark}, Dan P. and {Aloisi}, Alessandra and {Amor{\'\i}n}, Ricardo O. and {Arellano-C{\'o}rdova}, Karla Z. and {Bayliss}, Matthew and {Bordoloi}, Rongmon and {Brinchmann}, Jarle and {Charlot}, St{\'e}phane and {Chevallard}, Jacopo and {Clark}, Ilyse and {Erb}, Dawn K. and {Feltre}, Anna and {Gronke}, Max and {Hayes}, Matthew and {Henry}, Alaina and {Hernandez}, Svea and {Jaskot}, Anne and {Jones}, Tucker and {Kewley}, Lisa J. and {Kumari}, Nimisha and {Leitherer}, Claus and {Llerena}, Mario and {Maseda}, Michael and {Mingozzi}, Matilde and {Nanayakkara}, Themiya and {Ouchi}, Masami and {Plat}, Adele and {Pogge}, Richard W. and {Ravindranath}, Swara and {Rigby}, Jane R. and {Sanders}, Ryan and {Scarlata}, Claudia and {Senchyna}, Peter and {Skillman}, Evan D. and {Steidel}, Charles C. and {Strom}, Allison L. and {Sugahara}, Yuma and {Wilkins}, Stephen M. and {Wofford}, Aida and {Xu}, Xinfeng and {Classy Team}},
        title = "{The COS Legacy Archive Spectroscopy Survey (CLASSY) Treasury Atlas}",
      journal = {\apjs},
         year = 2022,
        month = aug,
       volume = {261},
       number = {2},
          eid = {31},
        pages = {31},
          doi = {10.3847/1538-4365/ac6c03},
archivePrefix = {arXiv},
       eprint = {2203.07357},
 primaryClass = {astro-ph.GA},
       adsurl = {https://ui.adsabs.harvard.edu/abs/2022ApJS..261...31B}
}

@ARTICLE{Liu:2025,
       author = {{Liu}, Boyuan and {Sibony}, Yves and {Meynet}, Georges and {Bromm}, Volker},
        title = "{Signatures of Rapidly Rotating Stars with Chemically Homogeneous Evolution in the First Galaxies}",
      journal = {\apjl},
         year = 2025,
        month = feb,
       volume = {980},
       number = {2},
          eid = {L30},
        pages = {L30},
          doi = {10.3847/2041-8213/adb151},
archivePrefix = {arXiv},
       eprint = {2412.02002},
 primaryClass = {astro-ph.GA},
       adsurl = {https://ui.adsabs.harvard.edu/abs/2025ApJ...980L..30L}
}

@ARTICLE{Hirschauer:2024,
       author = {{Hirschauer}, Alec S. and {Crouzet}, Nicolas and {Habel}, Nolan and {Lenki{\'c}}, Laura and {Nally}, Conor and {Jones}, Olivia C. and {Bortolini}, Giacomo and {Boyer}, Martha L. and {Justtanont}, Kay and {Meixner}, Margaret and {{\"O}stlin}, G{\"o}ran and {Wright}, Gillian S. and {Azzollini}, Ruyman and {Blommaert}, Joris A.~D.~L. and {Brandl}, Bernhard and {Decin}, Leen and {Nayak}, Omnarayani and {Royer}, Pierre and {Sargent}, B.~A. and {van der Werf}, Paul},
        title = "{Imaging of I Zw 18 by JWST. I. Detecting Dusty Stellar Populations}",
      journal = {\aj},
         year = 2024,
        month = jul,
       volume = {168},
       number = {1},
          eid = {23},
        pages = {23},
          doi = {10.3847/1538-3881/ad4967},
archivePrefix = {arXiv},
       eprint = {2403.06980},
 primaryClass = {astro-ph.GA},
       adsurl = {https://ui.adsabs.harvard.edu/abs/2024AJ....168...23H}
}

@ARTICLE{Vilchez:1998,
       author = {{V{\'\i}lchez}, Jos{\'e} M. and {Iglesias-P{\'a}ramo}, Jorge},
        title = "{Bidimensional Spectroscopic Mapping and Chemical Abundances of the Star-forming Dwarf Galaxy I ZW 18}",
      journal = {\apj},
         year = 1998,
        month = nov,
       volume = {508},
       number = {1},
        pages = {248-261},
          doi = {10.1086/306374},
       adsurl = {https://ui.adsabs.harvard.edu/abs/1998ApJ...508..248V}
}

@ARTICLE{Mintz:2025,
       author = {{Mintz}, Abby and {Telford}, O. Grace and {Kirby}, Evan N. and {Chisholm}, John and {McQuinn}, Kristen B.~W. and {Berg}, Danielle},
        title = "{A Spectroscopic Survey of Metal-Poor OB Stars in Local Dwarf Galaxy NGC 3109}",
      journal = {arXiv e-prints},
         year = 2025,
        month = mar,
          eid = {arXiv:2503.19020},
        pages = {arXiv:2503.19020},
          doi = {10.48550/arXiv.2503.19020},
archivePrefix = {arXiv},
       eprint = {2503.19020},
 primaryClass = {astro-ph.SR},
       adsurl = {https://ui.adsabs.harvard.edu/abs/2025arXiv250319020M}
}

@ARTICLE{Roy:2025,
       author = {{Roy}, A. and {Krumholz}, M.~R. and {Salvadori}, S. and {Meynet}, G. and {Ekstr{\"o}m}, S. and {Vink}, J.~S. and {Sander}, A.~A.~C. and {Sutherland}, R.~S. and {Paul}, S. and {Pallottini}, A. and {Sk{\'u}lad{\'o}ttir}, {\'A}.},
        title = "{Strong nebular He II emission induced by He$^{+}$ ionizing photons escaping through the clumpy winds of massive stars}",
      journal = {\aap},
         year = 2025,
        month = apr,
       volume = {696},
          eid = {A29},
        pages = {A29},
          doi = {10.1051/0004-6361/202553697},
       adsurl = {https://ui.adsabs.harvard.edu/abs/2025A&A...696A..29R}
}

@ARTICLE{Klessen:2023,
       author = {{Klessen}, Ralf S. and {Glover}, Simon C.~O.},
        title = "{The First Stars: Formation, Properties, and Impact}",
      journal = {\araa},
         year = 2023,
        month = aug,
       volume = {61},
        pages = {65-130},
          doi = {10.1146/annurev-astro-071221-053453},
archivePrefix = {arXiv},
       eprint = {2303.12500},
 primaryClass = {astro-ph.CO},
       adsurl = {https://ui.adsabs.harvard.edu/abs/2023ARA&A..61...65K}
}

@ARTICLE{Lecroq:2024,
       author = {{Lecroq}, Marie and {Charlot}, St{\'e}phane and {Bressan}, Alessandro and {Bruzual}, Gustavo and {Costa}, Guglielmo and {Iorio}, Giuliano and {Spera}, Mario and {Mapelli}, Michela and {Chen}, Yang and {Chevallard}, Jacopo and {Dall'Amico}, Marco},
        title = "{Nebular emission from young stellar populations including binary stars}",
      journal = {\mnras},
         year = 2024,
        month = jan,
       volume = {527},
       number = {3},
        pages = {9480-9504},
          doi = {10.1093/mnras/stad3838},
archivePrefix = {arXiv},
       eprint = {2312.08432},
 primaryClass = {astro-ph.GA},
       adsurl = {https://ui.adsabs.harvard.edu/abs/2024MNRAS.527.9480L}
}

@ARTICLE{Dorozsmai:2024,
       author = {{Dorozsmai}, Andris and {Toonen}, Silvia and {Vigna-G{\'o}mez}, Alejandro and {de Mink}, Selma E. and {Kummer}, Floris},
        title = "{Stellar triples with chemically homogeneously evolving inner binaries}",
      journal = {\mnras},
         year = 2024,
        month = feb,
       volume = {527},
       number = {4},
        pages = {9782-9809},
          doi = {10.1093/mnras/stad3819},
archivePrefix = {arXiv},
       eprint = {2307.04793},
 primaryClass = {astro-ph.SR},
       adsurl = {https://ui.adsabs.harvard.edu/abs/2024MNRAS.527.9782D}
}

@ARTICLE{Riley:2021,
       author = {{Riley}, Jeff and {Mandel}, Ilya and {Marchant}, Pablo and {Butler}, Ellen and {Nathaniel}, Kaila and {Neijssel}, Coenraad and {Shortt}, Spencer and {Vigna-G{\'o}mez}, Alejandro},
        title = "{Chemically homogeneous evolution: a rapid population synthesis approach}",
      journal = {\mnras},
         year = 2021,
        month = jul,
       volume = {505},
       number = {1},
        pages = {663-676},
          doi = {10.1093/mnras/stab1291},
archivePrefix = {arXiv},
       eprint = {2010.00002},
 primaryClass = {astro-ph.SR},
       adsurl = {https://ui.adsabs.harvard.edu/abs/2021MNRAS.505..663R}
}

@ARTICLE{Vigna-Gomez:2025,
       author = {{Vigna-G{\'o}mez}, A. and {Grishin}, E. and {Stegmann}, J. and {Olejak}, A. and {Popa}, S.~A. and {Liu}, B. and {Rajamuthukumar}, A.~S. and {van Son}, L.~A.~C. and {Bobrick}, A. and {Dorozsmai}, A.},
        title = "{Prompt stellar and binary black hole mergers in tight triples: Insights from chemically homogeneous evolution}",
      journal = {\aap},
         year = 2025,
        month = jul,
       volume = {699},
          eid = {A272},
        pages = {A272},
          doi = {10.1051/0004-6361/202554680},
archivePrefix = {arXiv},
       eprint = {2503.17006},
 primaryClass = {astro-ph.SR},
       adsurl = {https://ui.adsabs.harvard.edu/abs/2025A&A...699A.272V}
}

@ARTICLE{Cannon:1916,
       author = {{Cannon}, Annie Jump and {Pickering}, Edward Charles},
        title = "{Spectra having bright lines}",
      journal = {Annals of Harvard College Observatory},
         year = 1916,
        month = jan,
       volume = {76},
       number = {3},
        pages = {19-42},
       adsurl = {https://ui.adsabs.harvard.edu/abs/1916AnHar..76...19C}
}

@ARTICLE{Payne:1930,
       author = {{Payne}, Cecilia H.},
        title = "{Classification of the O Stars}",
      journal = {Harvard College Observatory Bulletin},
         year = 1930,
        month = oct,
       volume = {878},
        pages = {1-6},
       adsurl = {https://ui.adsabs.harvard.edu/abs/1930BHarO.878....1P}
}

@ARTICLE{Edlen:1932,
       author = {{Edlen}, B.},
        title = "{Highly ionised carbon, nitrogen, and oxygen in Wolf-Rayet stars}",
      journal = {The Observatory},
         year = 1932,
        month = apr,
       volume = {55},
        pages = {115-116},
       adsurl = {https://ui.adsabs.harvard.edu/abs/1932Obs....55..115E}
}

@ARTICLE{Beals:1933,
       author = {{Beals}, C.~S.},
        title = "{Classification and temperatures of Wolf-Rayet stars}",
      journal = {The Observatory},
         year = 1933,
        month = jun,
       volume = {56},
        pages = {196-197},
       adsurl = {https://ui.adsabs.harvard.edu/abs/1933Obs....56..196B}
}

@INPROCEEDINGS{Barlow:1982,
       author = {{Barlow}, M.~J. and {Hummer}, D.~G.},
        title = "{The WO Wolf-rayet stars.}",
    booktitle = {Wolf-Rayet Stars: Observations, Physics, Evolution},
         year = 1982,
       editor = {{De Loore}, C.~W.~H. and {Willis}, A.~J.},
       series = {IAU Symposium},
       volume = {99},
        month = jan,
        pages = {387-392},
       adsurl = {https://ui.adsabs.harvard.edu/abs/1982IAUS...99..387B}
}
%

\appendix

\section{Line luminosities}\label{sec:append}

Figures~\ref{fig:allopt} and \ref{fig:allUV} show the various optical and UV emission lines discussed in Sect.~\ref{sec:literature}. 

\begin{figure}[!h]\centering
	\includegraphics[width=\wi\columnwidth,page=18,angle=270]{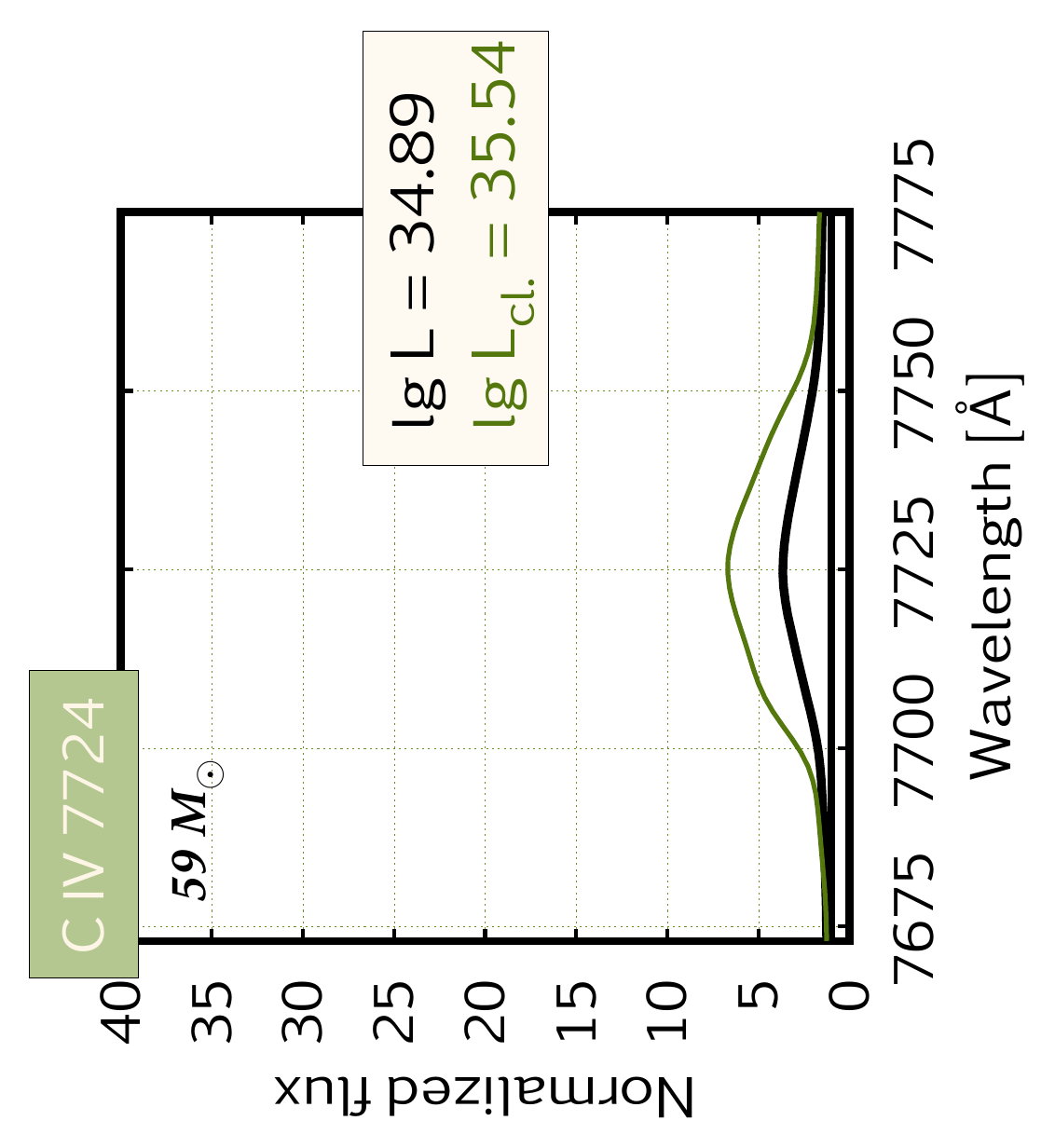}
	\includegraphics[width=\wi\columnwidth,page=18,angle=270]{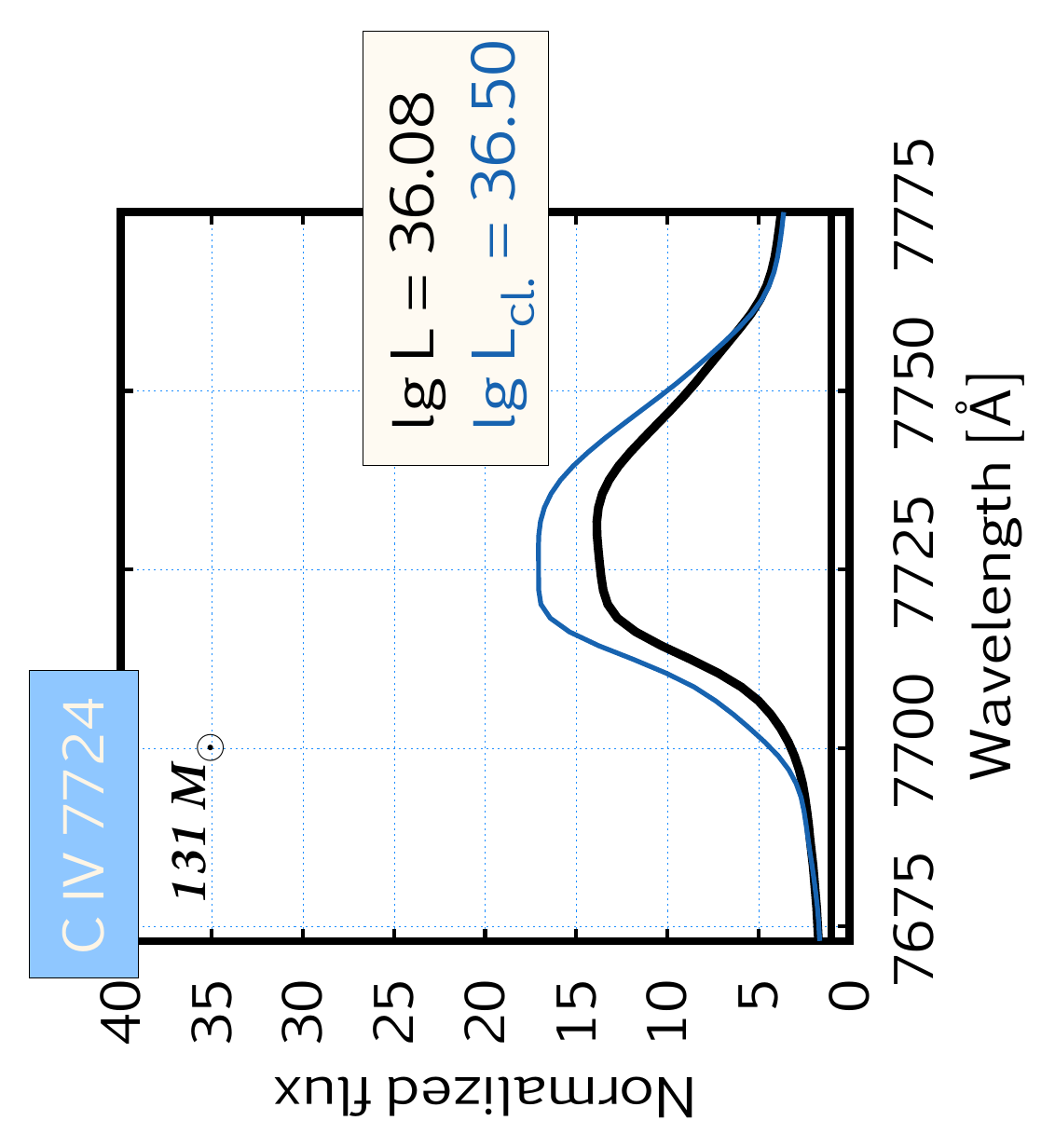}
	\includegraphics[width=\wi\columnwidth,page=17,angle=270]{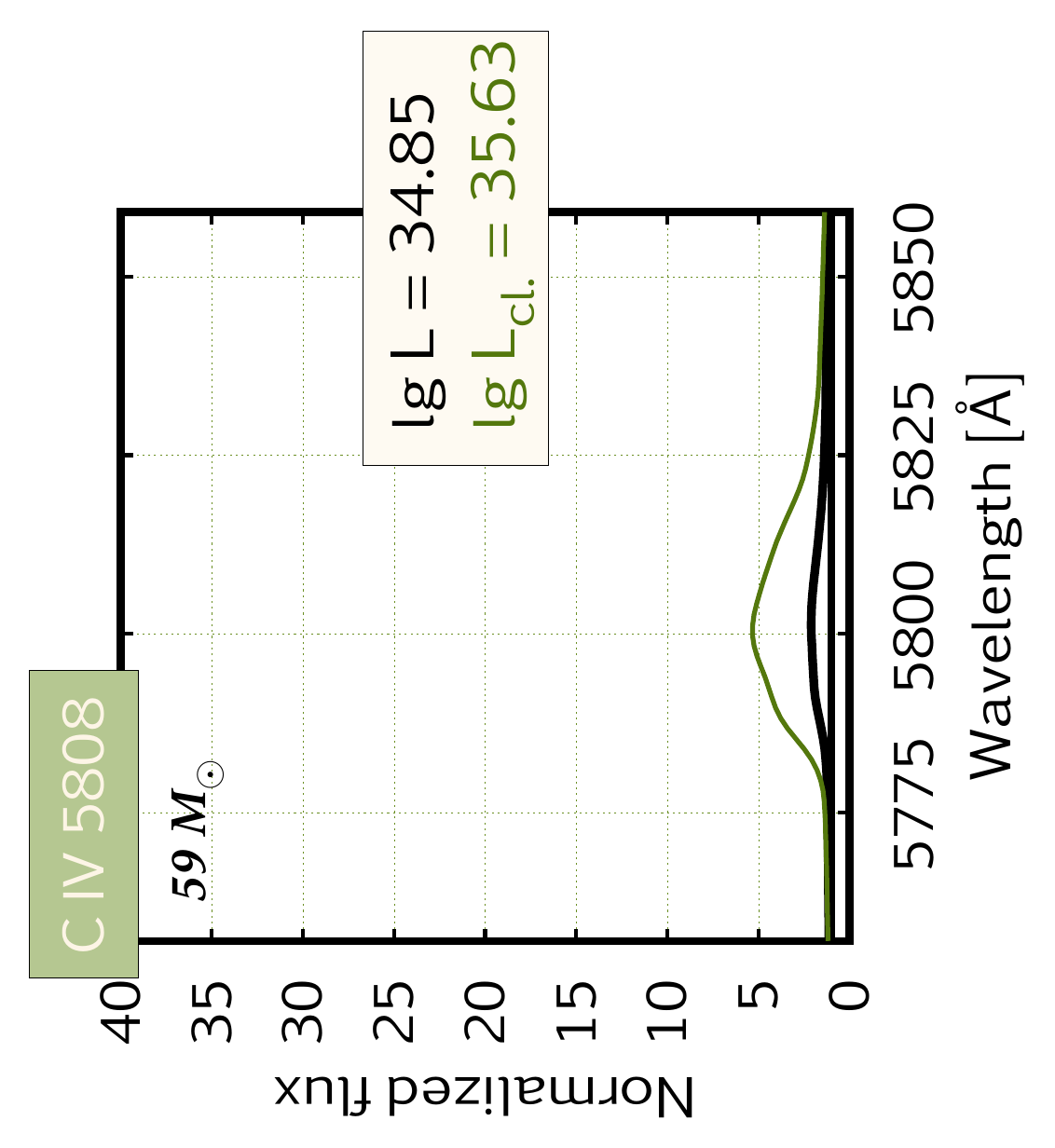}
	\includegraphics[width=\wi\columnwidth,page=17,angle=270]{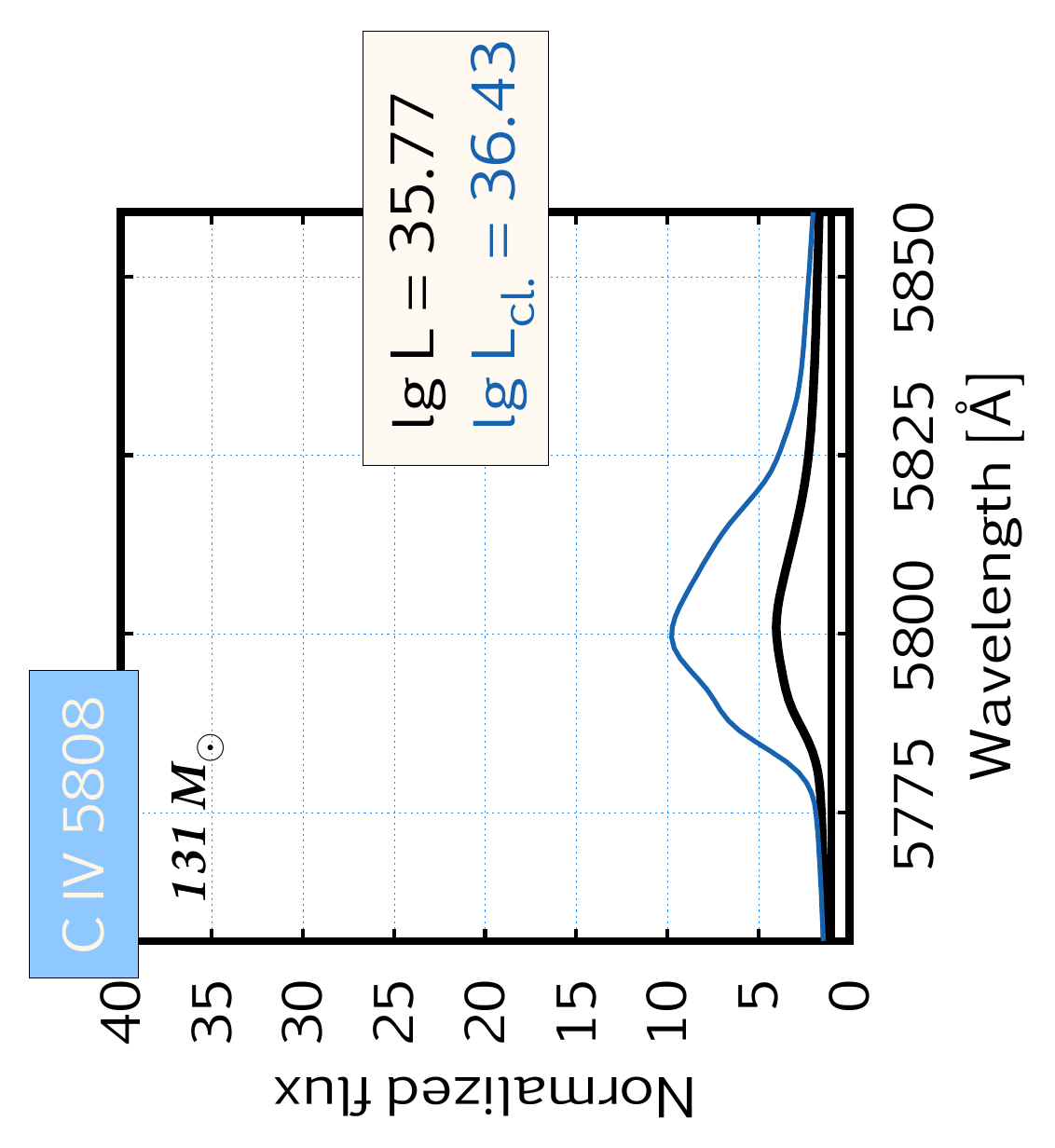}
	\includegraphics[width=\wi\columnwidth,page=14,angle=270]{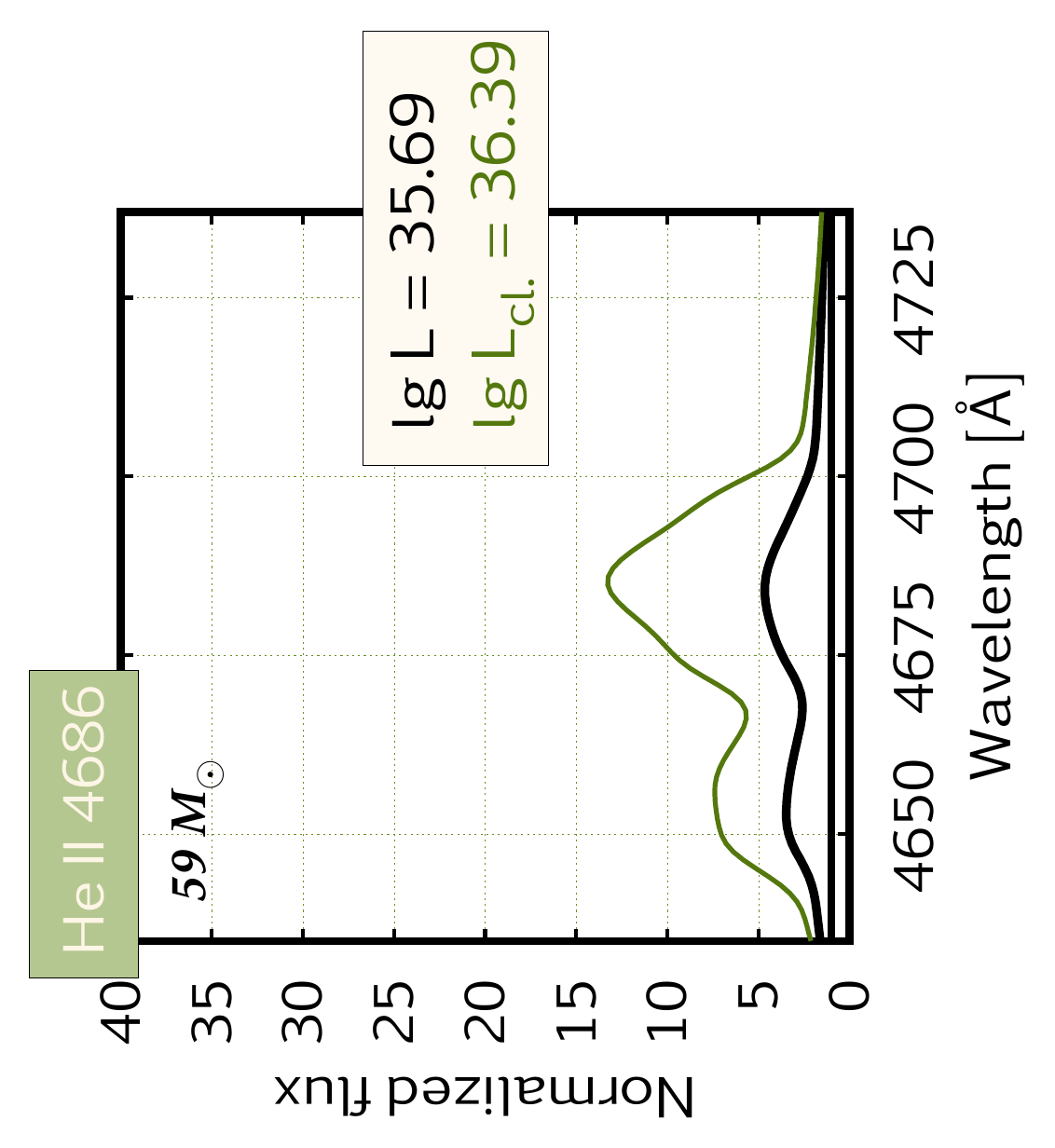}
	\includegraphics[width=\wi\columnwidth,page=14,angle=270]{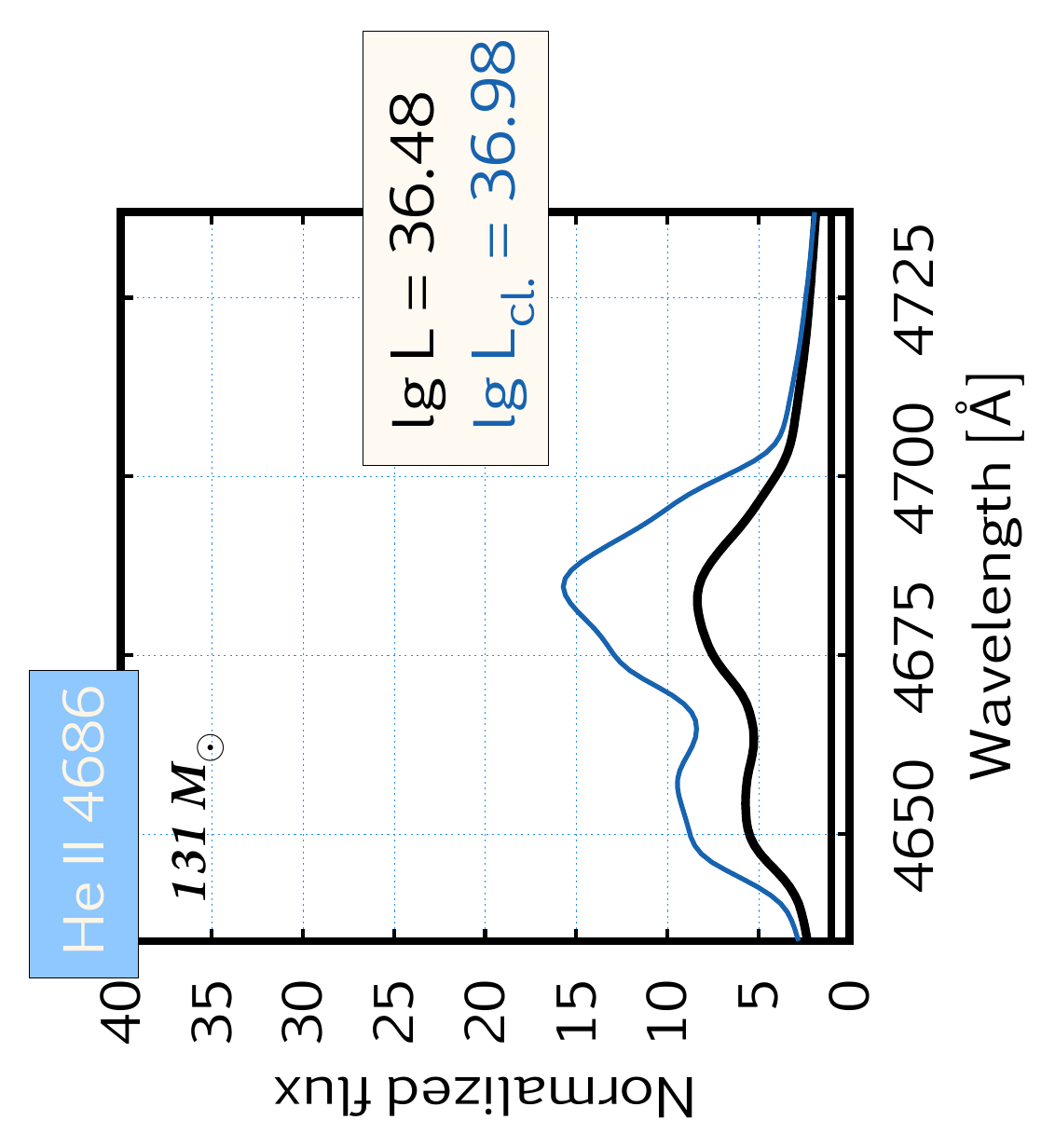}
	\includegraphics[width=\wi\columnwidth,page=9,angle=270]{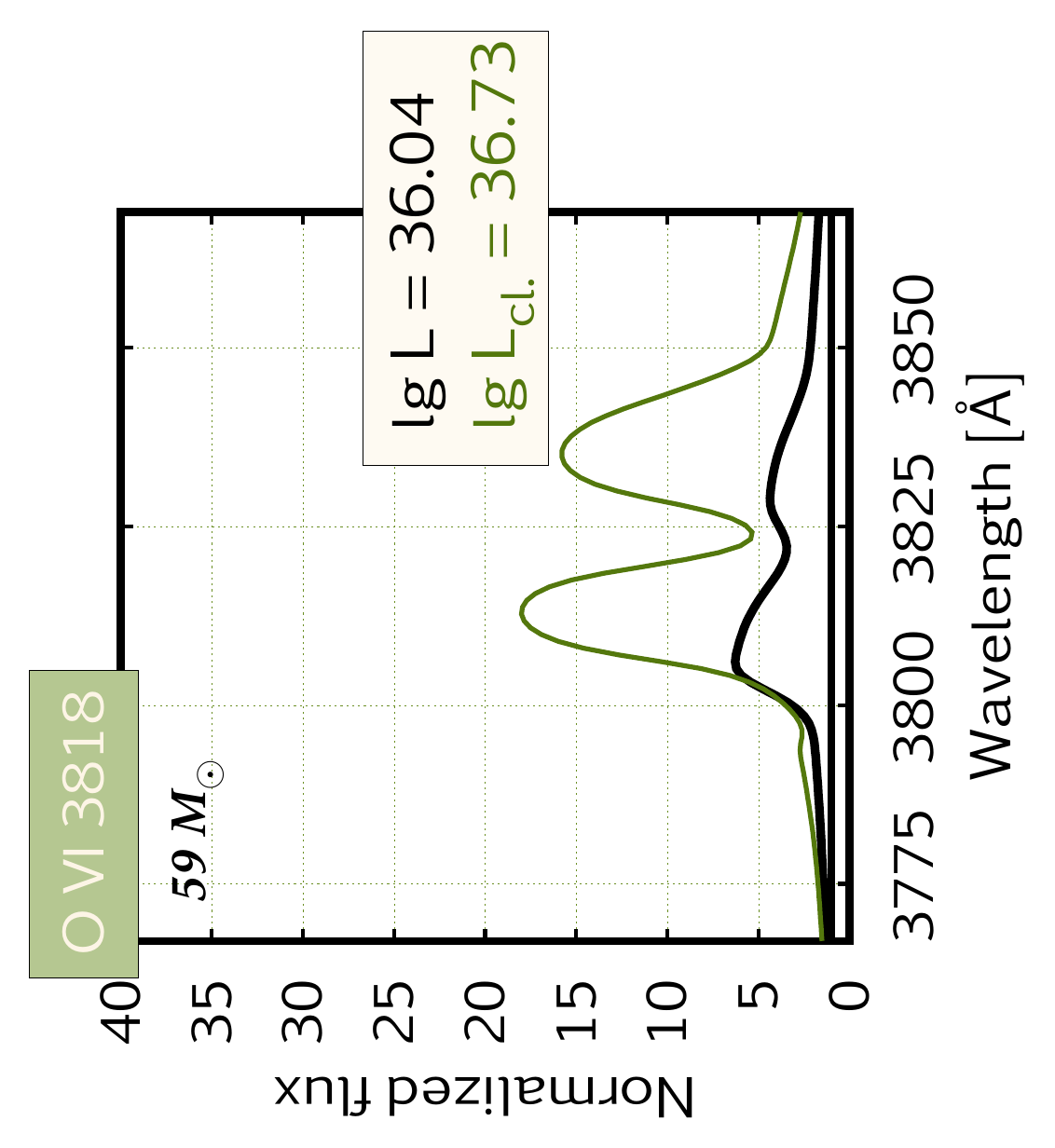}
	\includegraphics[width=\wi\columnwidth,page=9,angle=270]{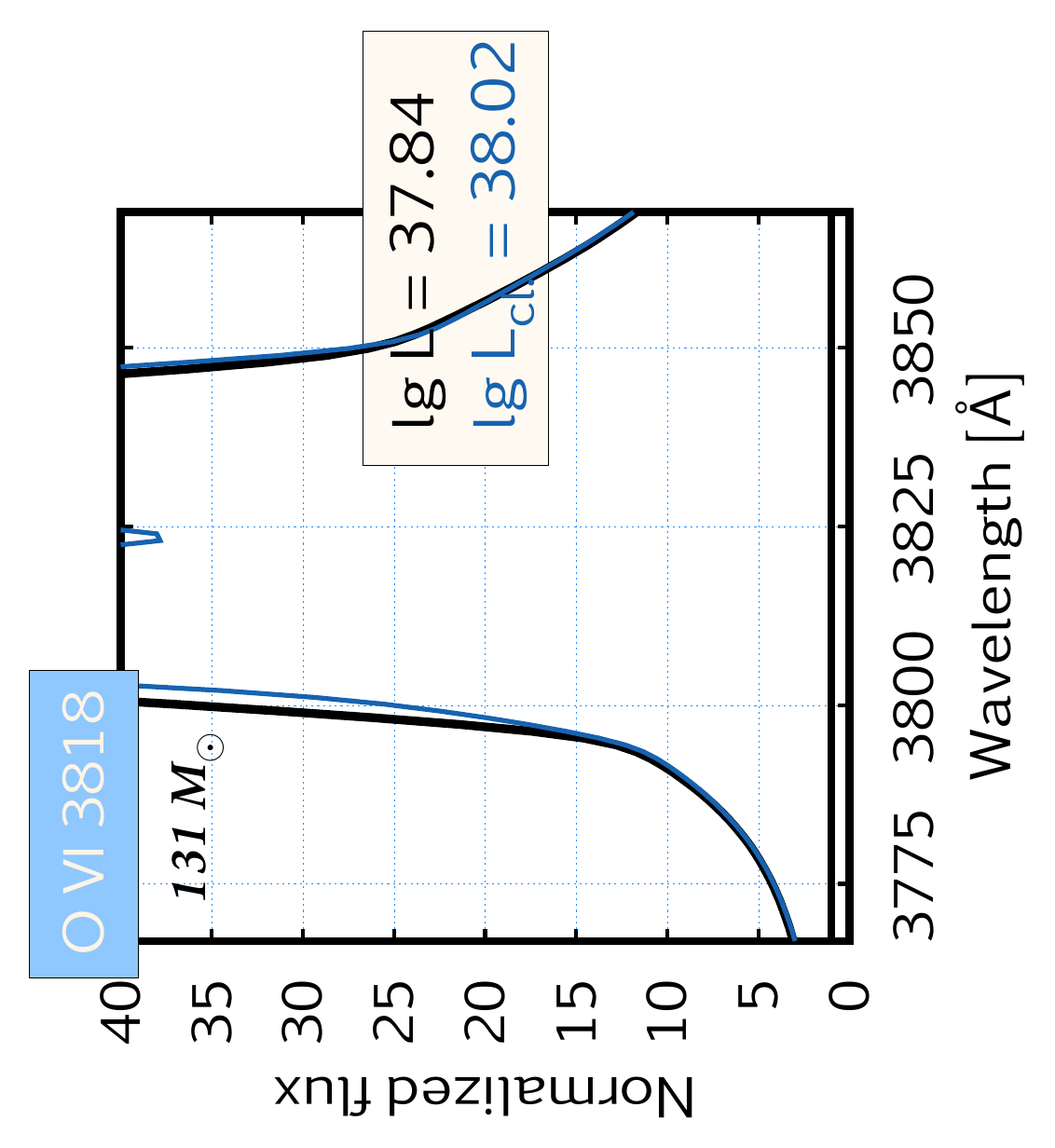}
	\includegraphics[width=\wi\columnwidth,page=8,angle=270]{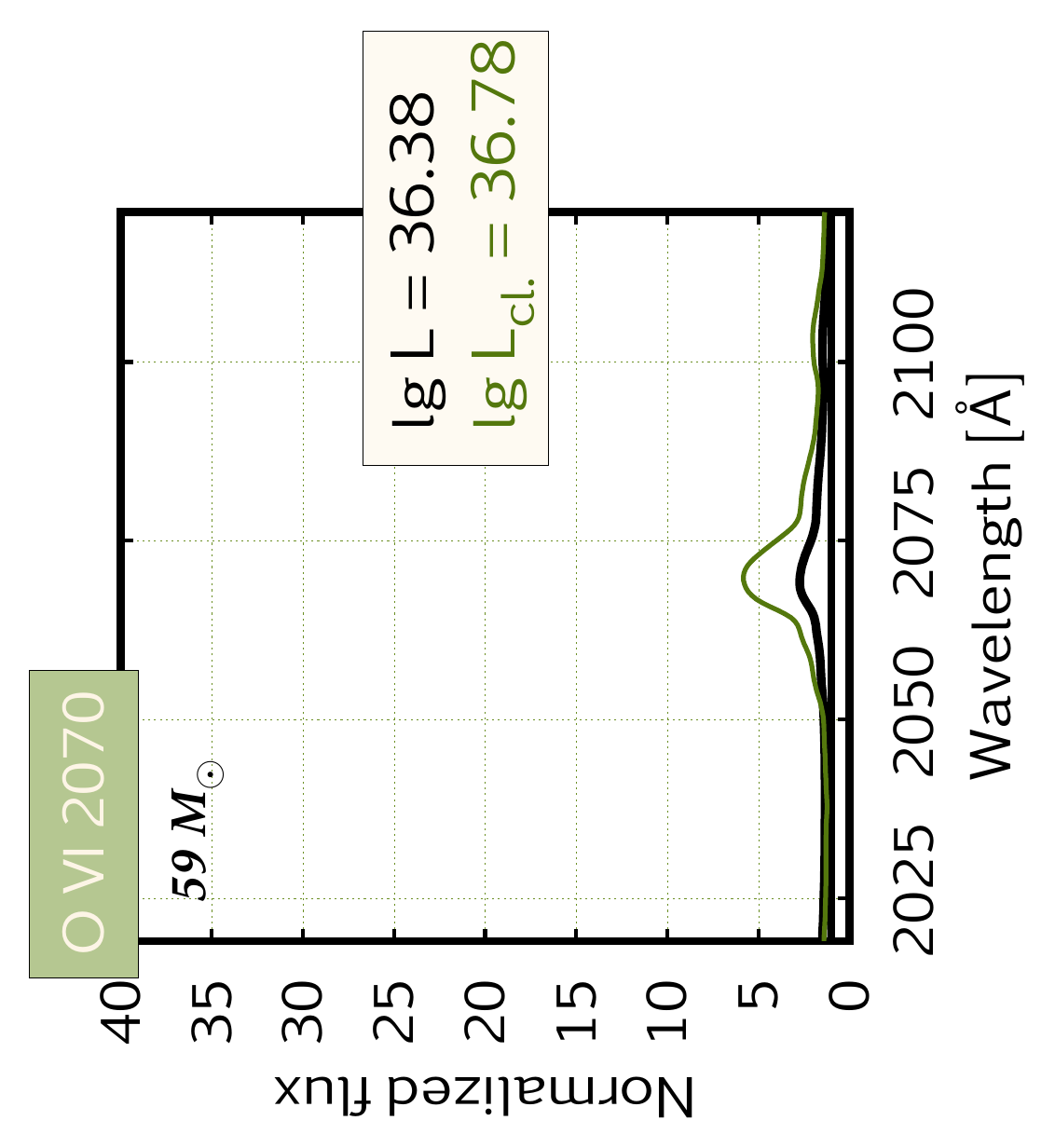}
	\includegraphics[width=\wi\columnwidth,page=8,angle=270]{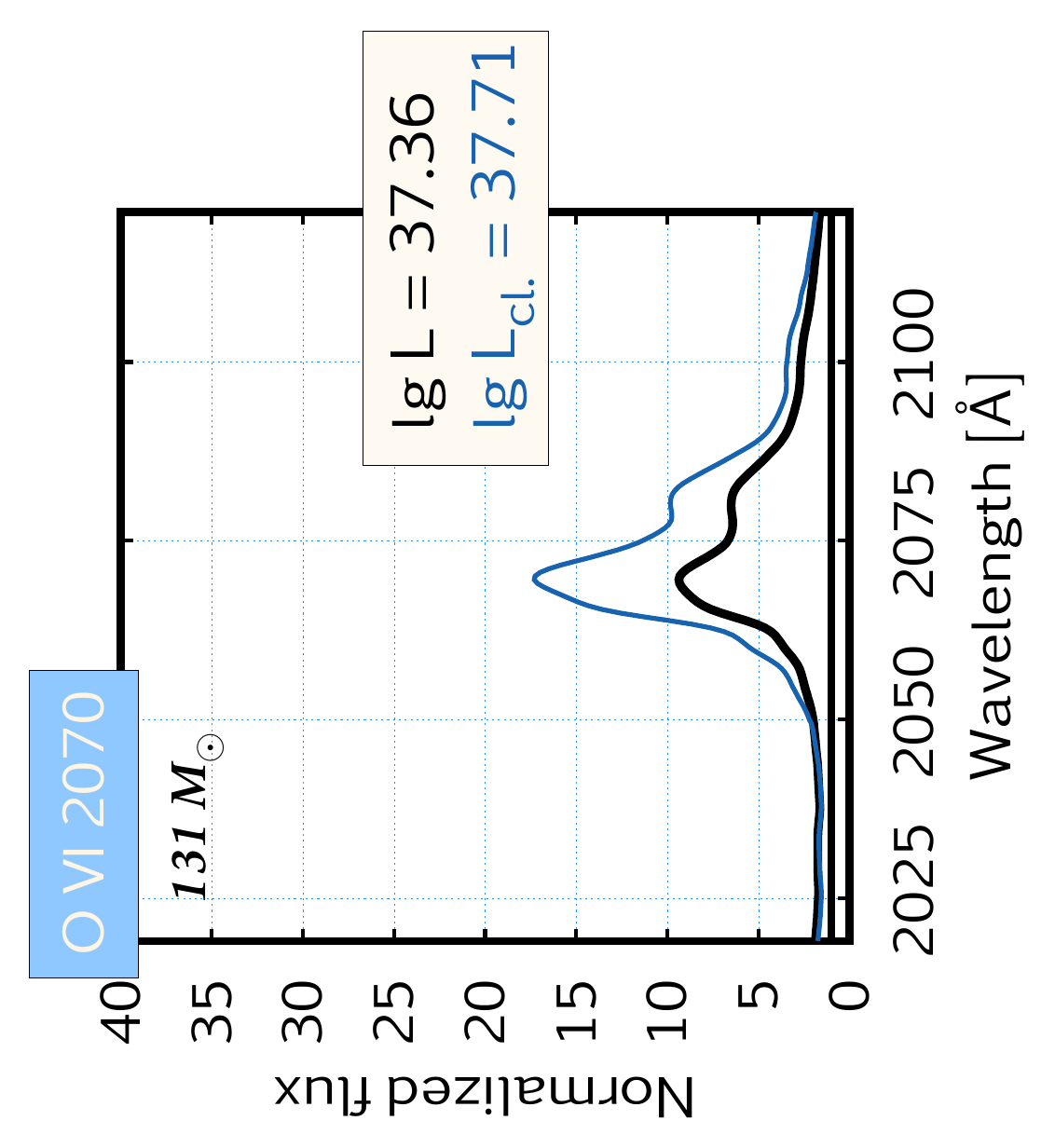} 
	\caption{Optical emission lines 
	predicted by our massive (M$_{ini}$~$=$~59~M$_{\odot}$) and very massive (M$_{ini}$~$=$~131~M$_{\odot}$), chemically homonegeously evolving, late-post-main-sequence models (classified as WO~stars in this phase by \citetalias{Kubatova:2019}; during the main-sequence evolutions, they were classified as early O~type stars). In our fiducial population (Sect.~\ref{sec:population}), one or two such very massive stars are present with $>$100~M$_{\odot}$; in our alternative population (Sect.~\ref{sec:main}), about six such stars with $\sim$40~M$_{\odot}$ and one with $\sim$80~M$_{\odot}$ are present, accounting for the measured \ion{He}{ii} ionizing flux.
	For the explanation of the legends, see Fig.~\ref{fig:Civ}. Note that we always apply the unclumped models' predictions (i.e. black line).}\label{fig:allopt}
\end{figure}

\begin{figure}\centering
	\includegraphics[width=\wi\columnwidth,page=7,angle=270]{pics/ok-59_pMS_D1_Md_4.7_Fe2_17}
	\includegraphics[width=\wi\columnwidth,page=7,angle=270]{pics/ok-131_pMS_D1_Md_4.23_Fe2_17}
	\includegraphics[width=\wi\columnwidth,page=5,angle=270]{pics/ok-59_pMS_D1_Md_4.7_Fe2_17}
	\includegraphics[width=\wi\columnwidth,page=5,angle=270]{pics/ok-131_pMS_D1_Md_4.23_Fe2_17}
	\includegraphics[width=\wi\columnwidth,page=4,angle=270]{pics/ok-59_pMS_D1_Md_4.7_Fe2_17}
	\includegraphics[width=\wi\columnwidth,page=4,angle=270]{pics/ok-131_pMS_D1_Md_4.23_Fe2_17}
	\includegraphics[width=\wi\columnwidth,page=3,angle=270]{pics/ok-59_pMS_D1_Md_4.7_Fe2_17}
	\includegraphics[width=\wi\columnwidth,page=3,angle=270]{pics/ok-131_pMS_D1_Md_4.23_Fe2_17}
	\includegraphics[width=\wi\columnwidth,page=2,angle=270]{pics/ok-59_pMS_D1_Md_4.7_Fe2_17}
	\includegraphics[width=\wi\columnwidth,page=2,angle=270]{pics/ok-131_pMS_D1_Md_4.23_Fe2_17}
	\includegraphics[width=\wi\columnwidth,page=1,angle=270]{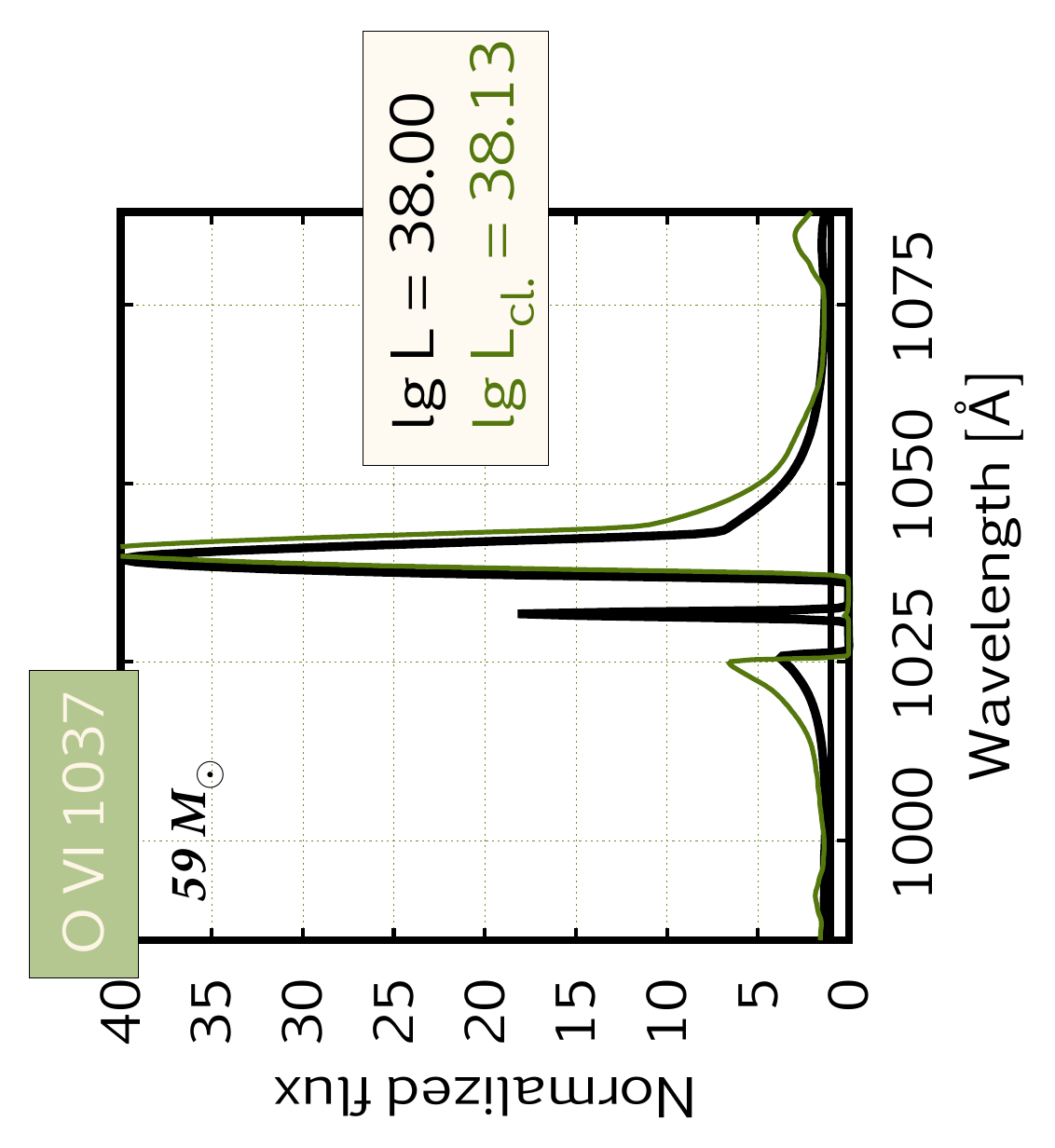}
	\includegraphics[width=\wi\columnwidth,page=1,angle=270]{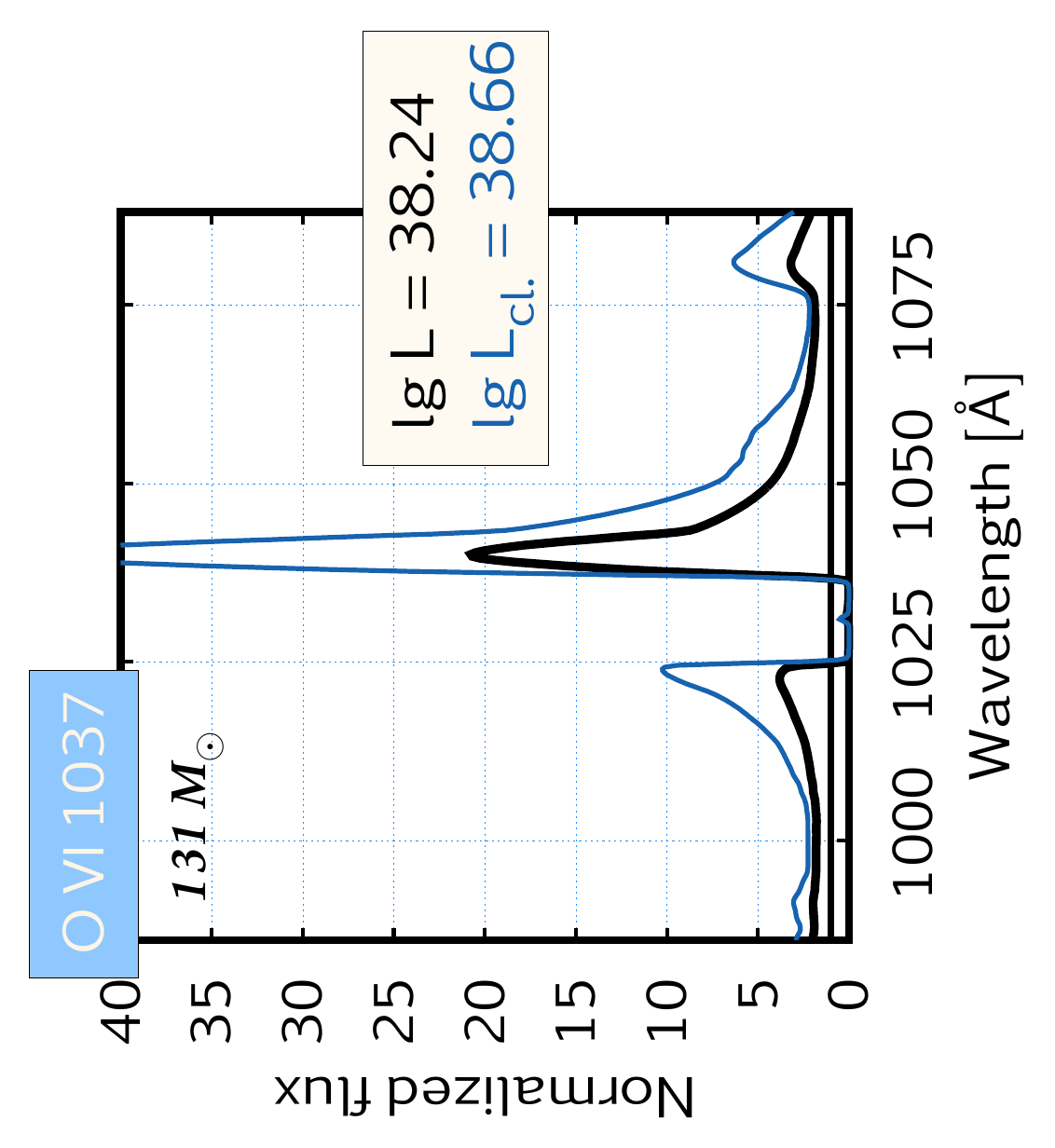}	
	\caption{UV emission lines predicted by our models. See Fig.~\ref{fig:allopt}. Note that we always apply the unclumped models' predictions (i.e. black line), {which is a conservative assumption}.
	}\label{fig:allUV}
\end{figure}

\onecolumn

\section{Ionizing flux predicted by our \PoWR\ stellar atmosphere models}

\begin{table}[h!]
\caption{Number of ionizing photons predicted by our models in the Lyman, \ion{He}{i} and \ion{He}{ii} continua, as well as the predicted hardness ratio I$_{4686}$/I$_{H\beta}$~$=$~A$\cdot$Q(\ion{He}{II})/Q(\ion{H}{I}), explained in Sect.~\ref{sec:hardness}. 
}\small
 \begin{tabular}{lccc|cccc|cccc}
 \hline\hline
 & & & & \multicolumn{4}{c|}{$D = 1$ (smooth wind)} & \multicolumn{4}{c}{$D=10$ (clumped wind, for reference)} \\ 
 star [M$_{\odot}$] & $\log \dot{M}$ & $\log T_{\mathrm{eff}}$ & $\log \frac{L}{L_\odot}$ & $\log Q_{\ion{H}{i}}$ & $\log Q_{\ion{He}{i}}$ & $\log Q_{\ion{He}{ii}}$ & I$_{4686}$/I$_{H\beta}$ & $\log Q_{\ion{H}{i}}$ & $\log Q_{\ion{He}{i}}$ & $\log Q_{\ion{He}{ii}}$ & I$_{4686}$/I$_{H\beta}$ \\ 
\hline
 20-0.28 [20] & -8.5 & 4.58 & 4.68 & 48.23 & 47.50 & 41.91 & $< 0.01$ & 48.18 & 47.32 & 42.03 & $< 0.01$ \\
 20-0.5 [20] & -7.8 & 4.65 & 4.97 & 48.70 & 48.20 & 43.88 & $< 0.01$ & 48.68 & 48.16 & 43.77 & $< 0.01$ \\
 20-0.75 [19.8] & -6.9 & 4.74 & 5.29 & 49.12 & 48.74 & 44.66 & $< 0.01$ & 49.11 & 48.73 & 44.66 & $< 0.01$ \\
 20-0.98 [19.2] & -5.8 & 4.88 & 5.58 & 49.45 & 49.19 & 46.87 & $< 0.01$ & 49.45 & 49.19 & 45.10 & $< 0.01$ \\
 20-pMS [16.8] & -5.5 & 5.08 & 5.67 & 49.47 & 49.32 & 48.52 & 0.20 & 49.50 & 49.34 & 48.61 & 0.22 \\
 59-0.28 [58.9] & -7.0 & 4.74 & 5.75 & 49.58 & 49.20 & 45.59 & $< 0.01$ & 49.57 & 49.20 & 45.25 & $< 0.01$ \\
 59-0.5 [58.7] & -6.7 & 4.79 & 5.94 & 49.80 & 49.46 & 46.26 & $< 0.01$ & 49.79 & 49.45 & 45.74 & $< 0.01$ \\
 59-0.75 [58.3] & -5.8 & 4.84 & 6.13 & 50.00 & 49.71 & 47.22 & $< 0.01$ & 49.99 & 49.70 & 46.79 & $< 0.01$ \\
 59-0.98 [55.3] & -4.7 & 4.92 & 6.29 & 50.08 & 49.94 & 49.31 & 0.30 & 50.10 & 49.95 & 49.24 & 0.24 \\
 59-pMS [49.4] & -4.7 & 5.14 & 6.34 & 50.06 & 49.94 & 49.44 & 0.42 & 50.09 & 49.95 & 49.32 & 0.30 \\
 131-0.28 [130.8] & -6.2 & 4.76 & 6.29 & 50.14 & 49.78 & 46.79 & $< 0.01$ & 50.14 & 49.77 & 46.24 & $< 0.01$ \\
 131-0.5 [129.9] & -5.9 & 4.79 & 6.42 & 50.28 & 49.95 & 47.19 & $< 0.01$ & 50.28 & 49.94 & 46.58 & $< 0.01$ \\
 131-0.75 [126.8] & -4.9 & 4.84 & 6.57 & 50.45 & 50.15 & 48.12 & $< 0.01$ & 50.44 & 50.15 & 46.10 & $< 0.01$ \\
 131-0.98 [112.5] & -4.27 & 4.93 & 6.69 & 50.53 & 50.32 & 49.23 & 0.09 & 50.53 & 50.32 & 48.95 & 0.05 \\
  131-e-pMS [106.7] & -4.09* & 5.14* & 6.70* & 50.42* & 50.32* & 49.80* & 0.42* & $-$ & $-$ & $-$ & $-$ \\
  131-m-pMS [101.8] & -4.16* &  5.14* &  6.69* & 50.42* & 50.31* & 49.79* & 0.41* & $-$ & $-$ & $-$ & $-$ \\
 131-(l)-pMS [93.3] & -4.23 & 5.14 & 6.68 & 50.41 & 50.26 & 49.71 & 0.35 & 50.46 & 50.27 & 49.55 & 0.21 \\ \hline
 \multicolumn{1}{c}{} & \multicolumn{9}{c}{Models with reduced mass loss, see \citetalias{Kubatova:2019} (for reference only, not used in population synthesis)} \\ \hline
 20-0.28 & -10.5 & \multicolumn{2}{c|}{\emph{same as above}} & 48.24 & 47.52 & 41.81 & $< 0.01$ & 48.19 & 47.34 & 40.84 & $< 0.01$ \\
 20-0.5 & -9.8 & & & 48.70 & 48.20 & 43.84 & $< 0.01$ & 48.68 & 48.16 & 42.41 & $< 0.01$ \\
 20-0.75 & -8.9 & & & 49.12 & 48.74 & 45.48 & $< 0.01$ & 49.11 & 48.73 & 43.75 & $< 0.01$ \\
 20-0.98 & -7.8 & & & 49.45 & 49.19 & 46.66 & $< 0.01$ & 49.44 & 49.19 & 46.09 & $< 0.01$ \\
 20-pMS & -7.5 & & & 49.48 & 49.32 & 48.48 & 0.17 & 49.49 & 49.32 & 48.41 & 0.14 \\
 59-0.28 & -9.0 & & & 49.58 & 49.20 & 45.67 & $< 0.01$ & 49.57 & 49.20 & 44.62 & $< 0.01$ \\
 59-0.5 & -8.7 & & & 49.80 & 49.46 & 46.13 & $< 0.01$ & 49.79 & 49.45 & 44.96 & $< 0.01$ \\
 59-0.75 & -7.8 & & & 49.99 & 49.71 & 46.71 & $< 0.01$ & 49.99 & 49.70 & 45.86 & $< 0.01$ \\
 59-0.98 & -6.7 & & & 50.11 & 49.96 & 49.19 & 0.21 & 50.13 & 49.96 & 49.11 & 0.17 \\
 59-pMS & -6.7 & & & 50.09 & 49.96 & 49.39 & 0.35 & 50.10 & 49.97 & 49.34 & 0.30 \\
 131-0.28 & -8.2 & & & 50.14 & 49.78 & 46.59 & $< 0.01$ & 50.14 & 49.77 & 46.44 & $< 0.01$ \\
 131-0.5 & -7.9 & & & 50.28 & 49.95 & 46.84 & $< 0.01$ & 50.28 & 49.94 & 45.34 & $< 0.01$ \\
 131-0.75 & -6.9 & & & 50.44 & 50.15 & 47.60 & $< 0.01$ & 50.44 & 50.14 & 46.75 & $< 0.01$ \\
 131-0.98 & -6.27 & & & 50.55 & 50.32 & 48.91 & 0.04 & 50.55 & 50.31 & 48.57 & 0.02 \\
 131-pMS & -6.23 & & & 50.43 & 50.30 & 49.74 & 0.36 & 50.44 & 50.31 & 49.71 & 0.32 \\ \hline
 \end{tabular}\label{tab:Q}
\tablefoot{Models marked with * are newly computed for this work. Population synthesis is performed on Q$_{\ion{H}{i}}$ and Q$_{\ion{He}{ii}}$ separately (see Sect.~\ref{sec:population}) to arrive at the final hardness ratio reported in Table~\ref{tab:popsyn} which is representative for the whole population.
For convenience, we list basic stellar parameters such as actual mass [M$_{\odot}$], mass-loss rate [M$_{\odot}$\,yr$^{-1}$], $\log T_{\mathrm{eff}}/K$ and $\log L/L_\odot$ (also to be found in \citetalias{Kubatova:2019}, except for the new models); these are taken from the evolutionary models and used as input for the \PoWR\ atmosphere code (see the HR~diagram in Fig.~\ref{fig:classification}). The sets with reduced mass loss and a high wind clumping are here for reference; in the population synthesis, we use the models with nominal mass-loss rates and smooth winds.}
\end{table}

\section{Stellar parameters of the new \PoWR\ models computed for this work}

\begin{table}[h]
	\caption{{Main parameters of the newly computed stellar atmosphere models, see Sect~\ref{sec:newmodel}.}}
	{\footnotesize 
		\begin{tabular}{lllllllllllllll}
			\hline\hline
			\rule[0mm]{0mm}{4.0mm}
			$\mini$ & label & $\log\Teff$ & $\log{\lstar}$
			& $\log\mdot$ & \YS & \YC & \ion{C}{} & \ion{O}{} & \ion{N}{} & \rstar & \mstar & $\log g$ &
			$v_\mathrm{rot}$ \\ 
			$\hzav{\msun}$ & & $\hzav{\Kelvin}$ & $\hzav{\lsun}$ &
			$\hzav{\msunyr}$ & & & & & & $\hzav{\rsun}$ & $\hzav{\msun}$ &
			$\hzav{\cmss}$ & $\hzav{\kms}$ \\  
			\hline
			131 & early-pMS & 5.14 & 6.70 & -4.09 & 0.99 & 0.80 & 1.97$\cdot10^{-4}$ & 2.17$\cdot10^{-5}$& 1.03$\cdot10^{-3}$ &  3.87 & 106.7 & 5.29 & 979 ($\sim$0.8\,v$_{K}$) \\
			131 & mid-pMS & 5.14 & 6.69 & -4.16 & 0.86 & 0.50 & 1.23$\cdot10^{-1}$ & 1.36$\cdot10^{-2}$& 2.30$\cdot10^{-3}$ &  3.82 & 101.8 & 5.28 & 793 ($\sim$0.4\,v$_{K}$) \\
			\hline
		\end{tabular}
	}
	\label{tab:newmodels}
\tablefoot{The columns $Y_{\mathrm{S}}$, C, N, and O show surface mass fractions of helium, carbon, nitrogen, and oxygen, respectively. v$_{K}$ denotes the Keplerian critical rotational velocity. Clumping factors are set at D~$=$~1. Both models are undergoing core-helium-burning (i.e., post-main-sequence evolution, see $Y_{\mathrm{C}}$). The parameters of all other models applied in this work are given in Table~1 of \citetalias{Kubatova:2019}.}
\end{table}

\end{document}